\documentclass[11pt,reprint,aps, pre]{revtex4-2}
\usepackage{graphicx}
\usepackage{amsmath,bm}
\usepackage{wrapfig}
\usepackage{cancel}
\usepackage{circuitikz}
\usepackage{hyperref}

\usepackage[capitalize]{cleveref}

\usepackage{xcolor}

\newcommand*\diff{\mathop{}\!\mathrm{d}}
\newcommand{\lD}{\lambda_{\mathrm{D}}}

\newcommand{\nn}{\nonumber\\}
\newcommand{\kB}{k}

\newcommand{\Laplace}[1]{\widehat{#1}}
\newcommand{\LaplaceNondim}[1]{\overline{#1}}
\newcommand{\ddx}{\partial_x}
\newcommand{\ddxnondim}{\partial_{\xnondim}}
\newcommand{\ddt}{\partial_t}
\newcommand{\nondim}[1]{\widetilde{#1}}
\newcommand{\ddtnondim}{\partial_{\tnondim}}

\newcommand{\xnondim}{\nondim{x}}
\newcommand{\snondim}{\nondim{s}}
\newcommand{\tnondim}{\nondim{t}}
\newcommand{\Lnondim}{\nondim{L}}

\newcommand{\psinondim}{\nondim{\psi}}
\newcommand{\jnondim}{\nondim{j}}
\newcommand{\cnondim}{\nondim{c}}
\newcommand{\smallparam}{\epsilon}

\newcommand{\nondimfirst}[1]{\tilde{#1}^{(1)}}
\newcommand{\rme}{\mathrm{e}}

\newcommand{\epsr}{\varepsilon_{r}}

\newcommand{\PhiAmp}{\nondim{\Psi}_\mathrm{amp}}
\newcommand{\PsiAmp}{\Psi_\mathrm{amp}}
\newcommand{\PsiBias}{\Psi_\mathrm{bias}}
\newcommand{\Ubias}{U_\mathrm{bias}}

\newcommand{\Uamp}{U_\mathrm{amp}}
\newcommand{\nup}{\nu}

\newcommand{\zeroth}{^{(0)}}
\newcommand{\first}{^{(1)}}
\newcommand{\RealPart}{\mathrm{Re}}
\newcommand{\ImagPart}{\mathrm{Im}}
\newcommand{\ImZ}{\ImagPart\,Z}
\newcommand{\ReZ}{\RealPart\,Z}

\newcommand{\omclin}{\omega_\mathrm{c}^0}
\newcommand{\omc}{\omega_\mathrm{c}}

\DeclareUnicodeCharacter{2212}{-}

\begin{document}

\title{The impedance of a charged flat-plate electric double-layer capacitor} 

\author{Adrian L. Usler}
\email{adrian.usler@nmbu.no}
\author{David Fertig}
\email{david.fertig@nmbu.no} 
\author{Mathijs Janssen}
\email{mathijs.a.janssen@nmbu.no}
\affiliation{Institute of Physics, Norwegian University of Life Sciences, \AA s, Norway}
\date{\today}

\begin{abstract}
We calculate the impedance of a flat-plate electric double-layer (EDL) capacitor by means of Finite Element Method simulations of modified Poisson--Nernst--Planck equations.
In Nyquist representation, the impedance spectra show a slanted line at intermediate frequencies if the capacitor is biased by a voltage, $\Ubias\neq 0$, or if the cation and anion diffusion coefficients differ, $D_-\neq D_+$.
By inspecting the concentration perturbations in the relevant frequency range, we confirm that the slanted line is in both cases related to ambipolar salt diffusion. 
On the basis of our impedance data, we disprove two previously made claims: 1) that the width $R_\mathrm{sl}$ of the slanted-line region represents an EDL resistance; and 2) that the slope $k_\mathrm{sl}$ of the slanted line is a measure of the ratio of the diffusion and charging time scales, $\tau_\mathrm{diff}/\tau_\mathrm{c}$. 
For the quantitative analysis of flat-electrode EDL capacitor impedance, we propose instead two equivalent circuits, for the cases \mbox{$D_-\neq D_+$} and \mbox{$\Ubias\neq 0$}.
These two cases give rise to antisymmetric and symmetric salt perturbations, which are best described by a Warburg short and Warburg open element, respectively.
From our circuit analysis, we obtain quantitative relations that link the Warburg prefactors to the ambipolar diffusion coefficient, the chemical capacitance of the bulk electrolyte, and the differential charge efficiency of the EDLs.
We thus provide a theoretical framework that explains why the width of the slanted line saturates at large biases, why it vanishes for large ion packing fractions, and how the system's overall capacitance is limited by the finite amount of ions in a closed system.
\end{abstract}

\maketitle

\section{Introduction\label{sec:introduction}}
\begin{figure}
    \centering
    \begin{flushleft}
        {\textsf{(a)}}
    \end{flushleft}\vspace{-2em}
    \includegraphics[width=.8\linewidth]{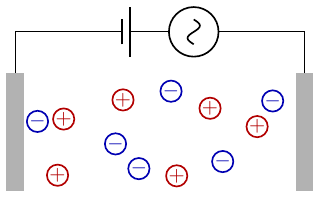}
    \begin{flushleft}
        {\textsf{(b)}}
    \end{flushleft}
    \includegraphics[width=\linewidth]{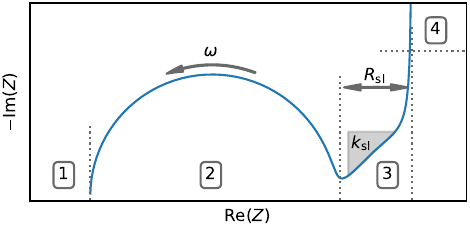}
    \caption{(a) Schematic of a flat-plate EDL capacitor subject to an oscillating voltage on top of a constant voltage bias. (b) Illustration of an impedance spectrum $Z(\omega)$ in Nyquist representation of a flat-electrode EDL capacitor subject to a voltage bias, showing four characteristic features, indicated here by numbers \textbf{1}--\textbf{4} and described in the introduction. Mei~\textit{et al.}~\cite{Mei_JPCC_2018} claimed that these four features are characteristic of EDL capacitors in general. The same shape of a Nyquist plot was found in flat-electrode simulations by Franceschetti and Macdonald~\cite{Franceschetti1979}. In these simulations, the slanted line (feature \textbf{3}) appeared if a voltage bias $\Ubias\neq 0$ was applied or if cation and anion diffusivities differed, $D_+\neq D_-$.}
    \label{fig:schematic_nyquist_four_features}
\end{figure}

Electric double layers (EDLs) that form at electrode--electrolyte interfaces underlie electrochemical devices for energy storage, capacitive mixing, and water desalinization.
Commercial EDL capacitors, also known as ultracapacitors or supercapacitors, employ porous electrodes to maximize energy and power densities~\cite{conway2013electrochemical}.
Conversely, theoretical~\cite{Macdonald1953,Franceschetti1979,Barbero2007, Mei_JPCC_2018, Barbero2018, Garcia2023,Garcia2025,Bazant_PRE_2004, Kilic_PRE_2007_2,janssen_pre_2018,Stout_PRE_2015,balu2018role,babel2018impedance,ma2022dynamic,pireddu2023frequency,pireddu2024impedance,barnaveli2024asymmetric,Palaia2025_edlc,Palaia2025_pnp,beunis_apl_2007,beunis2008dynamics} and experimental~\cite{beunis_apl_2007,beunis2008dynamics,Marzantowicz2008,nakamura2014structural,Kortschot2014,schuster2017electrochemical,schalenbach2021double,zhao_jpcc_2024,kutbay2025unveiling} studies of flat-electrode EDL capacitors provide fundamental insight into EDL formation.
Such flat-electrode capacitors, shown schematically in \cref{fig:schematic_nyquist_four_features}(a), are the focus of this article.

EDL capacitors charge by ion transport on various time and length scales: in and out of EDLs, through the bulk electrolyte, and, for porous electrodes,  through the porous structure.
The charging dynamics of an EDL capacitor can be probed by electrical measurements, among which electrochemical impedance spectroscopy~(EIS) \cite{Barsoukov, Orazem, Lasia} excels at separating processes by their different timescales.
Impedance is a complex-valued measure of a system's linear electrical response.
It is defined as 
\begin{equation}\label{eq:impedance}
    Z(s)=\frac{\Laplace{\delta U}(s)}{\Laplace{\delta I}(s)},
\end{equation}
where hats stand for Laplace transforms, defined as \mbox{$\Laplace{f}(s)\equiv \int_0^\infty f(t)\,\rme^{-s t}\diff t$}, and \mbox{$s\equiv\alpha+i\omega$} denotes complex frequency, most commonly considered with \mbox{$\alpha=0$}.
To measure $Z(\omega)$, one applies a small-amplitude voltage stimulus $\delta U(t)$, typically harmonic, and measures the current response $\delta I(t)$, or, vice versa, one applies a small-amplitude current $\delta I(t)$ and measures the voltage response $\delta U(t)$.
Either way, the amplitude of the stimulus must be sufficiently small to ensure linear response of the system around a steady reference state: $Z$ does not depend on the amplitude of the stimulus.
Conversely, $Z$ depends on the morphology, chemical composition, thermodynamic state and state of charge of the electrochemical system under consideration.
A system's behavior in different charging states can be probed by EIS if the voltage stimulus is applied on top of a constant bias voltage, $\Ubias$, so that \mbox{$U(t)=\Ubias+\delta U(t)$}~\cite{levie1967electrochemical,Ho_1980,lust2003electrochemical,lust2004influence,lust2004influence2,jurczakowski2004impedance,segalini2010qualitative,jokar2015synthesis,schalenbach2021double,liu2023nanoporous,zeng2024physical,hallemans2023electrochemical,li2026impedance}.
Unlike the amplitude $\Uamp$ of the stimulus $\delta U(t)$, $\Ubias$ need not be small.
Electrochemical impedance spectra at different $\Ubias$ can give insight, for instance, into a battery's state of health \cite{hallemans2023electrochemical}, or into the potential of zero charge of EDL capacitors \cite{liu2023nanoporous}.

In a much cited article, Mei~\textit{et al.}~\cite{Mei_JPCC_2018} analyzed impedance spectra of three EDL capacitors with porous electrodes.
In Nyquist representation (a parametric plot of $-\ImZ$ vs. $\ReZ$ for different $\omega$), these spectra exhibited four features, as illustrated in \cref{fig:schematic_nyquist_four_features}(b): \textbf{1}) a finite resistance in the high-frequency limit, \textbf{2}) a semi-circle at high frequencies, \textbf{3}) a slanted line at intermediate frequencies, and \textbf{4}) a vertical line at low frequencies.
To interpret these data, the authors calculated impedance spectra of a model flat-electrode EDL capacitor through continuum simulations of modified Poisson--Nernst--Planck (mPNP) equations \cite{Kilic_PRE_2007_2}.
The simulated spectra exhibited the same shape as the experimental spectra.
The authors claimed, therefore, that their simulations provided an unequivocal interpretation of experimental EDL capacitor impedance spectra.
In their interpretation, the four features represent \textbf{1})~the Ohmic resistance of the electrodes; \textbf{2})~ionic conduction through the bulk electrolyte; \textbf{3})~a competition of EDL formation and ion diffusion; and \textbf{4})~EDL formation.
There are two fundamental issues with the claimed generality of these interpretations:
first, the flat-electrode model missed an essential feature, the electrode porosity of the experimental system. 
Second, the simulations were performed for a single $\Ubias$, while the voltage bias used in the experiments, if one was applied at all, was not reported.

As to the first issue, impedance models specifically for porous electrodes are available in the literature~\cite{levie1967electrochemical,paasch1993theory}.
Electrical transmission lines, for instance, also give rise to a slanted line (region~\textbf{3}).
The transmission-line impedance is mathematically identical to the Warburg impedance, which is also commonly termed as ``diffusion impedance''~\cite{Huang2018}.
This similarity readily leads to misinterpretations of electromigrative transport through porous structures in terms of ion diffusion~\cite{Pedersen2023}.
Likewise, applying the interpretations of Mei~\textit{et al.} to measurements on porous electrodes is likely to lead to misinterpretations.
As to the second issue, we note that the impedance spectrum obtained at a single given \mbox{$\Ubias\neq 0$} is not representative of the impedance at a different $\Ubias$, particularly not at \mbox{$\Ubias=0$}.
This issue is the focus of this work.

The impedance of flat-electrode EDL capacitors was analyzed extensively in a little-recognized simulation work by Franceschetti and Macdonald~\cite{Franceschetti1979}, who found a slanted line (region~\textbf{3}) when a non-zero $\Ubias$ was applied, or when cation and anion diffusivities differed, $D_+\neq D_-$.
Only for both \mbox{$\Ubias=0$} and \mbox{$D_+= D_-$} was the slanted line absent, while features \textbf{1}, \textbf{2}, and \textbf{4} of \cref{fig:schematic_nyquist_four_features} remained.
Later numerical and analytical work also found region~\textbf{3} in cases with $\Ubias\neq 0$~\cite{Barbero2018,Garcia2023,Garcia2025} or $D_-\neq D_+$~\cite{Lelidis2005,Barbero2007}.
Approximations for the low-frequency impedance of a single flat electrode in contact with an electrolyte have also been derived analytically~\cite{Gunning1995,DeLacey1982}, but they agree neither with the analytical solution by Macdonald~\cite{Macdonald1953} nor with the numerical studies~\cite{Franceschetti1979,Barbero2007,Barbero2018,Mei_JPCC_2018,Garcia2023,Garcia2025}, and hence, we do not consider them further.

Most of the mentioned studies associate region~\textbf{3} with diffusion~\cite{Franceschetti1979, Barbero2007, Mei_JPCC_2018, Garcia2025}.
Franceschetti and Macdonald~\cite{Franceschetti1979} based their interpretation on the similarity of region~\textbf{3} to a Warburg impedance; both Mei \textit{et al.}~\cite{Mei_JPCC_2018} and López-García \textit{et al.}~\cite{Garcia2025} based their interpretation on the $L^2/D$ time scale of the process; Barbero and Lelidis pointed out that the process in the case $D_-\neq D_+$ is governed by the ambipolar diffusion coefficient~\cite{Barbero2007}.

The width $R_\mathrm{sl}$ and slope $k_\mathrm{sl}$ of region~\textbf{3} [see \cref{fig:schematic_nyquist_four_features}(b)] were suggested by Mei \textit{et al.} to be characteristic for the charging process.
They found a large $k_\mathrm{sl}$ (steep slope) when the diffusion time scale exceeded the time scale of charging through ohmic conduction, and a small $k_\mathrm{sl}$ (shallow slope) in the opposite case, 
and thus correlated $k_\mathrm{sl}$ with the ratio of time scales for diffusion and ohmic conduction.
We note that this correlation cannot hold general as, for \mbox{$\Ubias=0$}, the semicircle is followed by a vertical line, corresponding to $k_\mathrm{sl}\to\infty$, despite a finite ratio of time constants.
Mei \textit{et al.} further claimed that $k_\mathrm{sl}$ indicates whether the charging process is governed by EDL formation or by ion diffusion.
This interpretation is unjustified as a slow process need not be secondary.
The width $R_\mathrm{sl}$, in turn, was claimed by Mei \textit{et al.} to represent an EDL resistance.
This interpretation is in apparent conflict with the mentioned interpretations of region~\textbf{3} in terms of bulk ion diffusion in Refs.~\cite{Franceschetti1979, Barbero2007, Mei_JPCC_2018, Garcia2025}.
Moreover, it is not plausible that an EDL resistance should at all be observable in the impedance spectrum:
the EDL regions are typically much thinner than the bulk electrolyte, and more conductive, owing to a net accumulation of ions.

$R_\mathrm{sl}$ was further investigated in two studies by López-García \textit{et al.}~\cite{Garcia2023,Garcia2025},
who studied a range of relatively large $\Ubias$.
In this range, $R_\mathrm{sl}$ was virtually independent of $\Ubias$~\cite{Garcia2023} for open systems, that is, systems with a fixed bulk salt concentration.
Conversely, in closed systems (with a fixed amount of salt), $R_\mathrm{sl}$ increased strongly with increasing $\Ubias$, and so did the bulk-electrolyte resistance~\cite{Garcia2023}.
In such closed systems, EDL formation drives the salt concentration in the bulk electrolyte down, because the EDLs absorb a net amount of salt.
This coupling between the electrode potential and the bulk salt concentration will affect the impedance spectrum, and its effects will inherently be missed by an open-system approach.
In simulations of closed systems, in turn, the $\Ubias$-dependence of salt concentration in the bulk electrolyte makes it difficult to disentangle $\Ubias$-dependent changes in an impedance spectrum caused by changes in the EDLs from those caused by changes in the bulk electrolyte.

In summary, existing literature on the impedance of flat-electrode EDL capacitors contradict each other on its interpretation, and several open problems and questions remain: 
\mbox{(i) it} has not been conclusively demonstrated that region~\textbf{3} is, in the case $\Ubias\neq 0$, indeed due to diffusion, and it has not been explained how diffusion arises in this case;
\mbox{(ii) the} interpretation of $R_\mathrm{sl}$ as an EDL resistance is unjustified and implausible.
\mbox{(iii) the} correlation between $k_\mathrm{sl}$~\cite{Mei_JPCC_2018} and the ratio of diffusion and ohmic-conduction time constants has only been demonstrated in cases with \mbox{$D_-=D_+$}.
\mbox{(iv) in} closed systems, the influences of changing EDLs and a changing bulk salt concentration upon changing $\Ubias$ have not yet been disentangled.

To address these problems and questions, in this study, we re-analyze the impedance of flat-electrode EDL capacitors under a voltage bias through continuum Finite Element Method simulations.
To address problem (iv), we use a mixed grand canonical--canonical approach: for each $\Ubias$ we pre-equilibrate the simulation cell against a virtual salt reservoir, and then close the system for the subsequent impedance calculation.
This approach gives a clear view on the influence of changing EDL configurations on the impedance, unobstructed by $\Ubias$-dependent changes in the bulk electrolyte, while still including the effects of a finite amount of salt that would be missed by an open-system approach.
In nondimensional form, the governing equations of our system contain four dimensionless parameters, characterizing the applied voltage, electrode separation, diffusivity ratio, and ionic packing fraction.
We determine impedance spectra for a wide range of these parameters, giving a clearer and more systematic characterization than is possible by varying the nine dimensional parameters of our system separately.
Unlike earlier studies~\cite{Franceschetti1979, Barbero2007, Mei_JPCC_2018, Barbero2018, Garcia2023, Garcia2025}, we analyze the perturbations of ion-concentration profiles, $\Laplace{c}_\pm\first$, at different frequencies that are associated with characteristic points in the impedance spectrum.
With this large data set at hand, we address questions \mbox{(i)--(iii)}.

This paper is structured as follows:
in \cref{sec:model}, we introduce our model setup and governing equations and the numerical methods giving access to its impedance.
\Cref{sec:characteristic_quantities} introduces characteristic resistances, capacitance, timescales, and analytical impedance expressions, which we use to analyze our numerical impedance spectra, presented in \cref{sec:results}.
\Cref{sec:equivalent_circuit} presents equivalent circuits that fit the model's impedance data well.
We discuss how these circuits help with the interpretation of observed trends in the differential capacitance of the EDL capacitor.
The paper ends with further discussion in \cref{sec:discussion} and conclusions in \cref{sec:conclusions}.

\section{Model}\label{sec:model}
Our numerical simulations comprise two steps: first, we simulate equilibrium EDLs for a given applied bias voltage $\Ubias$; second, we solve linearized and Laplace-transformed mPNP equations, thus giving access to the system's response in the frequency domain, from which we calculate the impedance.

\subsection{Setup, transport equations, and boundary conditions in the time domain\label{sec:transport_equations}}
We consider a 1:1 electrolyte at a constant temperature $T$ between two parallel, flat electrodes.
Quantities pertaining to the cations and anions have subscripts $+$ and $-$, respectively, so ionic concentrations are denoted by $c_+$ and $c_-$, respectively.
We use a Cartesian coordinate system with the coordinate $x$ running perpendicular to the electrodes located at  \mbox{$x=\pm L$}.
Variations of $c_\pm$ parallel to the electrodes are not considered; this simplification reasonably describes cases in which the electrodes' surface area $A$ is large compared to the separation $2L$. 
For simplicity, we consider the  permittivity $\epsr\varepsilon_0$ to be constant, cations and anions to have the same effective radius $a$, and the cation and anion diffusivities $D_+$ and $D_-$ to be constant.

Neglecting electroconvection \cite{ratschow2025convection}, we model the electrostatic potential $\psi(x,t)$, ionic fluxes $j_\pm(x,t)$, and concentrations $c_\pm(x,t)$ in the above setup through the mPNP equations~\cite{Kilic_PRE_2007_2},
\begin{subequations}\label{eq:mPNP}
    \begin{align}
        \label{eq:poisson}
         \partial_x^2 \psi &= -\frac{e}{\epsr\varepsilon_0} (c_+ - c_-),\\
        \label{eq:continuity}    
          \ddt c_\pm &= -\ddx j_\pm,\\
        \label{eq:mpnp_fluxes}
        -j_\pm &= D_\pm \bigg[\ddx c_\pm \pm e\beta c_\pm\,\ddx\psi + \frac{a^3 c_\pm \ddx\left(c_+ + c_-\right)}{1-a^3 c_+-a^3 c_-}\bigg],
    \end{align}
\end{subequations}
where \cref{eq:poisson} is the Poisson equation, \cref{eq:continuity} is the continuity equation,
and \mbox{$\beta=1/(\kB T)$} is the inverse thermal energy, with $\kB$ the Boltzmann constant.
The first and second terms in the ionic fluxes [\cref{eq:mpnp_fluxes}] represent the regular Nernst--Planck expressions for diffusion and electromigration; the third term, introduced by Kilic~\emph{et al.}, accounts for finite ion size~\cite{Kilic_PRE_2007_2}. 

We consider the electrodes to be blocking, giving no-flux boundary conditions,
\begin{equation}\label{eq:noflux_bc}
    j_+(\pm L,t) = j_-(\pm L,t) = 0.
\end{equation}    
A time-varying voltage \mbox{$U(t)=\Ubias+\delta U(t)$} is applied to the electrodes, so
\begin{equation}\label{eq:potential_bcs}
    \psi(\pm L,t) = \pm\frac{U(t)}{2} =: \pm\Psi(t),
\end{equation}
where we have introduced the electrode potential \mbox{$\Psi(t)=\PsiBias+\delta\Psi(t)$}.
Note that we consider the potential to be set immediately at the electrode--electrolyte interface: our model does not account for electrode resistance or a Stern layer.
These effects add to the system's impedance through a series connection of a resistor and a capacitor, as demonstrated in \cref{sec:Stern_resistance} of the Supplemental Information (SI)~\cite{supplemental_information}.

\subsection{Equilibrium EDL configurations}\label{sec:equilibrium}
Before we can determine the system's impedance, we must find equilibrium ion concentration profiles for the given $\PsiBias$---that is, the reference profiles around which the potential and ion concentrations are perturbed by the small-amplitude voltage stimulus.
In electrochemical equilibrium, \cref{eq:mPNP,eq:potential_bcs} reduce to the modified Poisson--Boltzmann (mPB) equations~\cite{Borukhov1997,kornyshev2007double,Kilic_PRE_2007_1},
\begin{subequations}\label{eq:modified_poisson_boltzmann}
\begin{align}
    \label{eq:mpb_equation}
    \partial_{x}^2\psi &= \frac{2 e c_0}{\epsr\varepsilon_0} \frac{\sinh(e\beta\psi)}{1 + 2\nup\,\sinh^2(e\beta\psi/2)},\\
    \psi(\pm L) &= \pm\PsiBias.
\end{align}
\end{subequations}
For $2\nup\sinh^2{(e\beta\PsiBias/2})\ll 1$, steric exclusion between ions is negligible, and \cref{eq:mpb_equation} reduces to the Poisson--Boltzmann equation.
In \cref{eq:modified_poisson_boltzmann}, $c_0$ is the salt concentration of a virtual salt reservoir with which our setup is in electrochemical equilibrium.
That is, \cref{eq:modified_poisson_boltzmann} describes the EDLs in a grand-canonical ensemble.
When equilibrium profiles are calculated in this way from \cref{eq:modified_poisson_boltzmann}, the salt concentration outside the EDLs does not depend on $\PsiBias$.
Conversely, the reference equilibrium profiles can also be obtained by propagating initial concentration profiles \mbox{$c_\pm(x, 0)=c_0$} towards equilibrium using \cref{eq:mPNP}.
In this approach, taken in Refs.~\cite{Franceschetti1979,Garcia2023,Garcia2025}, EDL formation drives the salt concentration outside the EDLs down, because EDLs absorb a net amount of salt while the total number of ions of each species is conserved in the simulation domain.
As a consequence, within this approach the bulk electrolyte becomes more resistive with increasing $\PsiBias$, and the electrostatic screening length increases.
Here, we do the pre-equilibration grand-canonically, and the impedance calculation (described below) canonically.
This approach disentangles influences that are purely due to a changing EDL configuration from the mentioned variations in Debye length or bulk electrolyte resistance, and thus to get a clear view on how changes in the EDLs affect the impedance spectrum, and the slanted line (region~\textbf{3}) in particular.

An exact solution to \cref{eq:modified_poisson_boltzmann} is also an equilibrium solution of \cref{eq:mPNP}, since both theoretical frameworks derive from the same electrochemical potential. 
However, when we insert our numerical solutions into \cref{eq:mpnp_fluxes}, we observe small residual ion fluxes, which are simulation artifacts.
Since these residual fluxes are unrelated to the applied voltage perturbation $\delta U(t)$, they would compromise the calculation of the system's impedance.
To remove these residual fluxes, we use the equilibrium profiles obtained from \cref{eq:modified_poisson_boltzmann} as initial conditions for a subsequent mPNP calculation [\cref{eq:mPNP}] for the same applied $\PsiBias$.
This step affects the bulk concentration, but we verified for each parameter set that this change was not substantial.

\subsection{Calculating impedance spectra\label{sec:impedance_spectra}}
We obtain linearized transport equations by asymptotic expansions of the profiles $\psi(x)$ and $c_\pm(x)$ in the small parameter \mbox{$\smallparam:=e\beta\PsiAmp$}, with \mbox{$\PsiAmp:=\mathrm{max}\bm{(}\delta\Psi(t)\bm{)}$} being the stimulus amplitude.
Specifically, we take the pre-equilibrated profiles at $\PsiBias$ as a reference, around which we expand $\psi(x)$ and $c_\pm(x)$ as
\begin{align}
    \begin{aligned}
    \psi(x, t) &= \psi\zeroth(x) + \smallparam\,\psi\first(x,t) + \mathcal{O}(\smallparam^2),\\
    c_\pm(x, t)&= c_\pm\zeroth(x) + \smallparam\,c_\pm\first(x,t) + \mathcal{O}(\smallparam^2),
    \end{aligned}
    \label{eq:phi_n_expansions}
\end{align}
where the superscript $^{(n)}$ indicates the $n^\mathrm{th}$-order term of an expanded profile.
$\psi\zeroth(x)$ and $c_\pm\zeroth(x)$ are the equilibrium profiles at a given $\PsiBias$ (and \mbox{$\PsiAmp=0$}), described in \cref{sec:equilibrium}.
The linear response at sufficiently small $\epsilon$ is described by the first-order terms.
This is the regime for which the impedance is defined according to \cref{eq:impedance}.
We insert \cref{eq:phi_n_expansions} into the mPNP equations \eqref{eq:mPNP} and its boundary conditions \eqref{eq:noflux_bc} and \eqref{eq:potential_bcs} and, by taking partial derivatives with respect to $\PsiAmp$, obtain first-order perturbation equations [\cref{eq:nondim_firstorder_poisson,eq:nondim_firstorder_flux,eq:nondim_firstorder_boundary_conditions} of the SI~\cite{supplemental_information}]. 
As in Refs.~\cite{Franceschetti1979, Stout_PRE_2015, Song_PRE_2019}, Laplace transformations then bring the linearized equations from the time to the frequency domain, yielding
\begin{align}\label{eq:mpnp_first_laplace}
    \begin{aligned}
    \ddx^2\Laplace{\psi}\first &= -\frac{e}{\epsr\varepsilon_0}\left(\Laplace{c}_+\first - \Laplace{c}_-\first\right),\\
    s\,\Laplace{c}_\pm\first -\cancel{c_\pm\first(t=0)} &= -\ddx\Laplace{j}_\pm\first,
    \end{aligned}
\end{align}
where $j_\pm\first$ denotes the first-order perturbation of the fluxes $j_\pm$, and \mbox{$c_\pm\first(t=0)=0$} by definition.
The full form of the linearized, Laplace-transformed mPNP equations is specified in \cref{eq:pnp_lp_real_imag} in the SI~\cite{supplemental_information}. 

Next, we evaluate and Laplace-transform the first-order contributions to the no-flux boundary condition \cref{eq:noflux_bc},
\begin{align}
    \Laplace{j}_\pm\first(\pm L,s) &= 0
    \label{eq:mpnp_noflux_bc_first_laplace},
\end{align}
and the boundary condition \eqref{eq:potential_bcs} for the applied potential, 
\begin{align}
   \Laplace{\psi}\first(\pm L, s) &= \pm \Laplace{\delta\Psi}(s). 
   \label{eq:mpnp_bcs_first_laplace}
\end{align}
So far, we have not specified the voltage stimulus.
In the frequency domain, we choose $\Laplace{\delta\Psi}(s)=const.$, such that perturbations in Laplace space are equal-valued for all frequencies, giving good numerical stability.
In the time domain, this choice corresponds to a voltage pulse, which may seem unusual---it differs from most experiments, which employ harmonic potential stimuli \mbox{$\delta\Psi(t)=\PsiAmp\sin (\omega t)$} with different frequencies $\omega$.
However, impedance has also been  measured with different stimuli, such as pulses, white noise, step functions, and random step sequences~\cite{Pilla1970, Creason1973_3, Popkirov1992, Barsoukov2002, Sihvo2026}.
These non-harmonic stimuli bear the advantage that they contain all frequencies and thereby give access in principle to the entire impedance spectrum with a single measurement~\cite{VanLeeuwen1969, Pilla1970, Creason1973_3}.
Such an approach can be challenging in experiment, however, since one must compromise between an unfavorable signal-to-noise ratio for small $\PsiAmp$ and potential nonlinear contributions for larger $\PsiAmp$.
These limitations do not apply to our numerical calculation of impedance:
since the linearized transport problem yields by construction the linear response to an applied stimulus, the applied voltage stimulus $\Laplace{\delta\Psi}(s)$ need not be of small amplitude.

From the Laplace-transformed potential profile $\Laplace{\psi}\first(x, s)$ one obtains the Laplace-transformed surface charge $\Laplace{Q}\first(s)$ by Gauss's law,
\begin{equation}
    \label{eq:charge_laplace}
    \Laplace{Q}\first(s) = \pm\epsr\varepsilon_0\,A\ddx\Laplace{\psi}\first(\pm L, s).
\end{equation}
Knowledge of $\Laplace{Q}$, in turn, gives access to the Laplace transformed current
\begin{align}
    \label{eq:current_laplace}
    \Laplace{I}\first(s)=s\Laplace{Q}\first(s)-\cancel{Q\first(t=0)},
\end{align}
and, by combining \cref{eq:impedance,eq:mpnp_bcs_first_laplace,eq:charge_laplace,eq:current_laplace}, to the impedance 
\begin{align}
    \label{eq:impedance_from_simulations_dimensional}
    Z(s) = \frac{2\,\Laplace{\delta\Psi}(s)}{s\,\Laplace{Q}\first(s)}.
\end{align}

For the numerical solution, \cref{eq:mpnp_first_laplace,eq:mpnp_noflux_bc_first_laplace,eq:mpnp_bcs_first_laplace} were cast in nondimensional form.
The full differential equations and boundary conditions in nondimensional form are provided in \cref{eq:pnp_lp_real_imag,eq:pnp_lp_real_imag_bcs} in the SI~\cite{supplemental_information}.
These equations were solved numerically with the Finite Element Method, with the \textsc{FEniCS} framework~\cite{Logg2012automated, Logg2010dolfin, Alnaes2014ufl, Kirby2006ffc} for Python.

\section{Characteristic quantities of electrical response\label{sec:characteristic_quantities}}
In this section, we introduce characteristic resistances, capacitances, timescales, angular frequencies, and analytical impedance expressions for selected parameter settings.
We will use these quantities to analyze the simulation data presented in \cref{sec:results}.

\subsection{Resistance and capacitances}
The bulk electrolyte has a (geometric) capacitance
\begin{align}
    \label{eq:bulk_capacitance}
    C_\mathrm{bulk} &= \frac{A\,\epsr\varepsilon_0}{2\,L}
\end{align}
and a resistance \mbox{$R_\mathrm{bulk}=2L\,\varrho/A$}, where \mbox{$\varrho=\kB T /(2\,e^2\,c_0\,D_\mathrm{av})$} is the resistivity arising from the linearized Nernst-Planck equation, and \mbox{$D_\mathrm{av}=(D_++D_-)/2$} is the average diffusivity, so that
\begin{align}
    \label{eq:bulk_resistance}
    R_\mathrm{bulk} = \frac{L\,\kB T}{A\,e^2\,c_0\,D_\mathrm{av}}.
\end{align}
Within the mPB theory [\cref{eq:modified_poisson_boltzmann}], the EDL capacitance reads~\cite{Kilic_PRE_2007_1} 
\begin{align}
    \begin{aligned}
    C_\mathrm{EDL} &= \frac{\epsr\varepsilon_0\,A}{\lD}\frac{\left|\sinh{\left(e\beta\PsiBias\right)}\right|}{\left(1+\chi\right)\sqrt{\frac{2}{\nup}\ln{\left(1+\chi\right)}}},\\
    \chi &= 2\nup\sinh^2\left(\frac{e\beta\PsiBias}{2}\right),
    \end{aligned}
    \label{eq:differential_capacitance_mPB}
\end{align}
which includes the  Debye length of the 1:1 electrolyte, defined as
\begin{align}
    \lD = \sqrt{\frac{\epsr\varepsilon_0\,\kB T}{2e^2\,c_0}}.
\end{align}
When \mbox{$\nu\to 0$}, $C_\mathrm{EDL}$ reduces to the Gouy--Chapman capacitance,
\begin{align}
    \label{eq:differential_capacitance_PB}
    C_\mathrm{EDL}^\mathrm{GC} = \frac{\epsr\varepsilon_0\,A}{\lD}\cosh{\left(\frac{e\beta\PsiBias}{2}\right)}.
\end{align}
The prefactor in \cref{eq:differential_capacitance_mPB,eq:differential_capacitance_PB} is the differential EDL capacitance at \mbox{$\PsiBias=0$},
\begin{align}
    \label{eq:edl_capacitance_no_bias}
    C_\mathrm{EDL}^0 &= \frac{\epsr\varepsilon_0\,A}{\lD}.
\end{align}

\subsection{Timescales and angular frequencies}
Combination of $R_\mathrm{bulk}$ with $C_\mathrm{bulk}$ and $C_\mathrm{EDL}$ gives important time scales $\tau$ and the corresponding angular frequencies $\omega$: the Debye time and frequency
\begin{align}
    \label{eq:tau_debye}
    \omega_\mathrm{D}\equiv \frac{1}{\tau_\mathrm{D}}&=\frac{1}{R_\mathrm{bulk}\,C_\mathrm{bulk}}=\frac{D_\mathrm{av}}{\lD^2},
\intertext{and the time and frequency scale of EDL charging through ohmic conduction,}
    \label{eq:tau_rc_nonlin}
    \omc\equiv\frac{1}{\tau_\mathrm{c}} &= \frac{1}{R_\mathrm{bulk}\,C_\mathrm{EDL}}=\frac{D_\mathrm{av}}{\lD L\,\cosh{(e\beta\PsiBias/2})}.
\intertext{In the limit $\PsiBias\to 0$, the latter expression reduces to the time and frequency scale}
    \label{eq:tau_rc}
    \omclin\equiv\frac{1}{\tau_\mathrm{c}^0} &= \frac{1}{R_\mathrm{bulk}\,C_\mathrm{EDL}^0}=\frac{D_\mathrm{av}}{\lD L}.
\end{align}
$\tau_\mathrm{c}^0$ is often referred to as the ``$RC$ time'', to reflect that $\tau_\mathrm{c}^0$ is the response time of the EDL capacitance in series with the bulk electrolyte resistance.
However, this naming is misleading:
following the same logic, $\tau_\mathrm{D}$ could just as well be denoted as an ``$RC$ time'', since it is the response time of the geometric capacitance in parallel to the bulk electrolyte resistance.

Barbero and Lelidis~\cite{Barbero2007} pointed out that, for $D_-\neq D_+$, the impedance spectrum will show a contribution from ambipolar diffusion, and this contribution is governed by the characteristic diffusion frequency,
\begin{align}
    \label{eq:tau_diff_ambipolar}
    \omega_\mathrm{diff}&=\frac{1}{\tau_\mathrm{diff}} = \frac{D_\mathrm{amb}}{L^2},
\intertext{with the ambipolar diffusion coefficient}
    \label{eq:ambipolar_diffcoeff}
    D_\mathrm{amb}&=\frac{2\,D_- D_+}{D_- + D_+}.
\end{align}
For \mbox{$D_-=D_+$}, one finds \mbox{$D_\mathrm{amb}=D_\mathrm{av}=D_+=D_-$}.
That is, the characteristic frequency in \cref{eq:tau_diff_ambipolar} corresponds exactly to the $L^2/D$ time scale of diffusion that was evoked in the context of nonlinear response~\cite{Bazant_PRE_2004} and impedance at a finite voltage bias~\cite{Mei_JPCC_2018,Garcia2025}.
While Barbero and Lelidis~\cite{Barbero2007} were only concerned with cases in which \mbox{$D_-\neq D_+$}, the quantitative connection \mbox{$D_\mathrm{amb}=D_\mathrm{av}$} for \mbox{$D_+=D_-$} suggests that the ambipolar diffusion coefficient may be the relevant quantity for diffusive response also in cases with \mbox{$D_-=D_+$}.

\subsection{Analytical impedance expressions for zero voltage bias and point ions}
For \mbox{$\nup=0$}, the mPNP equations reduce to the PNP equations.
Given \mbox{$\Ubias=0$}, the governing equations and boundary conditions [\cref{eq:noflux_bc,eq:mPNP,eq:potential_bcs,eq:modified_poisson_boltzmann}] can then be solved analytically, yielding the impedance
\begin{align}\label{eq:uneq_impedance_maintext}
    \dfrac{Z(s)}{R_{\mathrm{bulk}}}&=\left(\dfrac{1+\xi}{2}\right)\left(\dfrac{\tanh\left(k_+\frac{L}{\lD}\right)-\gamma^3f\tanh\left(k_-\frac{L}{\lD}\right)}{\tau_\mathrm{D}s\frac{L}{\lD} k_+^3(1-f)}\right.\nn
    &+\left.\dfrac{k_+^2-1+f(\gamma^2-k_+^2)}{\tau_\mathrm{D}s k_+^2(1-f)}\right),
\end{align}
where \mbox{$k_{\pm}=\sqrt{\dfrac{\xi+(1+\xi)\tau_\mathrm{D}s\pm\sqrt{(1-\xi)^2(\tau_\mathrm{D}s)^2+\xi^2}}{2\xi}}$}, $\xi=D_-/D_+$, $\gamma = k_+/k_-$, and $f=\beta_-/\beta_+$ with $\beta_{\pm}=\dfrac{\xi\pm\sqrt{(1-\xi)^2(\tau_\mathrm{D}s)^2+\xi^2}}{\tau_\mathrm{D}s(\xi-1)}$,
as found by Macdonald~\cite{Macdonald1953}, and re-derived by us in \cref{sec:impedance_unequal_diffu}  of the SI~\cite{supplemental_information}.
For equal diffusivities, \cref{eq:uneq_impedance_maintext} reduces to 
\begin{align}
    \label{eq:Maconald_analytical_reiterated_maintext}
    Z(s) = \frac{R_\mathrm{bulk}}{1+\tau_\mathrm{D}s} + \frac{2\,\tanh{\left(\frac{L}{\lD}\sqrt{1+\tau_\mathrm{D}s}\right)}}{s\left(1+\tau_\mathrm{D}s\right)^{3/2}\,C_\mathrm{EDL}^0}.
\end{align}

\section{Simulation Results}\label{sec:results}

The governing mPNP equations and boundary conditions [\cref{eq:noflux_bc,eq:mPNP,eq:potential_bcs,eq:modified_poisson_boltzmann}] contain nine parameters: $L, c_0, \epsr, D_+, D_-, T, a, \PsiBias(\equiv\Ubias/2)$, and $\PsiAmp(\equiv\Uamp/2)$.
The impedance only depends on the first eight, as it is obtained from the system's linear response to $\PsiAmp$.
We reduce the number of parameters by nondimensionalizing \cref{eq:noflux_bc,eq:mPNP} (see \cref{sec:linearization_transport} of the SI~\cite{supplemental_information}). 
The resulting dimensionless governing equations \eqref{eq:mpnp_nondim} and \eqref{eq:boundary_condition_dimless} contain only four dimensionless numbers: the dimensionless bias electrode potential $e\beta\,\PsiBias$, the diffusivity ratio $D_-/D_+$, the packing fraction \mbox{$\nu=2a^3c_0$}, and the EDL separation parameter $L/\lD$.
Hence, rather than varying the nine dimensional parameters, we systematically study the model EDL capacitor's response by varying the above four dimensionless parameters.

\subsection{No applied bias, equal diffusivities\label{sec:impedance_nobias}}

\begin{figure}
    \centering
    \includegraphics[width=\linewidth]{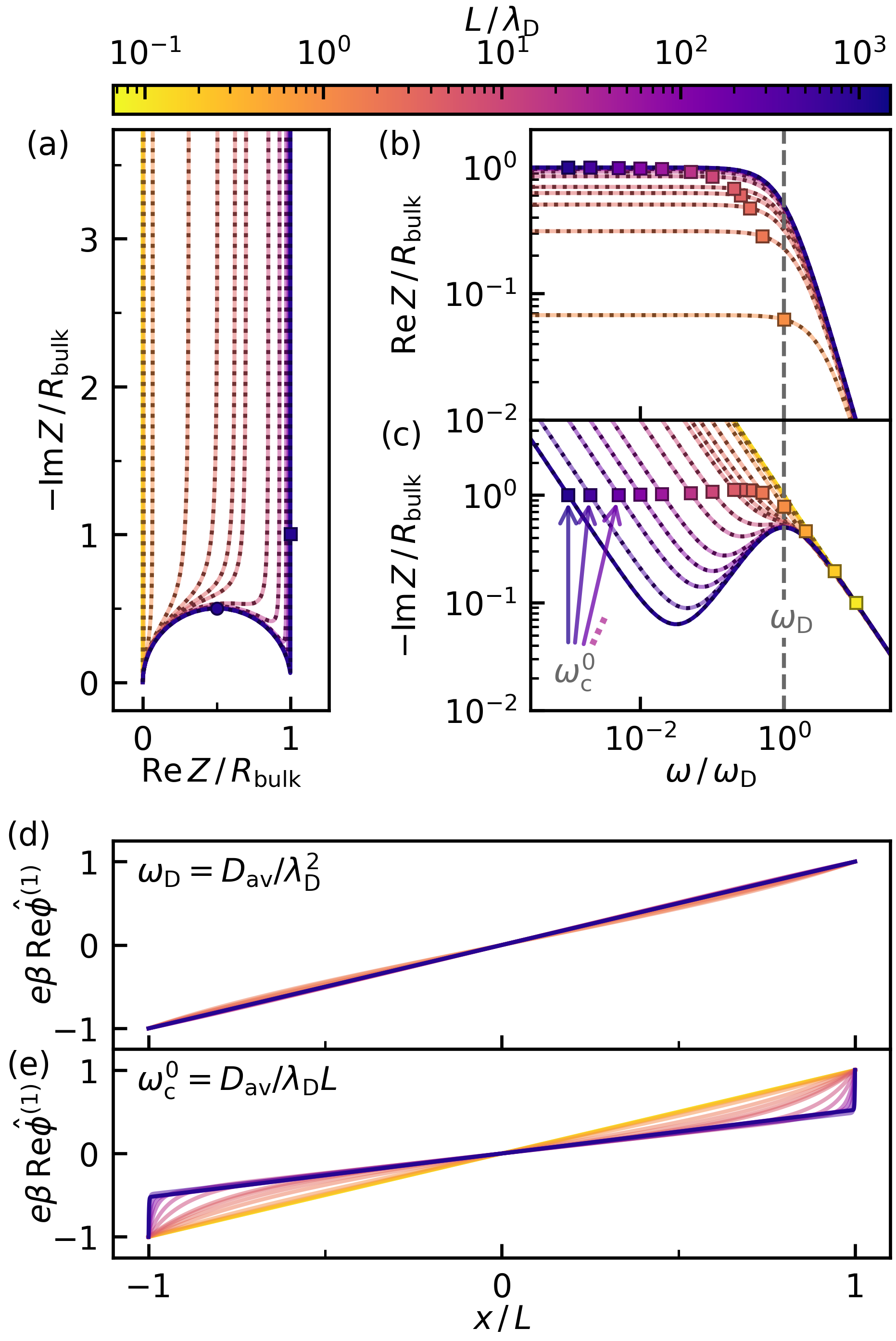}
    \caption{(a)--(c) Impedance spectra for \mbox{$D_+=D_-$}, \mbox{$\nu=0$}, \mbox{$\PsiBias=0$}, and various $L/\lD$, obtained by finite element method simulations of \cref{eq:pnp_lp_real_imag,eq:pnp_lp_real_imag_bcs} (full lines) and from \cref{eq:Maconald_analytical_reiterated_maintext} (dotted lines).  
    (a) Nyquist plots; (b) plots of $\ReZ$ vs. $\omega$; (c) plots of $-\ImZ$ vs. $\omega$. 
    $Z(\omega_\mathrm{D})$ is indicated with circular markers; $Z(\omclin)$ with square markers, in (a), only for the largest electrode separation \mbox{$L/\lD=1000$}.
    Spatial profiles of the electric potential perturbations in the frequency domain, $\Laplace{\psi}\first$, at $\omega_\mathrm{{D}}$~(d) and $\omclin$~(e).}
    \label{fig:impedance_spectra_nobias}
\end{figure}

We begin by examining the case of vanishing ion size (\mbox{$\nup=0$}), no applied voltage bias (\mbox{$\PsiBias=0$}), and equal ion diffusivities (\mbox{$D_-=D_+$}).
This case is well understood; it has been solved analytically both in the frequency~\cite{Macdonald1953} and time domain~\cite{janssen_pre_2018}.
It will, therefore, serve as a point of reference for the understanding of the parameter variations.
\mbox{\Cref{fig:impedance_spectra_nobias}(a)--(c)} presents impedance spectra for different EDL separations \mbox{$L/\lD$}, obtained from numerical solutions to \cref{eq:pnp_lp_real_imag,eq:pnp_lp_real_imag_bcs} (lines) and from the analytical solution \cref{eq:Maconald_analytical_reiterated_maintext} (dotted lines).
Here and in several figures below, we present the impedance $Z$ in different ways: (a)~in the plane of the real vs. minus the imaginary part of impedance, ~$\ReZ$ vs. $-\ImZ$, for various angular frequency $\omega$ (a so-called Nyquist plot); (b)~as~$\ReZ$ vs. $\omega$, and; (c)~as~$-\ImZ$, vs. $\omega$.
We indicate the impedance at the linear charging frequency $\omclin$ [\cref{eq:tau_rc}] by square markers, and the Debye frequency $\omega_\mathrm{D}$ [\cref{eq:tau_debye}] by a circular marker in (a) and a gray dashed line in (b) and (c).
In panel~(a), the markers are shown only for the largest electrode separation.

In \cref{fig:impedance_spectra_nobias}(a)--(c), our numerical data overlap perfectly with \cref{eq:Maconald_analytical_reiterated_maintext} for all \mbox{$L/\lD$} and $\omega$. 
We see that none of the spectra in \cref{fig:impedance_spectra_nobias}(a) has a slanted line (feature \textbf{3} in \cref{fig:schematic_nyquist_four_features}) and that the impedance spectra approach $0$ at high frequencies for all $L/\lD$.
Adding a finite electrode resistance to our model shifts the Nyquist plot along the $\ReZ$ axis (region~\textbf{1} in \cref{fig:schematic_nyquist_four_features}), as we demonstrate in \cref{sec:Stern_resistance} of the SI~\cite{supplemental_information} through an exact calculation; hence, we did not consider this influence explicitly in the simulations.

For large electrode separations, \mbox{$L\gg\lD$}, \cref{fig:impedance_spectra_nobias}(a) exhibits a clearly separated semicircle at high frequencies and vertical line at low frequencies (regions~\textbf{2} and \textbf{4} in \cref{fig:schematic_nyquist_four_features}).
The diameter of the semicircle equals the bulk electrolyte resistance.
For the smaller electrode separations, the semicircle and the vertical line merge.
A bulk electrolyte resistance can then no longer be easily obtained by visual inspection of the Nyquist plot.
The top of the semicircle is located at the Debye frequency, $\omega_\mathrm{D}$ (circular marker), and the point at which \mbox{$\ReZ=-\ImZ$} marks the linear charging frequency, $\omclin$ [\cref{eq:tau_rc}] (square marker).

\Cref{fig:impedance_spectra_nobias}(b) and~(c) show that $\omclin$ appears at lower angular frequency (scaled to $\omega_\mathrm{D}$) with increasing electrode separation \mbox{$L/\lD$}.
For $\omega<\omclin$, $\ReZ$ reaches a plateau [\cref{fig:impedance_spectra_nobias}(b)], while $-\ImZ$ vs.~$\omega$ [\cref{fig:impedance_spectra_nobias}(c)] shows scaling that is characteristic of capacitor charging, according to \mbox{$-\ImZ=1/(\omega C)$}.
Taken together, \cref{fig:impedance_spectra_nobias}(b) and~(c) thus show that, for \mbox{$\omega<\omclin$}, the system's impedance approaches that of a resistor (bulk electrolyte) and capacitor (EDLs) in series.
For large electrode separations \mbox{$L/\lD\gg 1$}, we see that the different frequencies are well-separated, \mbox{$\omclin\ll\omega_\mathrm{D}$}, which is reflected in the clear separation of the semicircle and the vertical line in the Nyquist plot.
In turn, for \mbox{$L/\lD\sim 1$}, the frequency scales are not well separated, \mbox{$\omclin\sim\omega_\mathrm{D}$}, which is reflected in the merging of the vertical line and the semicircle in the Nyquist plot [\cref{fig:impedance_spectra_nobias}(a)]. 
For even smaller electrode separations, $L/\lD<1$, we see that \mbox{$\omclin<\omega_\mathrm{D}$}, and the semicircle is no longer discernible in the Nyquist plot.

The significance of the Debye frequency, $\omega_\mathrm{D}$, and the linear charging frequency, $\omclin$, can be understood in more depth by inspecting the perturbations of the electric potential profiles at these frequencies, see \cref{fig:impedance_spectra_nobias}(d) and~(e).
At $\omega_\mathrm{D}$, the potential drops almost exclusively over the bulk electrolyte.
At $\omclin$, a substantial part of the voltage stimulus drops over the EDLs and the electric field in the bulk electrolyte is thus substantially reduced.
These observations are consistent with Bazant, Thornton, and Ajdari~\cite{Bazant_PRE_2004}, who pointed out that EDLs charge on the scale of $\tau_\mathrm{c}$.
Conversely, Garcia and coworkers~\cite{Garcia2025} claimed that the electric field within the bulk vanishes on the time scale of $\tau_\mathrm{D}$, which is contradicted by our data.

\subsection{Varying bias voltage\label{sec:impedance_var_bias}}

\begin{figure}
    \centering
    \includegraphics[width=\linewidth]{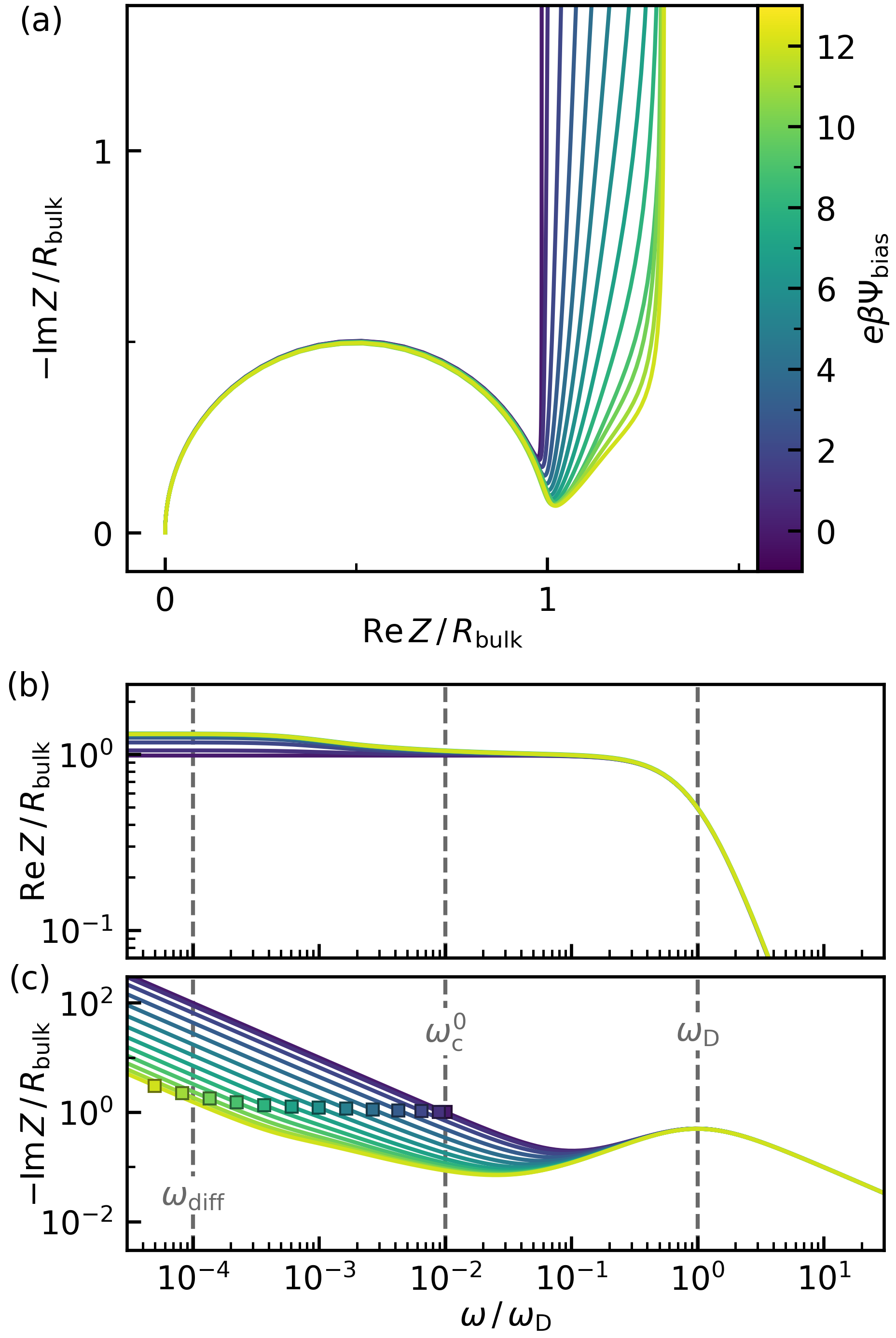}
    \caption{Impedance spectra obtained for various applied bias potentials \mbox{$e\beta\PsiBias=0, 1, 2, \hdots, 12$} (purple to yellow), with \mbox{$L/\lD=100$}, \mbox{$D_+=D_-$}, and \mbox{$\nup=0$}.
    (a)~Nyquist plots; (b)~plots of $\ReZ$ vs. $\omega$; (c)~plots of $-\ImZ$ vs. $\omega$.
    Dashed vertical lines indicate the characteristic frequencies $\omega_\mathrm{D}$~[\cref{eq:tau_debye}], $\omclin$~[\cref{eq:tau_rc}], and $\omega_\mathrm{diff}$ [\cref{eq:ambipolar_diffcoeff}].
    }
    \label{fig:impedance_spectra_varbias}
\end{figure}

\Cref{fig:impedance_spectra_varbias} shows impedance spectra for point ions (\mbox{$\nup=0$}), nonoverlapping EDLs (\mbox{$L/\lD=100$}) and bias potentials $e\beta\PsiBias$ between 0 and~12.
For all $\PsiBias$, we observe a semicircle  (region~\textbf{2}) of diameter $R_\mathrm{bulk}$ and a vertical line (region~\textbf{4}) in the Nyquist plot [\cref{fig:impedance_spectra_varbias}(a)], as we did for the \mbox{$\PsiBias=0$} case discussed in \cref{sec:impedance_nobias},
shown here in purple.
For nonzero $\PsiBias$, we further observe a slanted line (region~\textbf{3}) at intermediate frequencies, consistently with the findings of Ref.~\cite{Franceschetti1979}.
With increasing $\PsiBias$, the slanted line becomes straighter and its slope decreases towards \mbox{$k_\mathrm{sl}\approx 1$}.
Simultaneously, increasing $\PsiBias$ leads to an increasing low-frequency limit of $\ReZ$ [see \cref{fig:impedance_spectra_varbias}(a) and~\ref{fig:impedance_spectra_varbias}(b)], up to $e\beta\PsiBias\approx 4$, from which point onward $\ReZ(\omega\to 0)$ saturates at $R_\mathrm{bulk}+R_\mathrm{sl}\approx(4/3) R_\mathrm{bulk}$.
That is, $R_\mathrm{sl}$ is virtually independent of $\PsiBias$ beyond $e\beta\PsiBias\approx4$, which is remarkable as the ion concentrations in the EDLs still drastically changes with increasing $\PsiBias$.
For instance, the counter-ion concentration at \mbox{$e\beta\PsiBias=12$} is higher than at \mbox{$e\beta\PsiBias=8$} by a factor $\rme^4\approx 55$, and the co-ion concentration lower by the same factor.
As enormous concentration changes in the EDL do not affect $R_\mathrm{sl}$, Mei ~\textit{et al.}'s claim that $R_\mathrm{sl}$ represents an EDL resistance~\cite{Mei_JPCC_2018} cannot be correct.

\Cref{fig:impedance_spectra_varbias}(c) shows that $-\ImZ$ vs. $\omega$ reaches capacitive scaling ($-\ImZ\propto 1/\omega$) at sufficiently low frequencies for all considered $\PsiBias$.
For \mbox{$\PsiBias=0$}, where we observed the capacitive regime already in \cref{fig:impedance_spectra_nobias}(b), the capacitive regime appears at frequencies up to the linear charging frequency, $\omclin$ [\cref{eq:tau_rc}].
With increasing $\PsiBias$, this regime only holds up to lower frequencies, following the trend of the nonlinear charging frequency, $\omc$ [\cref{eq:tau_rc_nonlin}], shown by square markers, which decreases strongly with increasing $\PsiBias$.
While the linear charging frequency $\omclin$ is much larger than the diffusion frequency $\omega_\mathrm{diff}$, the nonlinear charging frequency $\omc$ can at sufficiently large $\PsiBias$ become smaller than the diffusion frequency, $\omega_\mathrm{diff}$.
In \cref{fig:impedance_spectra_varbias}(c), $\omc$ crosses the diffusion frequency $\omega_\mathrm{diff}$  between \mbox{$e\beta\PsiBias=10$} and~11.

Summarizing, for small voltage bias, the slanted-line region has a steep slope \mbox{$k_\mathrm{sl}\gg 1$} and $\omc\gg\omega_\mathrm{diff}$.
For large voltage bias, the slanted-line region has slope \mbox{$k_\mathrm{sl}\approx 1$} and \mbox{$\omc<\omega_\mathrm{diff}$}.
While these findings support Mei~\textit{et al.}'s claim about the connection between $k_\mathrm{sl}$ and $\omc/\omega_\mathrm{diff}$, they are restricted to the considered range of parameters, notably to \mbox{$D_-=D_+$}.

\subsection{Varying electrode separation at finite bias\label{sec:impedance_varL}}

\begin{figure}
    \centering
    \includegraphics[width=\linewidth]{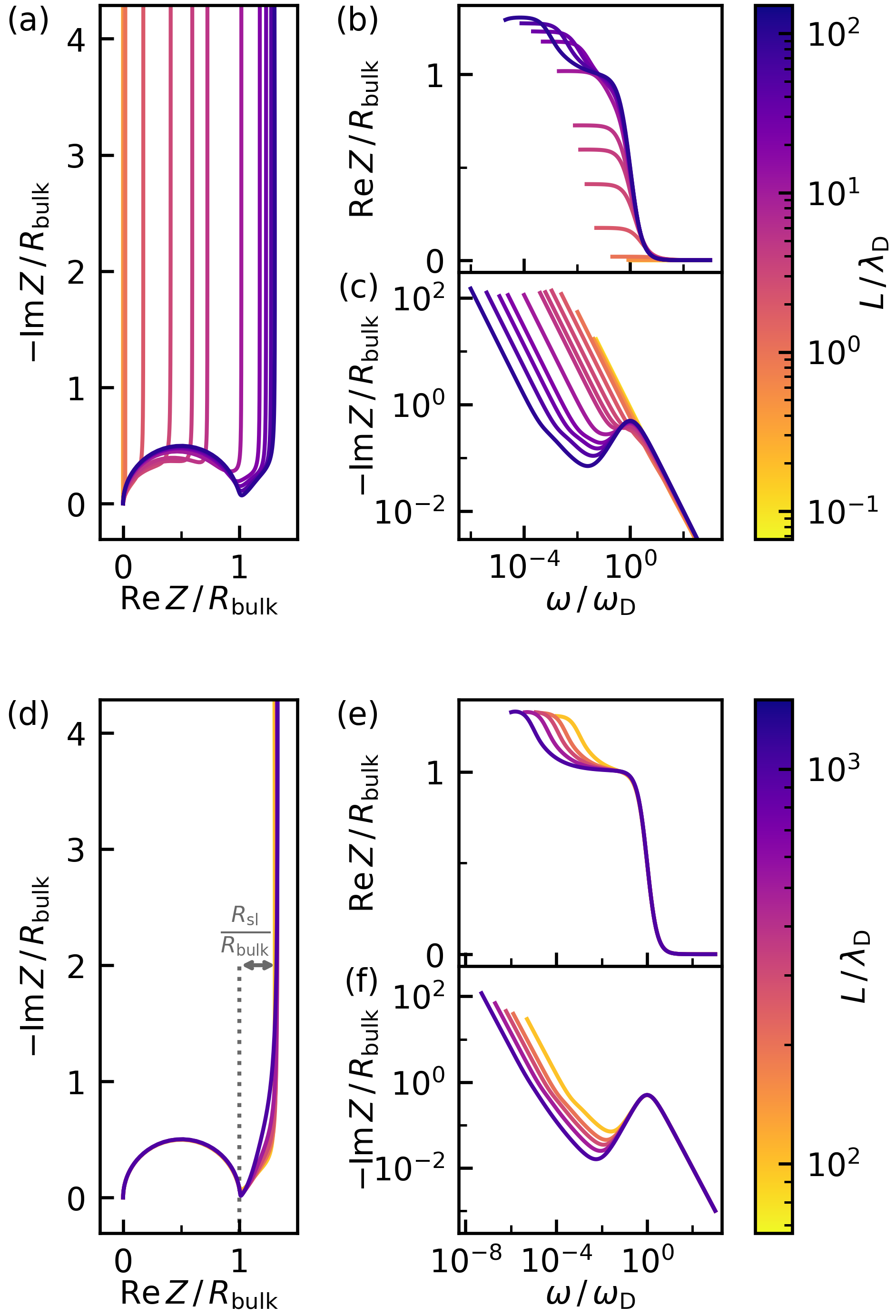}
    \caption{Impedance spectra obtained for various electrode separations $L\,/\,\lD$, given \mbox{$e\beta\PsiBias=12$}:
    (a),~(b),~and~(c)~small to large separations, $0.1\leq L/\lD\leq 100$;
    (d),~(e),~and~(f)~large to very large separations, $100\leq L/\lD\leq 1000$.
    (a)~and~(d)~Nyquist plots; (b)~and~(e)~plots of $\ReZ$ vs. $\omega$; (c)~and~(f)~plots of $-\ImZ$ vs. $\omega$.
    }
    \label{fig:impedance_spectra_varL}
\end{figure}

\Cref{fig:impedance_spectra_varL} shows impedance spectra for electrode separations $0.1\leq L/\lD\leq 100$ (a)--(c) and $L/\lD\geq 100$ (d)--(f), given \mbox{$e\beta\PsiBias=12$}.
In the Nyquist plots \cref{fig:impedance_spectra_varL}(a) and~(d), the semicircle, slanted line, and vertical line (regions \textbf{2}--\textbf{4} in \cref{fig:schematic_nyquist_four_features}) are most clearly separated around \mbox{$L/\lD\approx 100$}.
For larger  $L/\lD$, the slanted and vertical lines merge;
for smaller $L/\lD$, the semi-circle and slanted line merge.
At \mbox{$L/\lD=10$}, the slanted-line region is no longer discernible.
At even smaller $L/\lD$, similarly to the case without a bias (\cref{fig:impedance_spectra_nobias}), the semicircle and vertical line merge.
While all spectra in \cref{fig:impedance_spectra_varbias} corresponded to a single value of $R_\mathrm{bulk}$, the spectra in \cref{fig:impedance_spectra_varL} span a wide range of $R_\mathrm{bulk}$ values, across which we consistently find that the diameter of the semicircle equals $R_\mathrm{bulk}$.

We see in \cref{fig:impedance_spectra_varL}(d) that the slanted line becomes steeper with increasing $L\,/\,\lD$, and approaches \mbox{$k_\mathrm{sl}=1$} at the smaller $L\,/\,\lD$.
In \cref{fig:impedance_spectra_varL}(b,~c,~e,~f), we see that the capacitive regime at low frequencies, with constant $\ReZ$ and $1/\omega$ scaling of $-\ImZ$, are reached at lower frequencies, the larger $L/\lD$.
Specifically, this regime is reached for \mbox{$\omega/\omega_\mathrm{D}\sim 10^{-4}$} for \mbox{$L/\lD=100$} and for \mbox{$\omega/\omega_\mathrm{D}\approx 10^{-6}$} for \mbox{$L/\lD=1000$}.
Consistently with our previous finding in \cref{fig:impedance_spectra_varbias}(c), this suggests that for large biases, at which \mbox{$\omc<\omega_\mathrm{diff}$}, the capacitive regime is reached at $\omega_\mathrm{diff}\propto 1/L^2$.

Mei \textit{et al.} found that a steep slope $k_\mathrm{sl}$ of the slanted line corresponded to cases with \mbox{$\omc>\omega_\mathrm{diff}$}, and a shallow slope corresponded to cases with \mbox{$\omc<\omega_\mathrm{diff}$}.
Our results in \cref{fig:impedance_spectra_varL}(d) are consistent with this trend.
On the other hand, we find the same \mbox{$R_\mathrm{sl}/R_\mathrm{bulk}$} for \mbox{$L/\lD=1000$} as for \mbox{$L/\lD=100$}.
That is, although the slanted line appears at a lower frequency for \mbox{$L/\lD=1000$} than for \mbox{$L/\lD=100$}, it appears to affect the low-frequency impedance to a comparable extent.
This indicates that the slope of the slanted line alone does not allow for a simple statement about the which charging mechanism is dominant---contradicting Mei \textit{et al.}, who suggested that the slope of the slanted line indicates whether the charging process is controlled by diffusion or by EDL formation~\cite{Mei_JPCC_2018}.

\begin{figure}
    \centering
    \includegraphics[width=\linewidth]{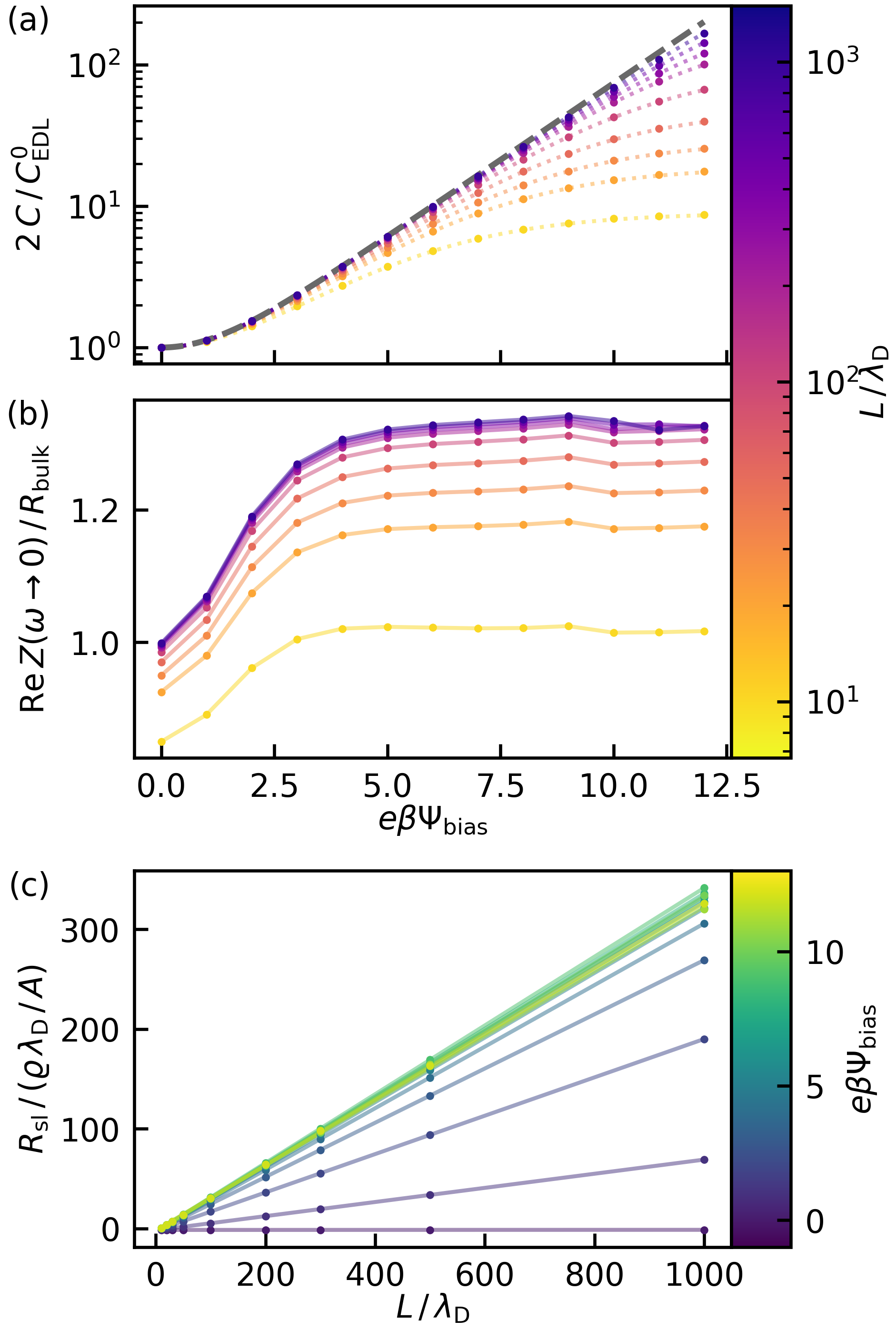}
    \caption{Characteristic quantities extracted from impedance spectra obtained for various electrode separations \mbox{$10\leq L\,/\,\lD\leq 1000$} and various bias voltages \mbox{$0\leq\PsiBias\leq 12$}. (a)~differential capacitance $C$ of the system, obtained from the simulated spectra as \mbox{$C=-\lim_{\omega\to0}1/[\omega\,\ImZ(\omega)]$}, and from the analytical description (dotted line), \cref{eq:C_N_maintext}, which we derive in \cref{sec:analytical_expressions_poisson_boltzmann}.
    (b)~low-frequency limit $\ReZ(\omega\to 0)$, estimated by taking $\ReZ$ at \mbox{$\omega=\omega_\mathrm{diff}$}, with connecting lines as a guide to the eye.
    (c)~slanted-line width $R_\mathrm{sl}$, estimated as $\ReZ(\omega\to 0)-R_\mathrm{bulk}$, with connecting lines as a guid to the eye.}
    \label{fig:limRe_limCap}
\end{figure}

\Cref{fig:limRe_limCap}(a) shows the system's differential capacitance $C$ vs. bias voltage,  with $C$ obtained from the impedance by \mbox{$C=-\lim_{\omega\to0}1/[\omega\,\ImZ(\omega)]$} (circular markers).
We show the Gouy--Chapman capacitance $C_\mathrm{EDL}$ [\cref{eq:differential_capacitance_PB}] (dotted line) for reference, as well as the differential capacitance of a closed system (dotted line), which we derive in \cref{sec:analytical_expressions_poisson_boltzmann} of the SI~\cite{supplemental_information}], and which for a fixed reference salt concentration $c_0$ reads
\begin{align}
    \label{eq:C_N_maintext}
    C = \frac{C_\mathrm{EDL}^0}{2}\,\frac{(L/\lD-1)\,\cosh\left(e\beta\Psi/2\right)+1}{L/\lD - 1 + \cosh\left(e\beta\Psi/2\right)}.
\end{align}
For all considered \mbox{$L/\lD$}, we find excellent agreement between the numerical data and \cref{eq:C_N_maintext}.
For the largest considered electrode separation, \mbox{$L/\lD=1000$}, the numerical solution also closely follows $C_\mathrm{EDL}$ [\cref{eq:differential_capacitance_PB}], except for voltage biases $e\beta\PsiBias>10$.
At sufficiently large biases, $C$ approaches a plateau value that is lower, the smaller $L/\lD$.
For \mbox{$\PsiBias\to 0$}, in turn, $C$ does not depend on $L/\lD$.

Physically, the plateau of the differential capacitance $C$ arises from the finite amount of ions in the system, which imposes an upper limit on how much charge may be accumulated in the EDLs.
Note that we pre-equilibrate our simulations in a way that leads to equal reference salt concentrations in the bulk electrolyte for all considered biases, as detailed in \cref{sec:model}.
If instead the amount of salt in the system was kept the same for all biases, the differential capacitance would go to zero for large biases, once most of the ions have been adsorbed by the EDLs.

In \cref{fig:limRe_limCap}(b) we show the bias dependence of the low-frequency limit of $\ReZ$ for various electrode separations $L/\lD$.
For all considered $L/\lD$,  \mbox{$\ReZ(\omega\to 0)$} increases with increasing bias and approaches an upper bound for $e\beta\PsiBias>4$.
For larger biases, \mbox{$\ReZ(\omega\to 0)$} becomes virtually independent of $\PsiBias$.

\Cref{fig:limRe_limCap}(c) plots the width $R_\mathrm{sl}$ of the slanted-line region vs. the electrode separations $L/\lD$, which we extract as \mbox{$R_\mathrm{sl}=\ReZ(\omega\to 0)-R_\mathrm{bulk}$} for $L/\lD\geq 10$.
For smaller $L/\lD$, the slanted line is not recognizable and hence $R_\mathrm{sl}$ is not well-defined.
Clearly, $R_\mathrm{sl}$ increases proportionally with the electrode separation $L/\lD$.
For $e\beta\PsiBias\leq 4$, the slope of the $R_\mathrm{sl}$ vs. $L/\lD$ plot increases with increasing $\PsiBias$.
For $e\beta\PsiBias>4$, in turn, $R_\mathrm{sl}$ does not change with $\PsiBias$. 
In this regime, $R_\mathrm{sl}$ can be seen as being purely a bulk-electrolyte property.

If $R_\mathrm{sl}$ were an EDL resistance, the found relationship $R_\mathrm{sl}\propto L/\lD$ would suggest that the concentration profiles in the EDL depend strongly on $L/\lD$.
However, \cref{fig:equilibrium_profiles} in the SI~\cite{supplemental_information} shows that concentration profiles in the EDLs are essentially the same for \mbox{$L/\lD=100$} and \mbox{$L/\lD=1000$}.
Hence, \cref{fig:limRe_limCap}(b) and~(c) demonstrate that $R_\mathrm{sl}$ cannot be an EDL resistance, unlike claimed by Mei \textit{et al.}.
Instead, $R_\mathrm{sl}$ must be a mixed measure that is governed by an interplay of the EDLs and the bulk electrolyte.

\subsection{Varying diffusivity ratio}\label{sec:impedance_vardiffrat}

\begin{figure}
    \centering
    \includegraphics[width=\linewidth]{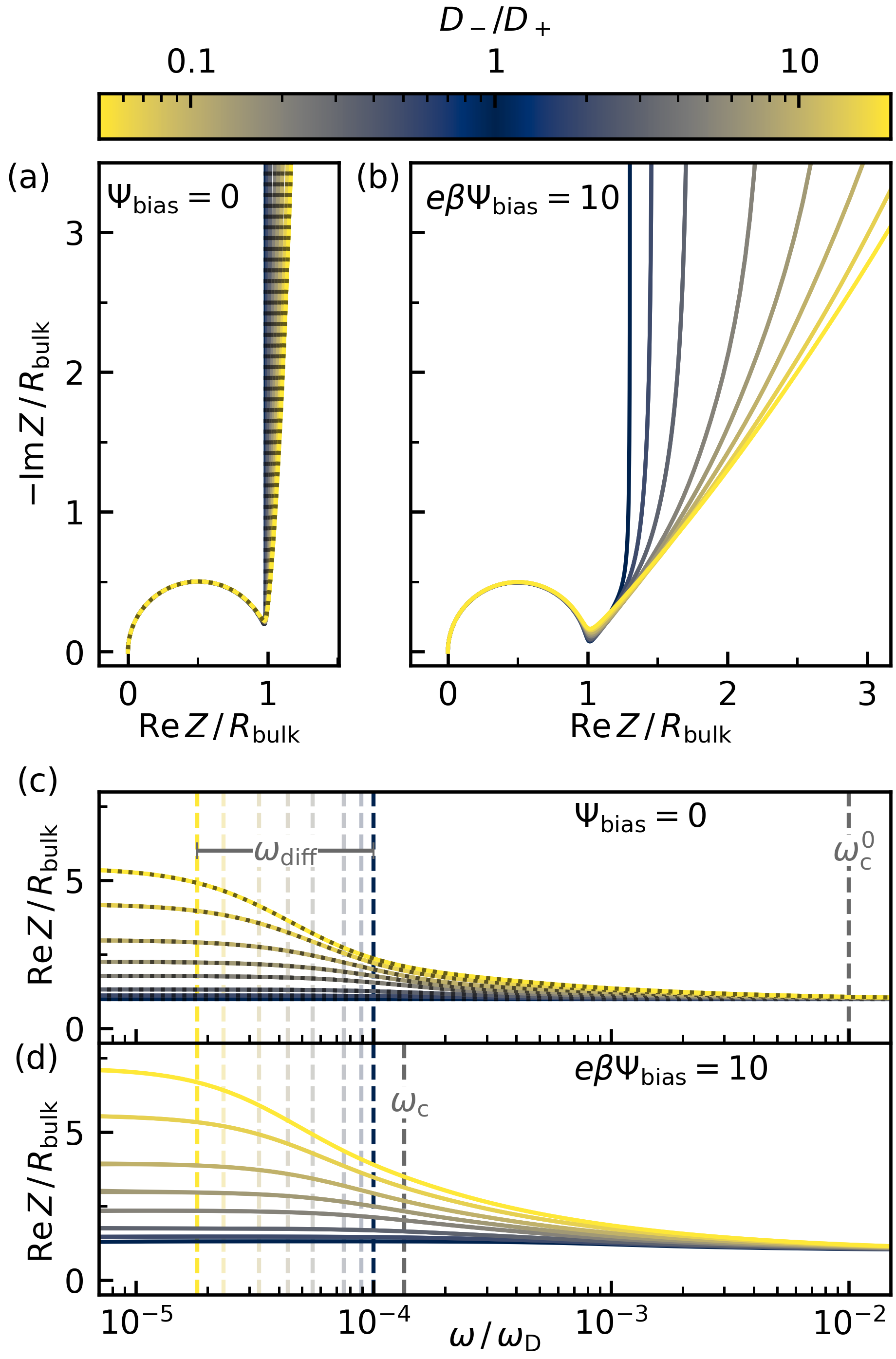}
    \caption{Impedance spectra for varying diffusivity ratios, $D_-/D_+$, with \mbox{$L/\lD=100$}, \mbox{$a=0$}. (a)~and~(b)~Nyquist plots of the simulated spectra, given (a)~\mbox{$\PsiBias=0$} and (b)~\mbox{$e\beta\PsiBias=10$}.
    (c)~and~(d)~plots of $\ReZ$ vs. $\omega$, given (c)~\mbox{$\PsiBias=0$} and (d)~\mbox{$e\beta\PsiBias=10$}.
    In (a)~and~(c), the analytical solution \cref{eq:uneq_impedance_maintext} by Macdonald~\cite{Macdonald1953} is shown for comparison by dotted lines.
    }
    \label{fig:impedance_spectra_var_diffrat}
\end{figure}

\Cref{fig:impedance_spectra_var_diffrat} shows impedance spectra for diffusivity ratios $D_-/D_+$ between $1/20$ and $20$, for \mbox{$\PsiBias=0$} [(a) and~(c)] and $\PsiBias=10$ [(b) and (d)], and \mbox{$L/\lD=100$} and \mbox{$\nu=0$}.
In all cases, same-colored lines overlap perfectly (not visible), so the normalized spectrum for a given $D_-/D_+$ and that for the inverse $D_-/D_+$, say, \mbox{$D_-/D_+=20$} and \mbox{$D_-/D_+=1/20$}, are equal, as expected.
In \cref{fig:impedance_spectra_var_diffrat}(a,~c), we also show the analytical expression \cref{eq:uneq_impedance_maintext} (dotted lines), with which our numerical data (colored lines) are in excellent agreement.

For \mbox{$\PsiBias=0$}, the spectra in the Nyquist representation [\cref{fig:impedance_spectra_var_diffrat}(a)] have a slanted-line region only when \mbox{$D_+\neq D_-$}, as pointed out previously by Barbero and Lelidis~\cite{Lelidis2005,Barbero2007}.
For \mbox{$e\beta\PsiBias=10$} [\cref{fig:impedance_spectra_var_diffrat}(b)], in turn, the slanted line is present also at \mbox{$D_-=D_+$}, as we saw in \cref{sec:impedance_var_bias,sec:impedance_varL}.
Irrespective of the bias, the relative width of the slanted line, $R_\mathrm{sl}/R_\mathrm{bulk}$, increases with increasing deviation of $D_-/D_+$ from unity.
Note that, when the slanted line is steep, one will likely underestimate $R_\mathrm{sl}$ from visual inspection of a Nyquist plot.
For \mbox{$D_-/D_+=20$}, for instance, we find \mbox{$R_\mathrm{sl}/R_\mathrm{bulk}\approx 6.1$} for \mbox{$e\beta\PsiBias=10$} [see \cref{fig:impedance_spectra_var_diffrat}(d)] and \mbox{$R_\mathrm{sl}/R_\mathrm{bulk}\approx 4.4$} for \mbox{$\PsiBias=0$} [see \cref{fig:impedance_spectra_var_diffrat}(d)], whereas visual inspection of the Nyquist plots [\cref{fig:impedance_spectra_var_diffrat}(a,~b)] falsely suggests that $R_\mathrm{sl}$ is much larger for \mbox{$e\beta\PsiBias=10$} than for \mbox{$\PsiBias=0$}.

Consistently with the results of Barbero and Lelidis~\cite{Lelidis2005,Barbero2007}, we find in \cref{fig:impedance_spectra_var_diffrat}(c) that the low-frequency plateau (\textit{i.e.} the vertical line in a Nyquist plot) is reached at $\omega_\mathrm{diff}$ [\cref{eq:tau_diff_ambipolar}] for \mbox{$\PsiBias=0$} and \mbox{$D_-\neq D_+$}.
That is, $\omega_\mathrm{diff}$ marks the lower frequency bound of the slanted-line region.
Notably, \cref{eq:tau_diff_ambipolar} correctly predicts that the low-frequency plateau shifts to lower relative frequencies, \mbox{$\omega/\omega_\mathrm{D}$}, the more \mbox{$D_-/D_+$} deviates from unity.
While the results of Ref.~\cite{Lelidis2005,Barbero2007} were limited to cases with \mbox{$\PsiBias=0$}, we find in \cref{fig:impedance_spectra_var_diffrat}(d) that $\omega_\mathrm{diff}$ also correctly predicts how the low-frequency plateau shifts to lower frequencies for \mbox{$\PsiBias\neq 0$}.

\begin{figure}
    \centering
    \includegraphics[width=\linewidth]{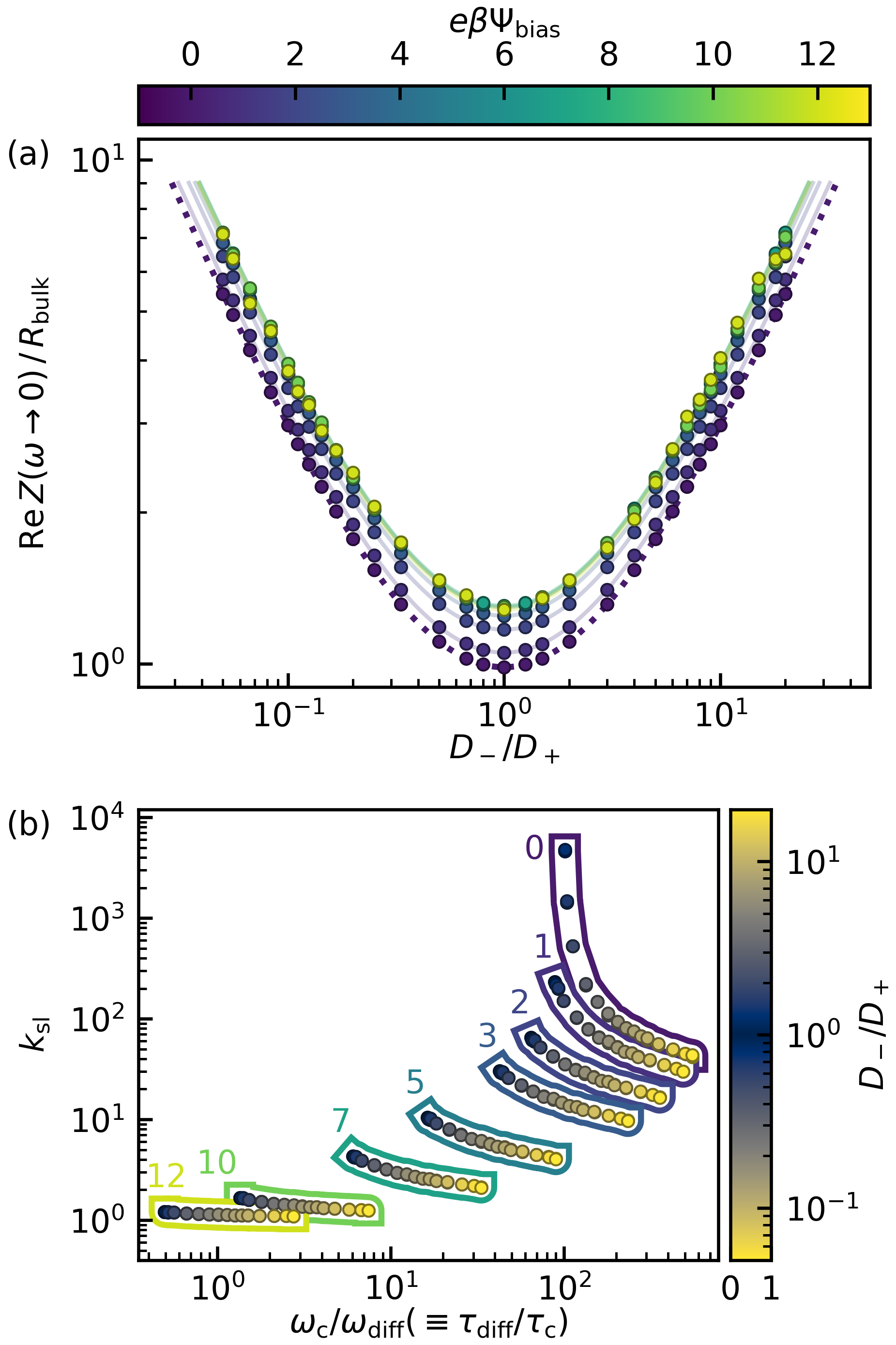}
    \caption{(a)~Plot of low-frequency limit of $\ReZ$ vs. $D_-/D_+$ for \mbox{$e\beta\PsiBias=0, 1, ..., 12$}. The analytical solution~\cite{Macdonald1953} [\cref{eq:analytical_ReZ_diffrat_lowfreq_dim}] is shown with dotted lines. Semitransparent lines show the same proportionality, \mbox{$\ReZ(\omega\to 0)\propto 1/D_\mathrm{amb}$}, rescaled by a $\PsiBias$-dependent prefactor to fit the data.
    (b)~Plot of the slope $k_\mathrm{sl}$ of the slanted line (region \textbf{3}) of simulated impedance spectra vs. the ratio of nonlinear charging frequency $\omc$ [\cref{eq:tau_rc_nonlin}] and diffusion frequency $\omega_\mathrm{diff}$ [\cref{eq:tau_diff_ambipolar}].
    $k_\mathrm{sl}$ was evaluated by linear regression of the Nyquist-plot data in the frequency range $[\sqrt{\omclin\,\min(\omc, \omega_\mathrm{diff})}, \omclin]$.}
    \label{fig:Rsl_and_ksl_vardiffrat}
\end{figure}

\Cref{fig:Rsl_and_ksl_vardiffrat}(a) shows the low-frequency limit \mbox{$\ReZ(\omega\to 0)$} for various $D_-/D_+$ and $\PsiBias$, given $L/\lD=100$.
For $\PsiBias=0$, the dependence of $\ReZ(\omega\to 0)$ on $D_-/D_+$ accurately matches an analytical expression [\cref{eq:analytical_ReZ_diffrat_lowfreq_dim} in the SI~\cite{supplemental_information}] that follows from \cref{eq:uneq_impedance_maintext}, and that scales as $\ReZ(\omega\to 0)\propto 1/D_\mathrm{amb}$.
For $\PsiBias\neq 0$, the data shows the same dependency on $D_\mathrm{amb}$, which we found by rescaling \cref{eq:analytical_ReZ_diffrat_lowfreq_dim} with a $\PsiBias$-dependent prefactor.

In \cref{fig:Rsl_and_ksl_vardiffrat}(b) we show the slope $k_\mathrm{sl}$ of the slanted line plotted against the ratio \mbox{$\omc/\omega_\mathrm{diff}$} of the nonlinear charging frequency [\cref{eq:tau_rc_nonlin}] and the diffusion frequency [\cref{eq:tau_diff_ambipolar}], again, for $L/\lD=100$.
If $k_\mathrm{sl}$ depended solely on this ratio of timescales, as Mei~\textit{et al.}~\cite{Mei_JPCC_2018} claimed, then all data should fall onto one curve.
\Cref{fig:Rsl_and_ksl_vardiffrat}(b) shows that this is not the case.
For a given $D_-/D_+$ and varying $\PsiBias$, we find the expected increase of $k_\mathrm{sl}$ with increasing \mbox{$\omc/\omega_\mathrm{diff}$}, as we found before in \cref{sec:impedance_var_bias,sec:impedance_varL}.
The trend is reversed, however, when considering a variation in \mbox{$D_-/D_+$} for a given $\PsiBias$: larger deviations of \mbox{$D_-/D_+$} from unity lead simultaneously to larger \mbox{$\omc/\omega_\mathrm{diff}$} and smaller $k_\mathrm{sl}$.
Clearly, $k_\mathrm{sl}$ is not suitable as a measure of the ratio of frequency (or time) scales $\omc/\omega_\mathrm{diff}$.

For $D_-\neq D_+$, Barbero and Lelidis attributed the slanted line to ambipolar diffusion ~\cite{Lelidis2005,Barbero2007}.
Likewise, ambipolar diffusion was linked to the transient response of a $D_-\neq D_+$-electrolyte to a voltage step ~\cite{Palaia2025_edlc}.
Conversely, in cases with \mbox{$\PsiBias\neq 0$} and \mbox{$D_-=D_+$}, the cause of the slanted line has not been clarified in the literature.
Some studies~\cite{Mei_JPCC_2018,Garcia2025} associated the slanted line with a characteristic frequency of the form \mbox{$D/L^2$}, with \mbox{$D=D_+=D_-$}.
It was not specified, however whether the diffusion coefficient $D$ therein was taken to be the average diffusion coefficient $D_\mathrm{av}$ or the ambipolar diffusion coefficient $D_\mathrm{amb}$.
\Cref{fig:impedance_spectra_var_diffrat}(d) and \cref{fig:Rsl_and_ksl_vardiffrat}(a) demonstrate clearly that the correct predictions for the characteristic frequency $\tau_\mathrm{diff}$ and the width $R_\mathrm{sl}$ of the slanted line will be obtained on the basis of $D_\mathrm{amb}$.
This suggests that the slanted line reflects the same physical phenomenon, ambipolar diffusion, whether it arises from \mbox{$D_-\neq D_+$} or from \mbox{$\PsiBias\neq 0$}.
To further substantiate this claim, we now inspect ionic concentration profiles.

\begin{figure*}
    \centering
    \includegraphics[width=\linewidth]{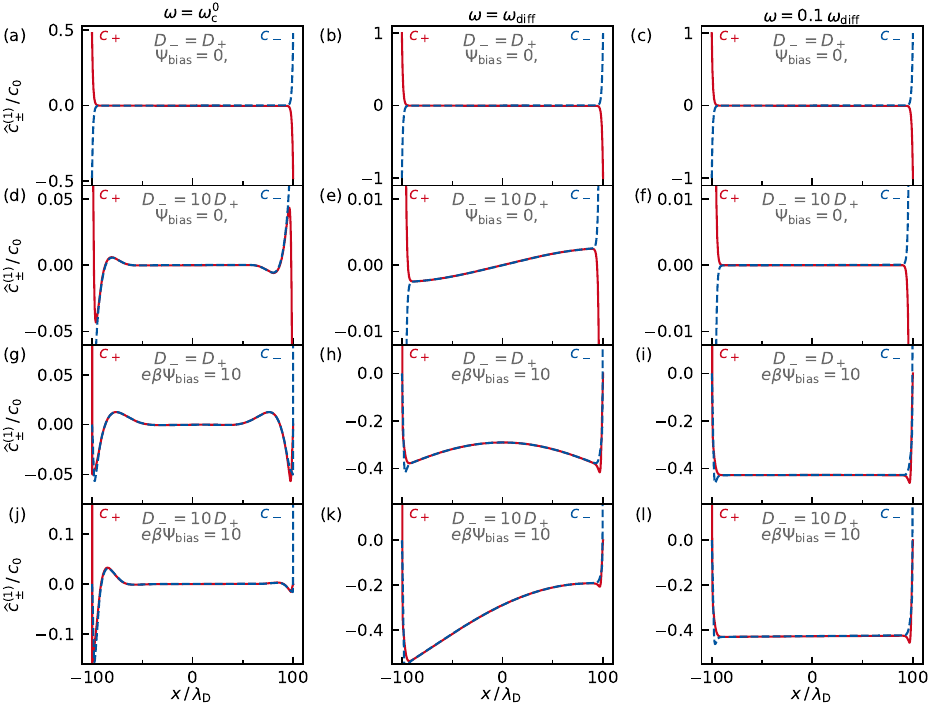}
    \caption{Profiles of Laplace-transformed concentration perturbations [see \cref{eq:phi_n_expansions}] of cations (solid red lines) and anions (dashed blue lines), given different \mbox{$\PsiBias$} and \mbox{$D_-/D_+$}.
    }
    \label{fig:perturbation_profiles}
\end{figure*}

\Cref{fig:perturbation_profiles} shows the concentration perturbations $\RealPart\big(\Laplace{c}_\pm\first\big)$ for \mbox{$L/\lD=100$} and \mbox{$e\beta\PsiBias=0$} (first two rows) and \mbox{$e\beta\PsiBias=10$} (third and fourth row), equal diffusivities (first and third row) and $D_-=10D_+$ (second and fourth row),
at three characteristic frequencies each: at the linear charging frequency~$\omclin$ (left column), at the diffusion frequency~$\omega_\mathrm{diff}$ (middle column), and at $\omega_\mathrm{diff}/10$ (right column).
The real part of the Laplace-transformed perturbation at a given frequency corresponds to the response in the positive half-cycle of an applied oscillating stimulus;
a more complete picture is offered by animations of the oscillating concentration profiles in the SI~\cite{supplemental_information}, obtained by transforming the Laplace-transformed profiles back to the time domain.
In all cases, \cref{fig:perturbation_profiles} shows increased counter-ion concentrations, relative to the reference concentration profile, and decreased co-ion concentrations in the EDLs, that is, up to a few $\lD$ from the electrodes.
In panels (a)--(c), no concentration perturbations are found at larger distances.
Panels (d)--(l), in turn, exhibit electroneutral concentration perturbations that reach much further: up to $\sim 50\lD$ at \mbox{$\omega=\omclin$} [see panels (d,~g,~j)], and throughout the entire bulk electrolyte at {$\omega=\omega_\mathrm{diff}$} and {$\omega=\omega_\mathrm{diff}/10$}.
The fact that the salt perturbations are electroneutral regardless of the choice of \mbox{$D_-/D_+$} or \mbox{$\Ubias$} can only be explained by a coupling of cation and anion transport through the electric field.
That is, both \mbox{$\PsiBias\neq 0$} and \mbox{$D_-\neq D_+$} give rise to ambipolar diffusion throughout the bulk electrolyte.

We discussed on the example of  \cref{fig:perturbation_profiles} that the electroneutral salt perturbations appeared only in cases with \mbox{$D_-\neq D_+$} and/or \mbox{$\Ubias\neq 0$}, that is, in those cases which also exhibited a slanted-line region in their impedance spectra.
Comparison of the left and middle columns of \cref{fig:perturbation_profiles} further reveals that electroneutral salt perturbations appear in the frequency range between $\omclin$ and $\omega_\mathrm{diff}$.
We identified before in \cref{fig:impedance_spectra_varbias}(c) and \cref{fig:impedance_spectra_var_diffrat}(c)~and~(d)~that this is the frequency range in which the slanted-line region is located.
With that, our results constitute clear evidence that the slanted-line region found in the simulated impedance spectra is the result of ambipolar salt diffusion.

Clearly, the similarity between the impedance spectra for \mbox{$\PsiBias\neq 0$} and \mbox{$D_-\neq D_+$} is not coincidental:
in both cases we related the slanted-line region to the occurrence of electroneutral salt perturbations.
A major difference between the two cases lies, however, in the symmetry of the salt perturbations.
For $\PsiBias= 0$ and $D_+\neq D_-$, the salt perturbations are antisymmetric [\cref{fig:perturbation_profiles}(e)];
for $\PsiBias\neq 0$ and $D_+= D_-$, the salt perturbations are symmetric [\cref{fig:perturbation_profiles}(h)];
for $\PsiBias\neq 0$ and $D_+\neq D_-$, the salt perturbations are asymmetric [\cref{fig:perturbation_profiles}(k)].

For \mbox{$D_-\neq D_+$} and \mbox{$\PsiBias=0$} the antisymmetric salt perturbation arises because because one ionic species is more mobile than the other, giving rise to a net salt flux in the bulk electrolyte, in response to an applied electric field.
A salt-concentration perturbation is then formed as the salt flux meets the blocking boundaries, from where the perturbation diffuses into the bulk electrolyte.
In contrast, for \mbox{$\PsiBias\neq 0$} and \mbox{$D_-=D_+$}, salt fluxes arise initially only within the EDLs, as we explain in \cref{sec:salt_charge_coupling}.

The symmetric salt perturbation for \mbox{$\PsiBias\neq 0$} and \mbox{$D_-=D_+$}, as well as the absence of such a perturbation for \mbox{$\PsiBias=0$}, can be understood from Gouy--Chapman theory, as follows:
at small electrode potentials, \mbox{$e\beta\Psi\ll 1$}, EDLs charge through an exchange of co-ions with counter-ions (``ion swapping'').
In this regime, the EDLs adsorb no salt, so the bulk salt concentration remains unaffected by the charging process.
At large electrode potentials, \mbox{$e\beta\Psi\gtrsim 4$}, where the co-ions are fully depleted, EDLs charge purely through counter-ion adsorption, so the charge accumulation in the EDLs translates to a net adsorption of salt.
The transition of the EDL charging mechanism from ion swapping at small $\Psi$ to counter-ion adsorption at large $\Psi$ can be quantified with the charge efficiency, which relates the excess amount of salt in the EDLs, $W$, to the charge $Q$ that the EDLs have acquired~\cite{Biesheuvel2010}.
More specifically, for the differential changes probed by impedance, the relevant quantity is the differential charge efficiency, \mbox{$e(\partial W/\partial Q)_{c_0}$}.
Gouy--Chapman theory predicts \mbox{$e(\partial W/\partial Q)_{c_0}=0$} for \mbox{$\Psi=0$}, and \mbox{$e(\partial W/\partial Q)_{c_0}\to 1$} in the limit of \mbox{$\Psi\to\infty$} (see \cref{sec:analytical_expressions_poisson_boltzmann}).
Practically, a differential charge efficiency close to unity is reached much sooner; for instance, within Gouy--Chapman theory one calculates \mbox{$e(\partial W/\partial Q)_{c_0}\approx 0.96$} for \mbox{$e\beta\Psi=4$}.
While the EDLs on either side of the domain acquire opposite charges, they adsorb equal amounts of salt, giving rise to salt depletion in the vicinity of either of the EDLs (cf.~\cite{Bazant_PRE_2004}).
This symmetric salt adsorption is reflected in the symmetric salt-diffusion profiles for \mbox{$\PsiBias\neq 0$} and \mbox{$D_-=D_+$}.
The differences between the perturbation profiles at high and low frequencies corresponds qualitatively to the early- and late-time response of a system to a voltage step (see ref.~\cite{Bazant_PRE_2004}).
At high frequencies (early times), the center of the domain is largely unaffected by the salt adsorption, whereas at low frequencies (late times), the salt depletion front has propagated throughout the bulk electrolyte, leading within it to a uniformly decreased salt concentration. 

The antisymmetric perturbation keeps the salt concentration in the center fixed; hence, the center of the domain acts effectively as a transmissive boundary.
Conversely, for the symmetric adsorption of salt into the EDLs, the salt concentration in the center decreases (in the positive half cycle) at low frequencies.

\subsection{Varying packing fraction at finite bias}
\begin{figure}
    \centering
    \includegraphics[width=\linewidth]{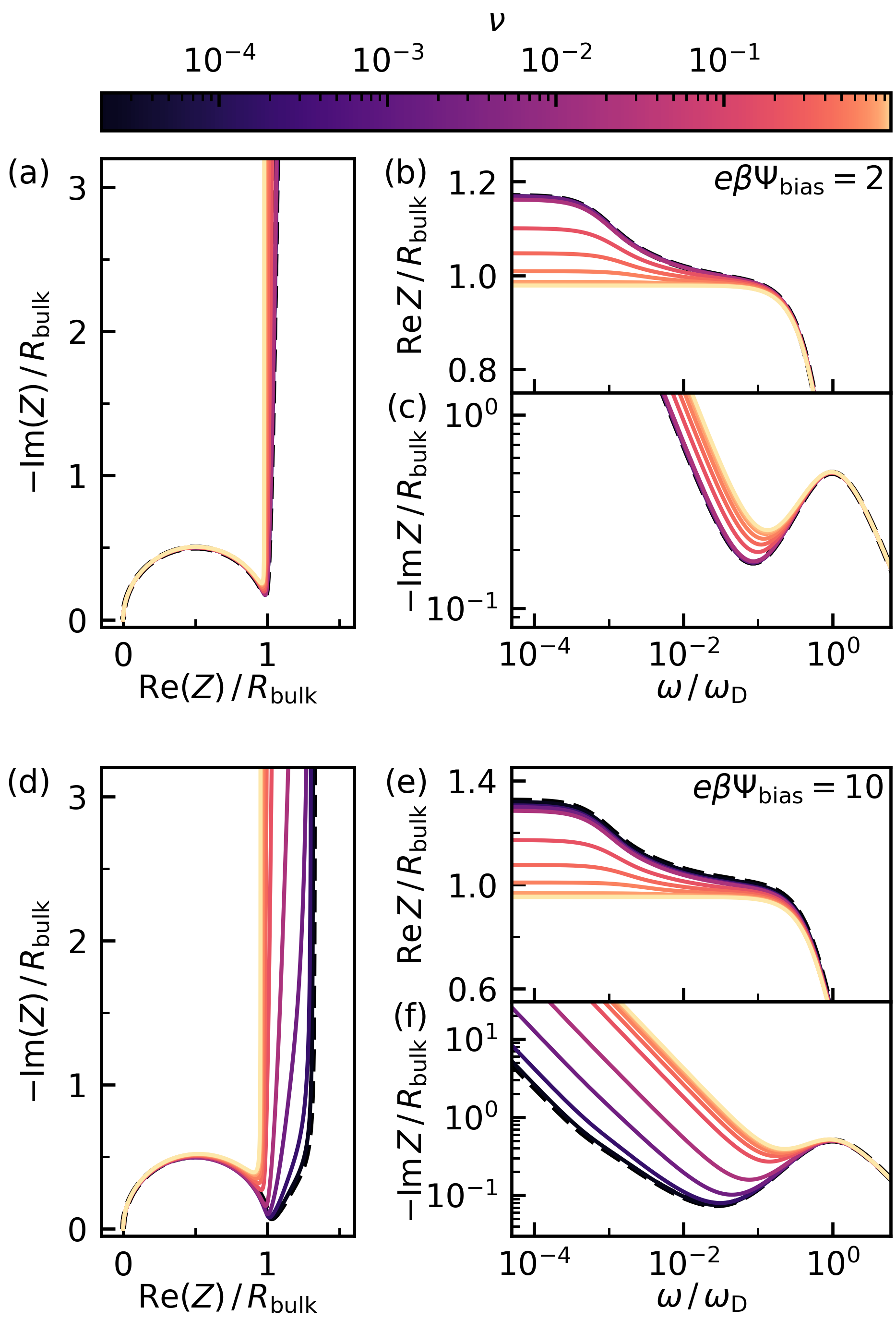}
    \caption{Impedance spectra for \mbox{$L/\lD=100$} and \mbox{$e\beta\PsiBias=0$} [(a), (b), and (c)] and \mbox{$e\beta\PsiBias=10$} [(d), (e), and (f)], for various packing fractions \mbox{$2\cdot 10^{-5}\leq \nup\leq 0.98$}, to illustrate the effects of steric exclusion. 
    For reference, the solutions obtained for \mbox{$\nup=0$} are shown as dashed lines.
    }
    \label{fig:impedance_spectra_mpnp}
\end{figure}

In \cref{fig:impedance_spectra_mpnp} we show impedance spectra for various packing fractions~\mbox{$2\cdot10^{-5}\leq \nup\leq 0.98$}, given \mbox{$e\beta\,\PsiBias=2$} [(a)--(c)] and \mbox{$e\beta\,\PsiBias=10$} [(d)--(f)].
The spectra for the smallest finite packing fraction, \mbox{$\nu=2\cdot 10^{-5}$}, are almost identical to the spectrum obtained for the PNP case of \mbox{$\nu=0$} (dashed lines).
With increasing $\PsiBias$, the slanted line becomes flatter and its width increases: for {$\nup=0.1$}, we find \mbox{$R_\mathrm{sl}/R_\mathrm{bulk}\approx 0.10$} for \mbox{$e\beta\PsiBias=2$} [\cref{fig:impedance_spectra_mpnp}(b)] and \mbox{$R_\mathrm{sl}/R_\mathrm{bulk}\approx 0.17$} for \mbox{$e\beta\PsiBias=10$} [\cref{fig:impedance_spectra_mpnp}(e)].
Hence, for $\nup\neq 0$, the slope $k_\mathrm{sl}$ and width $R_\mathrm{sl}$ of the slanted line change with increasing $\PsiBias$ in the same way as they did for $\nup=0$ [see~\cref{sec:impedance_var_bias,sec:impedance_vardiffrat}].

For a given $\PsiBias$, the slanted line becomes steeper with increasing $\nup$ and its width decreases.
At the largest values of \mbox{$\nup=0.98$}, the slanted line is no longer discernible; here, the impedance spectrum resembles the one found for \mbox{$\PsiBias=0$}.
We noted in \cref{sec:equilibrium} that steric-exclusion effects are negligible if \mbox{$2\nup\sinh^2{\left(e\beta\PsiBias/2\right)}\ll 1$}.
Indeed, at \mbox{$\PsiBias=0$}, we obtained identical spectra for all values of $\nup$ (not shown).
Further, comparison of \cref{fig:impedance_spectra_mpnp}(b) and~(e) shows that the decreasing width becomes noticeable already at smaller $\nup$ for \mbox{$e\beta\PsiBias=10$} than for \mbox{$e\beta\PsiBias=2$}.
For instance, at \mbox{$e\beta\PsiBias=2$}, the value of $R_\mathrm{sl}$ for \mbox{$\nup=10^{-3}$} is only \mbox{$\sim 0.5\,\%$} lower than that for \mbox{$\nup=0$}, whereas, at \mbox{$e\beta\PsiBias=10$}, this difference is \mbox{$\sim 8\,\%$}.

We find in \cref{fig:impedance_spectra_mpnp}(b,~c,~e,~f) that the onset of the slanted-line region is shifted to higher frequencies for increasing $\nup$.
This trend is consistent with the prediction of Kilic \textit{et al.}, who within the mPNP model derived for \mbox{$D_-=D_+=:D$} the diffusive time scale \mbox{$(1-\nup)L^2/D$}~\cite{Kilic_PRE_2007_2}.

Why does the slanted line disappear for $\nup\to 1$?
In the previous section, we identified that the slanted-line region in the impedance spectra is associated with ambipolar diffusion of salt, and it becomes more important with increasing $\PsiBias$ because of an increasing differential charge efficiency $e(\partial W/\partial Q)_{c_0}$ of the EDLs.
Given $\nup\neq 0$, the charge efficiency [\cref{eq:differential_charge_efficiency_mpb} in the SI~\cite{supplemental_information}] is reduced by a factor $(1-\nup)$ relative to the Gouy--Chapman charge efficiency. 
For $\nup$ close to unity, the differential charge efficiency approaches zero.
Physically, this limit means that the bulk electrolyte approaches close packing of the ions.
A differential increase in the electrode potential then does not lead to a substantial increase in the counter-ion concentration within the EDL, since it is densely packed with counter-ions.
Instead, the EDL charges by adsorbing counter-ions and expelling co-ions only at its outer edge, thereby becoming wider.
This charging mechanism corresponds to ion swapping, rather than counter-ion adsorption.

\section{Equivalent circuit analysis}\label{sec:equivalent_circuit}
Impedance spectra are most commonly analysed through fitted equivalent circuits.
We will demonstrate in this section how fitting an appropriate equivalent circuit to impedance data can help with the quantitative extraction of electrical parameters, and provides an intuition for the electrochemical processes that make up the system's electrical response.

\subsection{Formulation of the circuits: two types of Warburg elements\label{sec:formulation_circuit}}
We begin by motivating an equivalent circuit for the case with \mbox{$D_-=D_+$} and \mbox{$\Ubias=0$}, on the basis of Macdonald's analytical solution, \cref{eq:Maconald_analytical_reiterated_maintext}.
For the first term in the analytical solution, we identify an exact correspondence to an electrical circuit, as
\begin{align}
    \frac{R_\mathrm{bulk}}{1+\tau_\mathrm{D}s}
    \qquad\widehat{=}\qquad
\begin{circuitikz}[font=\Large,european resistors,scale=0.6,transform shape,baseline={([yshift=-1em]current bounding box.center)}]
\draw
(0,0) -- (0.8,0)
(0.8,0) -- (0.8,0.8)
         to[C,l=$C_\mathrm{bulk}$] (2.8,0.8)
         -- (2.8,0)
(0.8,0) -- (0.8,-0.8)
         to[R,l=$R_\mathrm{bulk}$] (2.8,-0.8)
         -- (2.8,0)
(2.8, 0) to (3.6, 0);
\end{circuitikz}.
\end{align}
To interpret the second term in a circuit picture, we evaluating it in the limit \mbox{$s\to 0$}, giving
\begin{align}
    \begin{aligned}
    &\frac{2\,\tanh{\left(\frac{L}{\lD}\sqrt{1+\tau_\mathrm{D}s}\right)}}{s\left(1+\tau_\mathrm{D}s\right)^{3/2}\,C_\mathrm{EDL}^0}
    \xrightarrow{s\to 0}
    \frac{2}{s C_\mathrm{EDL}^0}\\
    &-\frac{R_\mathrm{bulk}}{2}\left[\tanh^2\left(\frac{L}{\lD}\right)-1 +\frac{3\lD}{L}\tanh\left(\frac{L}{\lD}\right)\right]\\
    &+\mathcal{O}(s^2).
    \end{aligned}
\end{align}
Here, we identify the first summand, $2/(s\,C_\mathrm{EDL}^0)$, as the capacitive response of the two EDLs.
The second summand, formally a negative ohmic resistance, provides a correction to the bulk resistance $R_\mathrm{bulk}$, by accounting for the finite extent of the EDLs.
For sufficiently large $L/\lD$, this correction may be approximated as $3\lD R_\mathrm{bulk}/(2 L)$, and thus we define the effective bulk resistance
\begin{align}
    \label{eq:Rbulk_eff}
    R_\mathrm{bulk}^\mathrm{eff} := \frac{\left(L-\frac32\lD\right)\kB T}{A e^2 c_0 D_\mathrm{av}}.
\end{align}
On the basis of the above considerations, we propose to approximate \cref{eq:Maconald_analytical_reiterated_maintext} by the circuit \mbox{$C_\mathrm{EDL}^0(C_\mathrm{bulk},~R_\mathrm{bulk}^\mathrm{eff})C_\mathrm{EDL}^0$},
which converges to the analytical solution in the low-frequency limit, and correctly predicts the shape of the impedance spectrum throughout the entire frequency range.

Having established a reference equivalent circuit for the impedance in absence of salt diffusion, we now turn to the influence of salt diffusion on the impedance.
Diffusive contributions to an impedance spectrum are most commonly represented by a Warburg element. 
To describe systems of a finite extent, there are two types of Warburg element: first, the \textit{Warburg short} element, the impedance of which is given by
\begin{align}
    Z_{Ws} = B_{Ws}\frac{\tanh{\sqrt{i\omega\tau_{Ws}}}}{\sqrt{i\omega\tau_{Ws}}},
\end{align}
with the prefactor $B_{Ws}$ and the characteristic time scale $\tau_{Ws}$.
Second, the \textit{Warburg open} element, the impedance of which is given by
\begin{align}
    Z_{Wo} = B_{Wo}\frac{\coth{\sqrt{i\omega\tau_{Wo}}}}{\sqrt{i\omega\tau_{Wo}}},
\end{align}
with the prefactor $B_{Wo}$ and the characteristic time scale $\tau_{Wo}$.
In a circuit picture, the Warburg short element corresponds to an electrical transmission line that is short-circuited at its end.
It is the solution in the frequency domain for a diffusion problem with a transmissive boundary.
The Warburg open element, in contrast, corresponds to a transmission line that is terminated in an open circuit, preventing current flow at the end boundary.
It is the solution for a diffusion problem with a blocking boundary.

Which Warburg impedance is appropriate in which case?
We found in \cref{fig:perturbation_profiles} that the cases \mbox{$D_-\neq D_+$} and \mbox{$\Ubias\neq 0$} differ qualitatively in the symmetry of the salt perturbation:
\mbox{$D_-\neq D_+$} gives rise to an antisymmetric salt perturbation that vanishes for \mbox{$\omega\to 0$}, whereas \mbox{$\Ubias\neq 0$} gives rise to a symmetric salt perturbation that remains finite for \mbox{$\omega\to 0$}.
These symmetries translate directly to the two types of Warburg element:
In the antisymmetric case, we find \mbox{$\Laplace{c}\first(x=0)=0$} for all frequencies, so the center of the cell acts effectively as a transmissive boundary.
In the symmetric case, in turn, we find \mbox{$\partial\Laplace{c}\first/\partial x=0$} at $x=0$, that is, the center of the cell acts effectively as a blocking boundary.
We deduce that the cases with \mbox{$D_-\neq D_+$} and \mbox{$\Ubias=0$}, in which we found the antisymmetric salt perturbations, should be described by a Warburg short element, whereas the cases with \mbox{$D_-=D_+$} and \mbox{$\Ubias\neq 0$}, in which we found the symmetric salt perturbations, should be described by a Warburg open element.

\begin{figure}
    \centering
    \includegraphics[width=.7\linewidth]{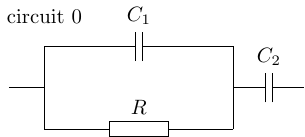}\vspace{1em}
    \includegraphics[width=.7\linewidth]{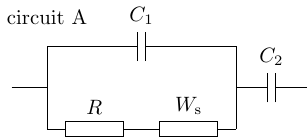}\vspace{1em}
    \includegraphics[width=.7\linewidth]{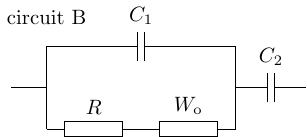}
    \caption{Proposed equivalent circuits for the description of flat-electrode EDL capacitor impedance. Circuit~0 describes cases with \mbox{$D_-=D_+$} and \mbox{$\Ubias=0$}, circuit~A describes cases with \mbox{$D_-\neq D_+$} and \mbox{$\Ubias=0$}, and circuit~B describes cases with \mbox{$D_-=D_+$} and \mbox{$\Ubias\neq 0$}.
    Circuit 0 was motivated by asymptotic analysis of Macdonald's analytical solution~\cite{Macdonald1953}, \cref{eq:Maconald_analytical_reiterated_maintext}. Circuits~A and B were constructed on the basis of circuit~0 by adding a Warburg short or Warburg open element, respectively, to account for salt diffusion on a finite length scale.}
    \label{fig:equivalent_circuits_overview}
\end{figure}

On the basis of these considerations, we propose the circuits shown in \cref{fig:equivalent_circuits_overview}, with circuit~0 for cases with \mbox{$D_-=D_+$} and \mbox{$\Ubias=0$}, circuit~A for cases with \mbox{$D_-\neq D_+$} and \mbox{$\Ubias=0$}, and circuit~B for cases with \mbox{$D_-=D_+$} and \mbox{$\Ubias\neq 0$}.
Note that a circuit similar to circuit~A was previously proposed by Franceschetti and Macdonald~\cite{Franceschetti1979} for their equivalent circuit fits.
In their circuit, $C_2$ was placed in the lower branch of the parallel subcircuit, rather than in series with it.
Our circuit~A has the advantage that for \mbox{$\omega\to 0$} its overall capacitance converges to $C_2$, which may thus be identified with the EDL capacitor's overall differential capacitance.
In order to fit spectra with both \mbox{$D_-\neq D_+$} and \mbox{$\Ubias\neq 0$}, we expect that an appropriate equivalent circuit will need to include a Warburg short and a Warburg open element on parallel electric paths.
The identification of the appropriate circuit for such cases is left for future research.

\subsection{The width $R_\mathrm{sl}$ of the slanted line\label{sec:equivalent_circuit_Rsl}}
Throughout \cref{sec:results}, we studied in detail the width $R_\mathrm{sl}$ of the slanted line, which had previously been misinterpreted as an EDL resistance, and which we found to be a mixed measure that reflects the interplay of salt adsorption by the EDLs and ambipolar diffusion in the bulk.
To analyse how $R_\mathrm{sl}$ is described in the scope of our proposed equivalent circuits, we consider the low-frequency limits of the involved Warburg elements.
In circuit~A, proposed for \mbox{$\Ubias=0$} and \mbox{$D_-\neq D_+$}, salt diffusion is represented by a Warburg short element, for which we find in the low-frequency limit the impedance
\begin{align}
    \label{eq:lowfreq_warburg_short}
    Z_{Ws}&\xrightarrow{\omega\to 0} B_{Ws}+\mathcal{O}(\omega).
\end{align}
In circuit~B, proposed for \mbox{$\Ubias\neq 0$} and \mbox{$D_-=D_+$}, salt diffusion is represented by a Warburg open element, for which we find in the low-frequency limit
\begin{align}
    \label{eq:lowfreq_warburg_open}
    Z_{Wo}&\xrightarrow{\omega\to 0} \frac{B_{Wo}}{3}+\frac{B_{Wo}}{i\omega\tau_{Wo}} + \mathcal{O}(\omega).
\end{align}
Assuming that $C_1$ and $C_2$ represent $C_\mathrm{bulk}$ and $C_\mathrm{EDL}/2$, respectively, it will hold for $L/\lD\gg 1$ that $C_2\gg C_1$.
In that case, the equivalent circuit will predict an impedance spectrum that exhibits (in Nyquist representation) a semicircle of diameter $R$, followed by a slanted line of width \mbox{$R_\mathrm{sl}=B_{Ws}$} for circuit~A, and \mbox{$R_\mathrm{sl}=B_{Wo}/3$} for circuit~B.
We thus note that, within our circuit analysis, $R_\mathrm{sl}$ is a measure of the Warburg prefactor.

\subsection{The influence of chemical capacitance\label{sec:equivalent_circuit_Cchem}}
We saw in \cref{fig:limRe_limCap}(a) that the system's differential capacitance approached a plateau for large $\Ubias$, and laid out that the plateau value corresponds to the finite amount of salt in the closed system.
We discussed in \cref{sec:impedance_vardiffrat} that this influence will only affect the impedance spectrum if the EDL charging involves a net adsorption of salt.
Whether and to which degree this is the case can be quantified with the charge efficiency, \mbox{$e(\partial W/\partial Q)_{c_0}$}.
To translate these qualitative considerations to an equivalent circuit picture, we consider again the analytical Poisson--Boltzmann expression for the differential capacitance $C$ in a closed system, \cref{eq:C_N_maintext}.
By Taylor-expanding this expression in $\lD/L$ and truncating at first order, we derive in \cref{sec:analytical_expressions_poisson_boltzmann} for sufficiently large $L/\lD$ the approximation
\begin{align}
    \label{eq:approx_C_N_maintext}
    \frac{1}{C}&\approx \frac{2}{C_\mathrm{EDL}} + \frac{\tanh^2(e\beta\PsiBias/2)}{C_\mathrm{chem}},
\end{align}
with the \textit{chemical capacitance}~\cite{Jamnik2001} of the bulk electrolyte being defined as
\begin{align}
    C_\mathrm{chem} := e^2 V_\mathrm{bulk} \left(\frac{\partial c_0}{\partial\mu}\right),
\end{align}
with the volume \mbox{$V_\mathrm{bulk}=A\cdot 2L$} of the EDL capacitor, the bulk salt concentration $c_0$, and the chemical potential $\mu$ of salt in the bulk electrolyte.
$C_\mathrm{chem}$ is a measure of how readily a certain amount of salt may be taken out of the bulk electrolyte.
The prefactor to $1/C_\mathrm{chem}$ in \cref{eq:approx_C_N_maintext}, in turn, is the squared charge efficiency of a Gouy--Chapman EDL (see \cref{eq:differential_charge_efficiency_gc} in the SI~\cite{supplemental_information}).

On the basis of \cref{eq:approx_C_N_maintext}, we motivate that, in an equivalent-circuit picture, the differential capacitance of the EDL capacitor may be understood as the series combination of two capacitances: first, the Gouy--Chapman capacitance of the EDLs, and second, an \textit{effective} chemical capacitance, which is given by
\begin{align}
    \label{eq:Cchem_eff}
    C_\mathrm{chem}^\mathrm{eff} := e^2\left(\frac{\partial W}{\partial Q}\right)_{c_0}^{-2}\,C_\mathrm{chem}.
\end{align}
We note that this description goes beyond the electrical characteristics derived by Bazant~\cite{Bazant_PRE_2004} for the thin-double-layer limit, in which the electrolyte's chemical capacitance is formally infinite and its geometric capacitance is negligible. 
Taking this limit, our expression \cref{eq:C_N_maintext} for the differential capacitance converges to $C_\mathrm{EDL}/2$, and Bazant's charging time \mbox{$\tau_\mathrm{c}=R_\mathrm{bulk}\,C_\mathrm{EDL}$} is recovered.

From the influence of the charge efficiency in \cref{eq:Cchem_eff}, we can deduce that an equivalent circuit for \mbox{$\Ubias\neq 0$} will need to include the effects of bulk chemical capacitance, whereas the impedance for \mbox{$\Ubias=0$} will be unaffected by chemical capacitance.
How does this relate to the proposed circuits~A and~B?
We showed in \cref{eq:lowfreq_warburg_short} that the low-frequency limit of the Warburg short element is purely resistive, so circuit~A describes the influence of ambipolar salt diffusion without modifications to the system's overall capacitance.
In contrast, we showed in \cref{eq:lowfreq_warburg_open} that the Warburg open element behaves, for \mbox{$\omega\to 0$}, as the series combination of a resistor and a capacitor, with capacitance \mbox{$\tau_{Wo}/B_{Wo}$} for \mbox{$\omega\to 0$}.
We thus postulate \mbox{$\tau_{Wo}/B_{Wo}=C_\mathrm{chem}^\mathrm{eff}$}.
Provided that $\tau_{Wo}=\tau_\mathrm{diff}$, we can thereby express $B_{Wo}$ purely in terms of known properties of the EDLs and the bulk electrolyte:
\begin{align}
    B_{Wo} = \frac{\tau_\mathrm{diff}\,e^2}{C_\mathrm{chem}}\left(\frac{\partial W}{\partial Q}\right)_{c_0}^2.
\end{align}
thereby relating the prefactor $B_{Wo}$ of the Warburg open element to the effective chemical capacitance.
Within the scope of the PNP (and PB) framework, the chemical capacitance is given by \mbox{$e^2\,V_\mathrm{bulk}\,\beta\,c_0/2$}, so we find
\begin{align}
    \label{eq:BWo_charge_efficiency_gc}
    \frac{B_{Wo}}{R_\mathrm{bulk}} = \frac{D_\mathrm{av}}{D_\mathrm{amb}}\,\tanh^2\left(\frac{e\beta\PsiBias}{2}\right).
\end{align}

\subsection{No bias, unequal diffusivities}
\begin{figure}
    \centering
    \includegraphics[width=\linewidth]{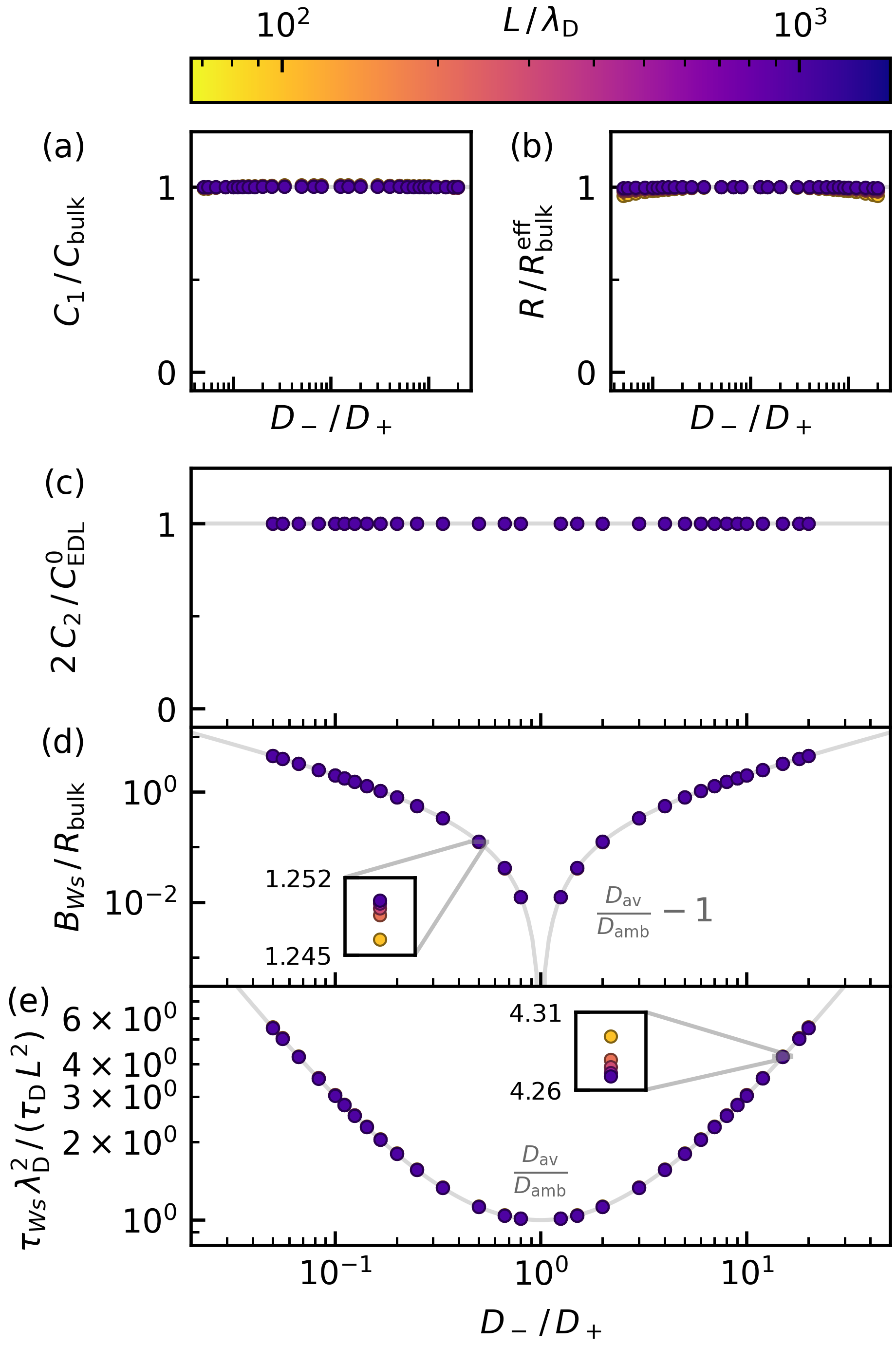}
    \caption{Circuit parameters $C_1$, $R$, $C_2$, $B_{Ws}$, and $\tau_{Ws}$, obtained by fitting equivalent circuit~A to impedance data obtained from simulations with various \mbox{$1/20\leq D_-/D_+\leq 20$} and \mbox{$100\leq L/\lD\leq 1000$}, given \mbox{$\PsiBias=0$} and \mbox{$\nup=0$}.
    }
    \label{fig:circuit_parameters_vardiffrat}
\end{figure}

We fitted circuit~A to simulated impedance spectra for various \mbox{$1/20\leq D_-/D_+\leq 20$} and \mbox{$100\leq L/\lD\leq 1000$}, given \mbox{$\Ubias=0$} and \mbox{$\nup=0$}.
\Cref{fig:circuit_parameters_vardiffrat} shows the resulting fitted values of (a)~$C_1$, (b)~$R$, (c)~$C_2$, (d)~$B_{Ws}$, and (e)~$\tau_{Ws}$.
We find good agreement between the fitted parameters with the expressions for $C_\mathrm{bulk}$ [\cref{eq:bulk_capacitance}], $R_\mathrm{bulk}$ [\cref{eq:bulk_resistance}], $C_\mathrm{EDL}^0$ [\cref{eq:edl_capacitance_no_bias}], respectively.
In \cref{fig:Rsl_and_ksl_vardiffrat}(a), we identified \mbox{$Z(\omega\to 0)=R_\mathrm{bulk}+R_\mathrm{sl}\propto 1/D_\mathrm{amb}$}.
Further, we identified for circuit~A that $R_\mathrm{sl}\approx B_{Ws}$ (see \cref{sec:equivalent_circuit_Rsl}).
Consistently, in \cref{fig:circuit_parameters_vardiffrat}(d) we find \mbox{$B_{Ws}/R_\mathrm{bulk}=D_\mathrm{av}/D_\mathrm{amb}-1$}.

As circuit~A fits excellently to the simulated data, our results contradict the claim made by Barbero~\cite{Barbero2017} that a Warburg impedance will arise only if electrodes are non-blocking.

\subsection{Applied bias, equal diffusivities}
\begin{figure}
    \centering
    \includegraphics[width=\linewidth]{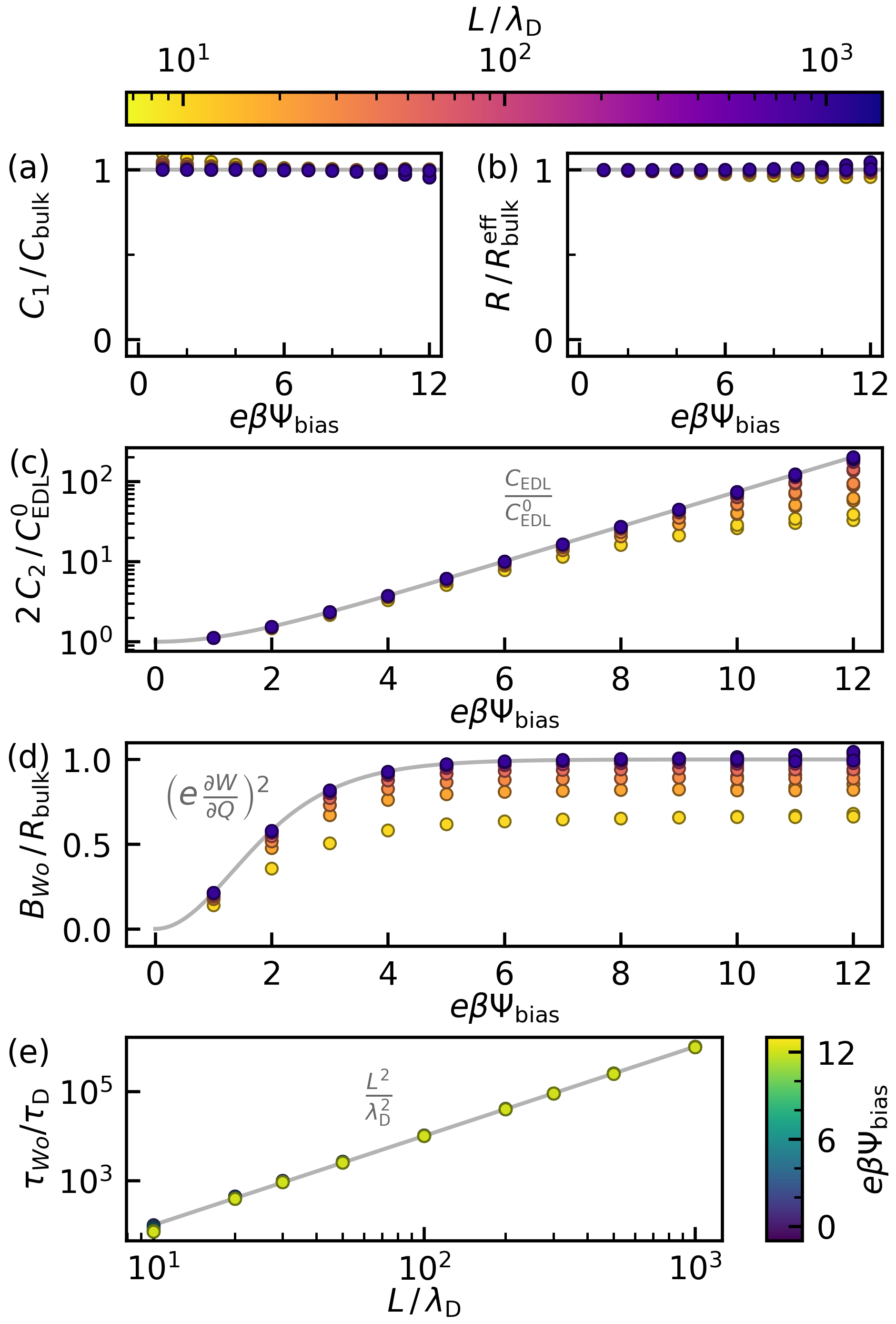}
    \caption{Impedance parameters obtained by fitting circuit~B [see \cref{fig:equivalent_circuits_overview}] to simulated impedance spectra for various \mbox{$\PsiBias$} and \mbox{$L/\lD$}, given \mbox{$D_-=D_+$} and \mbox{$\nup=0$}.
    }
    \label{fig:circuit_parameters_varbias}
\end{figure}

\Cref{fig:circuit_parameters_varbias} shows circuit parameters obtained from fits of circuit~B to simulated spectra for various \mbox{$1\leq e\beta\PsiBias\leq 12$} and \mbox{$10\leq L/\lD\leq 1000$}, given \mbox{$D_-=D_+$} and \mbox{$\nup=0$}.
Panels~(a) and~(b) show the fitted $C_1$ and $R$, which throughout the parameter range are well described by $C_\mathrm{bulk}$ [\cref{eq:bulk_capacitance}] and $R_\mathrm{bulk}^\mathrm{eff}$ [\cref{eq:Rbulk_eff}], respectively.
Panel~(c) shows the fitted $C_2$, which, for the largest $L/\lD$ is accurately described by the Gouy--Chapman capacitance [\cref{eq:differential_capacitance_PB}] up to the largest $\PsiBias$.
For smaller \mbox{$L/\lD$}, in turn, \cref{eq:differential_capacitance_PB} overestimates the fitted $C_2$.
Panel~(d) shows the bias dependence of the fitted Warburg prefactor $B_{Wo}$, which for the \mbox{$L/\lD\gtrsim 100$} is excellently described by \cref{eq:BWo_charge_efficiency_gc}.
We note that, through the connection \mbox{$R_\mathrm{sl}\approx B_{Wo}/3$}, we can now explain the bias-dependent trend of $R_\mathrm{sl}$: being a measure of \mbox{$e(\partial W/\partial Q)_{c_0}$}, $B_{Wo}$ is small for small $\PsiBias$ and increases with increasing $\PsiBias$, saturating for \mbox{$e\beta\PsiBias\gtrsim 4$}.
Lastly, we find in panel~(e) that the fitted Warburg time scale $\tau_{Wo}$ agrees closely with the definition in \cref{eq:tau_diff_ambipolar} for the characteristic diffusion time scale.

Franceschetti and Macdonald applied circuit~A also to cases with \mbox{$\PsiBias\neq 0$}.
They found the circuit to provide a good description of their impedance data in the case of a single-electrode setup with an open boundary.
In two-electrode setups, in turn, they obtained in this way different values of $\tau_{Ws}$ for different applied $\PsiBias$, which they interpreted in terms of a localization of diffusion effects.
Inspection of our concentration profiles in \cref{fig:perturbation_profiles} revealed that no such localization takes place.
Likewise, we find that, in our fits of circuit~B, $\tau_{Wo}$ is unaffected by $\PsiBias$ and corresponds consistently to the diffusion time scale.

\subsection{Applied bias, equal diffusivities, finite ion size}
\begin{figure}
    \centering
    \includegraphics[width=\linewidth]{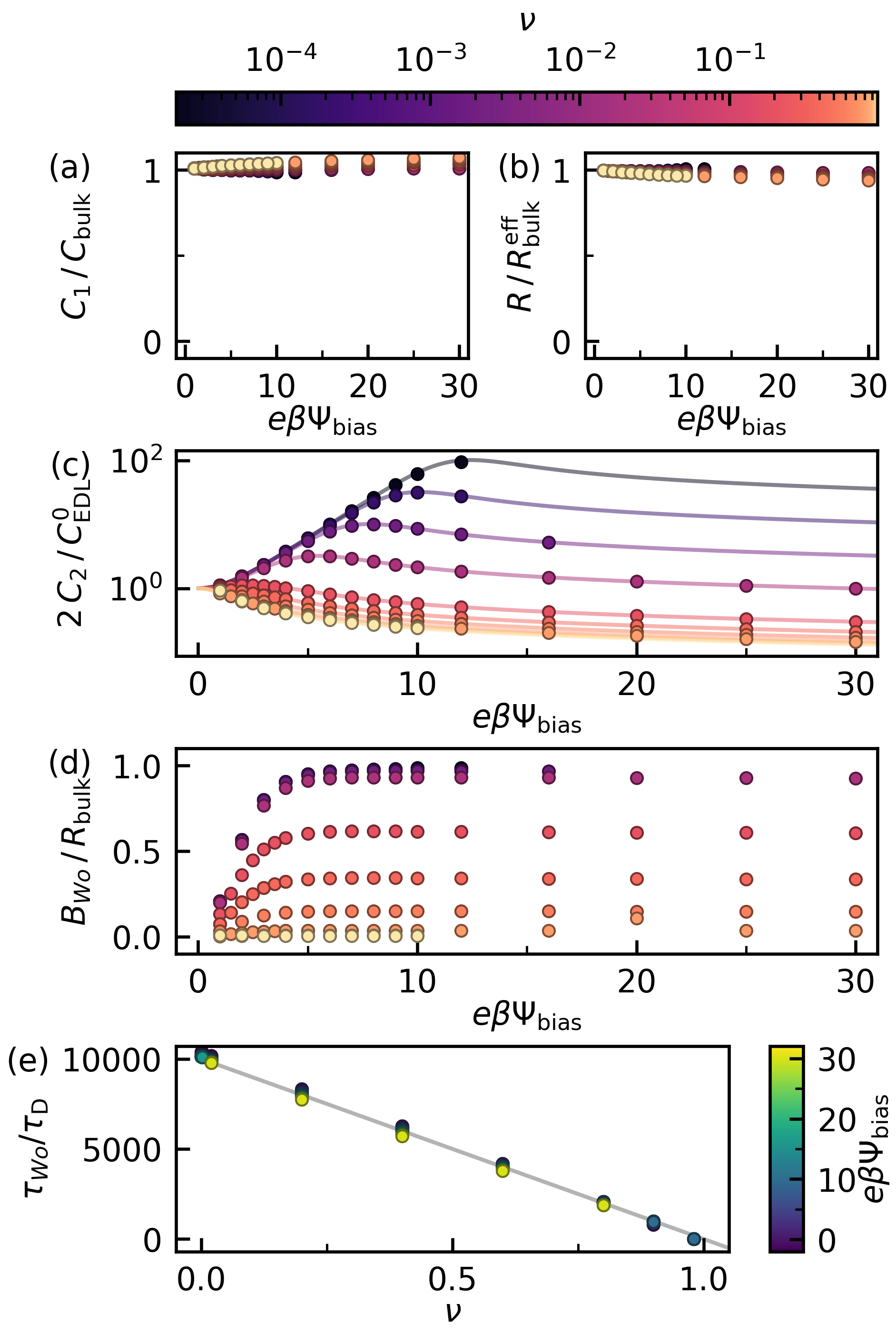}
    \caption{Impedance parameters obtained by fitting equivalent circuit B [see \cref{fig:equivalent_circuits_overview}] to simulated impedance spectra for various $\nup$ and $\PsiBias$, given \mbox{$L/\lD=100$} and \mbox{$D_+=D_-$}.
    }
    \label{fig:circuit_parameters_mpnp}
\end{figure}

\Cref{fig:circuit_parameters_mpnp} shows the circuit parameters obtained from fitting circuit~B to impedance data obtained for various \mbox{$2\cdot 10^{-5}\leq\nup\leq 0.98$} and applied biases $1\leq e\beta\PsiBias\leq 30$, with \mbox{$D_-=D_+$} and \mbox{$L/\lD=100$}.
Again, we find in panels~(a) and~(b) that $C_1$ is accurately described by $C_\mathrm{bulk}$ [\cref{eq:bulk_capacitance}], and $R$ by $R_\mathrm{bulk}^\mathrm{eff}$ [\cref{eq:Rbulk_eff}].
In panel~(c) we see that $C_2$ is excellently described by the EDL capacitance as derived in Ref.~\cite{Kilic_PRE_2007_1} for the modified Poisson--Boltzmann model [\cref{eq:differential_capacitance_mPB}.
Accordingly, we find that for small $\nup$, the obtained values of $C_2$ increase with increasing $\PsiBias$ at small $\PsiBias$, before reaching a maximum and decreasing at larger $\PsiBias$ (\textit{camel shape}).
For large $\nup$, the differential EDL capacitance decreases monotonically with increasing $\PsiBias$ (\textit{bell shape}).
Panel~(e) shows that $\tau_{Wo}$ is excellently described by the diffusive time scale \mbox{$(1-\nup)\,L^2/D$} that was derived in Ref.~\cite{Kilic_PRE_2007_2} for the mPNP model.

\section{Discussion}\label{sec:discussion}

\subsection{How salt fluxes affect impedance}
We established in \cref{sec:impedance_vardiffrat} that the slanted line is associated with ambipolar diffusion of salt throughout the bulk electrolyte.
Subsequently, our equivalent circuit analysis in \cref{sec:equivalent_circuit} confirmed that this phenomenon can be quantitatively described with Warburg elements; specifically, a Warburg short element for $D_-\neq D_+$ and a Warburg open element for $\Ubias\neq 0$.
In the literature, different scenarios were pointed out in which a diffusive feature will be observed in an impedance spectrum: notably, unequal ionic diffusivities, $D_-\neq D_+$~\cite{Barbero2007,Barbero2017}; an applied voltage bias, $\Ubias\neq 0$~\cite{Franceschetti1979,Garcia2025}, or semi-blocking boundaries~\cite{Brumleve1981, Jamnik1999, Jamnik2001, VanSoestbergen2010}.
But why does salt diffusion affect an impedance spectrum at all?
$Z$ is defined in terms of electric currents, which more naturally couple to ionic charge fluxes ($j_+-j_-$) than to salt fluxes ($j_++j_-$).

Bazant \textit{et al.} pointed out that salt diffusion arises in the nonlinear response of blocking-electrode systems because the EDLs at both electrodes adsorb equal amounts of salt, and that the involved symmetric salt fluxes cannot arise from electromigration~\cite{Bazant_PRE_2004}.
Consistently, López-García~\textit{et al.} argued that a diffusive flux must arise because co-ion and counter-ion concentrations will change by different amounts during EDL relaxation if $\Ubias\neq 0$---this is synonymous with a net change in the salt concentration.
These explanations leave no doubt as to \textit{why} diffusion must contribute to the charging process, but they do not address \textit{how} a salt flux arises in response to an applied electric field.

A finite salt flux corresponds to a mismatch of cation and anion fluxes, $j_-\neq -j_+$.
As pointed out by Barbero and Lelidis~\cite{Barbero2007}, such a mismatch can be caused by unequal cation and anion diffusivities, $D_-\neq D_+$.
A second possibility is that unequal cation and anion fluxes result from a difference in their local concentrations, $c_+\neq c_-$, as we demonstrate in \cref{sec:salt_charge_coupling} in the SI~\cite{supplemental_information}.
This corresponds to a non-zero local charge density, as found in an EDL.
This possibility was alluded to by Franceschetti and Macdonald~\cite{Franceschetti1979}, who suggested that the diffusion effects be associated with the surplus of one ionic species over the other.
A third way in which $j_-\neq -j_+$ can arise is through a boundary condition which breaks the antisymmetry \mbox{$j_-=-j_+$}.
This can be the case, for instance, for semi-blocking boundaries~\cite{Brumleve1981, Jamnik1999, Jamnik2001, VanSoestbergen2010}.

Mathematically, salt and charge evolution are decoupled only in systems with \mbox{$c_-=c_+$}, \mbox{$D_-=D_+$}, and boundary conditions that satisfy \mbox{$j_-=-j_+$}.
That is, salt diffusion affecting an EDL capacitor's impedance must be seen as the rule, rather than the exception.

\subsection{Ambipolar diffusion and its implications}
For the case of \mbox{$D_-\neq D_+$} and \mbox{$\PsiBias=0$}, Barbero~\textit{et al.} pointed out that a low-frequency plateau (the slanted line in a Nyquist plot) arises in the $\ReZ$ vs.~$\omega$ plot, and that it reflects ambipolar diffusion.
In this work, we found that the slanted line reflects ambipolar diffusion not only for \mbox{$D_-\neq D_+$}, but also when the slanted line appears as a result of \mbox{$\Ubias\neq 0$}, for \mbox{$D_-=D_+$}.
Both the characteristic frequency $\tau_\mathrm{diff}$ and the width $R_\mathrm{sl}$ of the slanted line were found to be determined by the ambipolar diffusion coefficient, $D_\mathrm{amb}$, throughout the considered parameter range.
In the commonly considered case of \mbox{$D_+=D_-$}, the average diffusion coefficient $D_\mathrm{av}$ and the ambipolar diffusion coefficient $D_\mathrm{amb}$ are equal.
As a consequence, previous treatments~\cite{Bazant_PRE_2004,Mei_JPCC_2018,Garcia2025} that presupposed \mbox{$D_+=D_-$} have not discriminated explicitly between $D_\mathrm{av}$ and $D_\mathrm{amb}$, but instead formulated quantities such as the diffusion time as $\tau_\mathrm{diff}=L^2D^{-1}$, in terms of an unspecified diffusivity \mbox{$D=D_+=D_-$}. 
We stress that, for more clarity, the diffusive time should be written as $\tau_\mathrm{diff}=L^2D_\mathrm{amb}^{-1}$.
If we consider EDL capacitor charging and take a system with \mbox{$D_-=D_+$} as a reference, it becomes clear in this way that replacing one of the ionic species with a faster-diffusing species will not accelerate the diffusive relaxation mode considerably, even though the electrolyte's ohmic resistance, being governed by $D_\mathrm{av}$, will drop.

The dependence of $\tau_\mathrm{diff}$ and $R_\mathrm{sl}$ on $D_\mathrm{amb}$ can further be employed to distinguish the influence of ambipolar diffusion on an impedance spectrum from possible other influences.
We mentioned in \cref{sec:introduction} that, for porous electrodes, a slanted-line region can arise from electromigration through the pore structure, rather than from diffusion.
Such a contribution can be described in terms of transmission-line models, which are in their mathematical form identical to the Warburg impedance associated with a diffusion effect.
As the electromigration through a pore structure is governed by $D_\mathrm{av}$, however, the slanted line will in such a case react differently to changes in \mbox{$D_-/D_+$} than in cases in which it arises from ambipolar diffusion.
In principle, this difference can be employed to distinguish between truly diffusive and pseudo-diffusive components in an impedance spectrum.
If the slanted line is due to ambipolar diffusion, $R_\mathrm{sl}/R_\mathrm{bulk}$ can become arbitrarily large for unequal diffusion coefficients, whereas $R_\mathrm{sl}/R_\mathrm{bulk}$ should be unaffected by the diffusivity ratio if the slanted line reflects electromigration.
Note, however, that any deviation from \mbox{$D_-=D_+$} will give rise to ambipolar diffusion, thus making the quantitative analysis more difficult.
The study of ambipolar diffusion in porous structure is left for future research.

\subsection{The false notion of an EDL resistance}
Our results demonstrated that, contrary to previous claims~\cite{Mei_JPCC_2018}, the width $R_\mathrm{sl}$ of the slanted line in a Nyquist plot does not reflect an EDL resistance.
The idea of an EDL resistance is appealing because it assigns a tangible name to an unintuitive quantity with units of resistance, and because there are several notable examples in which a sample's electrical resistance is indeed affected considerably by EDLs, or, in solids, by space-charge layers.
For instance, the presence of EDLs can strongly reduce the resistance of a porous electrode through surface conduction~\cite{Mirzadeh2014}.
The presence of space-charge layers in electroceramics, in turn, can either reduce~\cite{Tschope2001_1, Tschope2001_2} or increase~\cite{Verkerk1982,Vollman1994,Guo2002,Guo2003,Shirpour2012} the sample's electrical resistance.
The general rule is that a sample's macroscopic conductivity will be substantially affected by EDLs or space-charge layers if they constitute either highly conductive pathways in parallel to the bulk regions, or highly resistive paths in series with them~\cite{Maier1986}.
In a flat-electrode EDL capacitor, neither of these conditions is met: the EDLs are highly conductive, but they are electrically in series with the bulk electrolyte.
Hence, they do not contribute noticeably to the system's ohmic resistance, even though they can influence the system's electrical behavior drastically through charge accumulation and salt adsorption, as we have laid out in this work.

\subsection{In defense of equivalent circuits}
In \cref{sec:equivalent_circuit} we saw that equivalent circuits, despite being only an approximation to the charging problem, are highly useful to extract characteristic quantities, such as bulk-electrolyte resistance or EDL capacitance.
Fitting equivalent circuits constitutes one of three possible possible approaches to the interpretation of experimental impedance, and it is arguably the most commonly used one.
The second approach, discussed by Macdonald~\cite{Macdonald2013}, is to fit the solution, analytical or numerical, of a transport model such as PNP directly to impedance spectrum.
The third approach, discussed by Mei \textit{et al.}~\cite{Mei_JPCC_2018}, is to resort to visual inspection of Nyquist plots.

The second approach of fitting a transport model appears superior to equivalent circuit fits, both in terms of precision and physical insight.
But what if the transport model is inadequate?
A PNP model, for instance, cannot be fitted properly to data obtained for an electrolyte with a large packing fraction, since it cannot reproduce a decreasing EDL capacitance with increasing $\PsiBias$ (cf.~\cref{fig:circuit_parameters_mpnp}).
If one then adds auxiliary parameters, such as the Stern-layer width, to the model, one will likely misinterpret the observed trends.
In contrast, we saw in \cref{sec:equivalent_circuit} that the same circuit~(B) provided us with excellent fits for both PNP and mPNP data.
A circuit that properly represents the structure of the transport problem at hand will not be bound to a particular transport model.

As to the third approach, Mei \textit{et al.} claimed that a graphical inspection can alleviate the need for equivalent circuit fits altogether~\cite{Mei_JPCC_2018}.
Before least-squares fitting became a commonly available tool, the standard approach to extract quantities such as the bulk-electrolyte resistance from impedance spectra was by means of geometrical constructions.
Said constructions were, however, devised on the basis of equivalent circuits.
By interpreting the diameter of a semicircular arc as an ohmic resistance, one follows implicitly the logic of a parallel connection of a resistor and capacitor.
More generally, any subdivision of an impedance spectrum into a sum of constituent parts presupposes that these parts are electrically in series.
That is, extracting electrical quantities by visual inspection corresponds to an equivalent circuit fit with subjective error-weighting.
The major issue with a heuristic, visual approach, however, is that the form of the implicit equivalent circuit need not be stated.
As a consequence, conceptual errors may be hard to spot.
Notably, a slanted-line region as in \cref{fig:schematic_nyquist_four_features,fig:impedance_spectra_varbias,fig:impedance_spectra_var_diffrat} cannot be fitted with resistors and capacitors alone, unless they are arranged in an infinite transmission line.
An equivalent circuit analysis thus reveals that the width of the slanted-line region does not simply reflect the ohmic resistance of a particular component in the system, whereas this conflict goes unnoticed in a visual approach.

Regardless of how impedance data are analysed, reliable physical insight is not reached purely on the grounds of quantitative agreement.
Instead, it is constructed through characteristic dependencies or invariances: for instance, the bulk (geometric) capacitance will be within certain bounds independent of an applied bias, and the EDL capacitance independent of electrode separation (cf.~\cref{fig:circuit_parameters_varbias}).
Analysis of experimental data with equivalent circuits can help identify such trends without restricting the analysis to a particular transport model, or to an unjustified set of implicit assumptions.

\section{Conclusions}\label{sec:conclusions}
Through numerical simulations, we obtained impedance spectra for flate-plate EDL capacitors. 
We analyzed how the impedance spectra are affected by an applied voltage bias, electrode separation, ion diffusivity difference, and ion steric exclusion.
While our simulations do not explicitly include a Stern layer and a finite electrode resistance, we derived in \cref{sec:Stern_resistance} of the SI~\cite{supplemental_information} that these influences do not affect our conclusions. Our main findings are as follows.

First, we demonstrated that the slanted line appearing in our simulated spectra represents the ambipolar diffusion of salt.
This feature appears in the spectrum if a finite bias $\Ubias\neq 0$ is applied, or if cation and anion diffusivities are unequal, $D_+\neq D_-$.
We unify previous explanations~\cite{Franceschetti1979,Barbero2007,Garcia2025} of why a diffusive effect arises in impedance measurements by the common element that the antisymmetry \mbox{$j_-=-j_+$} of anion and cation fluxes is broken, whether it is in the bulk electrolyte, by $D_+\neq D_-$, locally in the EDLs, by $c_+\neq c_-$, or by flux boundary conditions that do not satisfy \mbox{$j_-=-j_+$}.

Second, we scrutinized previously formulated heuristic rules~\cite{Mei_JPCC_2018} for the width $R_\mathrm{sl}$ and slope of the slanted line.
We disproved both the notion that $R_\mathrm{sl}$ represents an EDL resistance and that $k_\mathrm{sl}$ is a measure of the ratio of time constants for diffusion and charging through ohmic conduction.

Third, we proposed equivalent circuits that accurately describe the spectra by representing salt diffusion for \mbox{$D_-\neq D_+$} and \mbox{$\Ubias\neq 0$} by a Warburg short and Warburg open element, respectively.
For the cases (i)~\mbox{$D_-\neq D_+$} and (ii)~\mbox{$\Ubias\neq 0$}, we showed that the prefactors of the Warburg elements, when normalized to the bulk-electrolyte resistance, are a measure of (i)~the mismatch of cation and anion diffusivities, (ii)~the relation between salt adsorption and charge accumulation in the EDLs, as quantified by the differential charge efficiency.
By highlighting the connection between the Warburg open element and chemical capacitance, we further demonstrated how the effects of a finite amount of ions in the bulk electrolyte may be formalized in an equivalent-circuit picture.

On the basis of our results, we suggest to utilize the different scaling with the diffusivity ratio of the ambipolar and average diffusion constants: truly diffusive features should be governed by $D_\mathrm{amb}$, wheras pseudo-diffusive contributions, known to arise from electromigration in porous electrodes, should be governed by $D_\mathrm{av}$.
Measuring porous-electrode impedance repeatedly with different electrolytes should thereby provide insight into the origins of a (pseudo-)diffusive impedance feature. 
Future theoretical work may explore how ambipolar diffusion influences the impedance as well as the nonlinear charging and discharging dynamics in porous electrodes.

 \section*{Acknowledgements}
This work was supported by a FRIPRO grant from The Research Council of Norway (Project No. 345079).

 \section*{Author Declarations}
The authors have no conflicts to disclose.

\section*{Data availability}
All simulation scripts and the generated data will be made available.

\bibliographystyle{apsrev4-2}
\bibliography{bibliography}

\clearpage
\renewcommand{\theequation}{{S}\arabic{equation}}\setcounter{equation}{0}
\renewcommand{\thefigure}{{S}\arabic{figure}}\setcounter{figure}{0}
\renewcommand{\thesection}{{S}\arabic{section}}\setcounter{section}{0} 
\renewcommand{\thetable}{{S}\arabic{table}}\setcounter{table}{0} 
\renewcommand{\theHequation}{Equation.\theequation}
\renewcommand{\theHfigure}{Figure.\thefigure}
\renewcommand{\theHsection}{Section.\thesection}
\renewcommand{\theHtable}{Table.\thesection}

\begin{widetext}
\section*{Supplemental Information for: ``The impedance of a charged flat-plate electric double-layer capacitor''}
{\centering Adrian L. Usler,  David Fertig, and Mathijs Janssen}

\section{Influence of a Stern layer and a finite electrode resistance on the impedance spectrum}\label{sec:Stern_resistance}
In the main text, we obtained impedance spectra for systems without electrode resistance and no Stern layers. 
These additional influences affect the impedance spectrum in a predictable way.
In fact, they enter the impedance spectrum simply as summand terms, as we will show below.
To this end, we derive a relation between the impedance $Z$ of the system, including electrodes and Stern layers, and the impedance $Z_\mathrm{d}$ of the diffuse layer.
The latter quantity corresponds to the impedance that we extract from our simulations.

We consider an electrolyte between two planar electrodes, with Stern layers of width $L_\mathrm{s}$ at the electrolyte--electrode interfaces and a finite electrode width $L_\mathrm{e}$ (see \cref{fig:schematic_electrode_stern_diffuse_layer}).
The boundaries between the Stern layers and the diffusive parts of the electric double layers (EDLs) are located at $x=\pm L$.
Note that, in this definition, we differ from Mei \textit{et al.}~\cite{Mei_JPCC_2018}, who included the Stern layer in the length $L$. Both descriptions are physically equivalent. 

\begin{figure}[hb]
    \centering
    \includegraphics[width=0.5\linewidth]{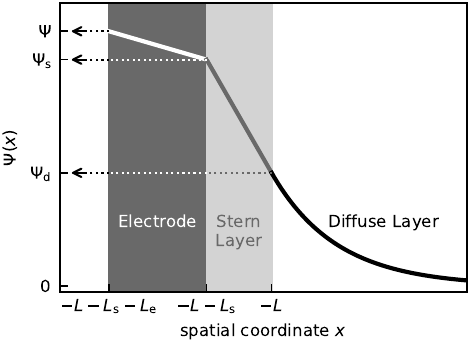}
    \caption{Illustration of the voltage drops over electrode, Stern layer, and diffuse layer.}
    \label{fig:schematic_electrode_stern_diffuse_layer}
\end{figure}

As described in \cref{sec:impedance_spectra} of the main text, we consider a small pulse voltage perturbation around the reference bias $U^\mathrm{bias}$, applied at the outer edges of the electrode,
\begin{equation}
    \psi\bm{(}x=\mp(L+L_{\mathrm{s}}+L_\mathrm{e}), t\bm{)} = \pm\left[\PsiBias + \delta\Psi(t)\right],
\end{equation}
The interior of the electrode is charge neutral; hence the potential is linear within the electrode, and hence the potential at the Stern layer--electrode interface is
\begin{align}
    \Psi_\mathrm{S}(t) :&= \psi(x=-L-L_{\mathrm{s}}, t)
    = \Psi(t) + L_\mathrm{e}\left.\ddx^-\psi\right|_{x=-L-L_{\mathrm{s}}},
\end{align}
where $\partial_x^-$ denotes the left-side partial derivative.
The current $I(t)$ that flows in the electrode is related to the electric field in the electrode by
\begin{equation}\label{eq:it}
    I(t) = -A\,\varrho_\mathrm{e}^{-1}\left.\ddx^-\psi\right|_{x=-(L+L_{\mathrm{s}})},
\end{equation}
where $\varrho_\mathrm{e}$ is the electrode resistivity.
Gauss' law at the electrode--Stern layer interface states that
\begin{align}
    Q(t)=-A\varepsilon_0\epsr\partial_x^+\psi|_{x=-(L+L_{\mathrm{s}})}.
\end{align}
where $\partial_x^+$ denotes the right-side spatial derivative.
$Q(t)$ represents the electronic charge accumulation at the electrode--Stern layer interface.
The electronic surface charge is built up at $x=-(L+L_{\mathrm{s}})$ through the electronic current, i.e. $Q(t)=\int I(t)\,\diff t$.
Differentiating with respect to $t$ yields
\begin{align}\label{eq:it2}
    I(t)=\partial_t Q(t)=-A\varepsilon_0\epsr\partial_t\partial_x^+\psi|_{x=-(L+L_{\mathrm{s}})}.
\end{align}
Combining \cref{eq:it,eq:it2} yields
\begin{align}\label{eq:currconv}
    \varepsilon_0\epsr\partial_t\partial_x^+\psi|_{x=-(L+L_{\mathrm{s}})}=\varrho_\mathrm{e}^{-1}\left.\ddx^-\psi\right|_{x=-(L+L_{\mathrm{s}})}.
\end{align}
\Cref{eq:currconv} is the current conservation at the electrode--Stern layer interface.
Laplace-transforming \cref{eq:currconv} gives
\begin{align}\label{eq:ddxminstern}
    \left.\ddx^-\Laplace{\psi}\right|_{x=-(L+L_{\mathrm{s}})}=\varepsilon_0\epsr\varrho_\mathrm{e}s\partial_x^+\Laplace{\psi}|_{x=-(L+L_{\mathrm{s}})}.
\end{align}

The Stern layer is charge neutral, and thus
\begin{align}\label{eq:diffpot}
    \Psi_\mathrm{D}(t) &:= \psi(x=-L, t) \nn
    &= \Psi_\mathrm{S}(t) + L_{\mathrm{s}}\left.\ddx^-\psi\right|_{x=-L}\nn
    &= \Psi(t) + L_\mathrm{e}\left.\ddx^-\psi\right|_{x=-(L+L_{\mathrm{s}})}  + L_{\mathrm{s}}\left.\ddx^-\psi\right|_{x=-L}.
\end{align}
As $\left.\ddx^-\psi\right|_{x=-L}=\partial_x^+\psi|_{x=-(L+L_{\mathrm{s}})}$, combining \cref{eq:ddxminstern} with the Laplace transform of \cref{eq:diffpot} gives
\begin{align}\label{eq:diffpotlapl}
 \Laplace{\delta\Psi}_{\mathrm{D}}(s)= \Laplace{\delta\Psi}(s)+(L_{\mathrm{s}}+\varepsilon_0\epsr\varrho_\mathrm{e}sL_{\mathrm{e}})\left.\ddx^-\Laplace{\psi}\right|_{x=-L}.  
\end{align}

The impedance $Z$ of the considered half cell is then given by using \cref{eq:ddxminstern,eq:diffpotlapl},
\begin{align}
    Z&=\dfrac{\Laplace{\delta\Psi}(s)}{\Laplace{\delta I}(s)}=\dfrac{\Laplace{\delta\Psi}_{\mathrm{D}}(s)-(L_{\mathrm{s}}+\varepsilon_0\epsr\varrho_\mathrm{e}sL_{\mathrm{e}})\left.\ddx^-\Laplace{\psi}\first\right|_{x=-L}}{-sA\varepsilon_0\epsr \left.\ddx^-\Laplace{\psi}\first\right|_{x=-L}}\nn
    &=L_{\mathrm{e}}\varrho_{\mathrm{e}}+\dfrac{L_\mathrm{s}}{sA\varepsilon_0\epsr}
    -\dfrac{\Laplace{\delta\Psi}_{\mathrm{D}}}{sA\varepsilon_0\epsr \left.\ddx^-\Laplace{\psi}\right|_{x=-L}} \nn
    &=R_{\mathrm{e}}+\dfrac{1}{sC_{\mathrm{s}}}+Z_{\mathrm{D}}.
\end{align}
The above expression is the impedance of a series connection of, first, a resistor (electrode), second, a capacitor (Stern layer), and third, an EDL capacitor with no electrode resistance or Stern layer thickness.
\Cref{fig:equivalent_circuit_Stern_and_Rs} shows this circuit.

\begin{figure}
    \centering
    \includegraphics[width=0.3\linewidth]{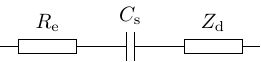}
    \caption{Equivalent circuit for a model parallel plate EDL capacitor with resistive electrodes and Stern layers at the electrode-electrolyte interface. }
    \label{fig:equivalent_circuit_Stern_and_Rs}
\end{figure}

\section{Linearization of the transport problem and transformation to the frequency domain}\label{sec:linearization_transport}

\begin{table}[htb]
    \centering
    \caption{Definitions of the nondimensional parameters}
    \begin{tabular}{cll}
    \textbf{dimensional} & \textbf{nondimensional} & \textbf{description} \\\hline
    $x$ & $\xnondim=x/\lD$ & spatial coordinate\\
    $t$ & $\tnondim=t\,D_+/\lD^2$ & time variable\\
    $\psi$ & $\psinondim=e\beta\psi$ & electric potential\\
    $c_\pm$ & $\cnondim_\pm=c_\pm/c_0$ & ion concentration\\
    $j_\pm$ & $\nondim{j}_\pm=j_{\pm}\lD/(c_0 D_+)$ & ion flux\\
    $L$ & $\Lnondim=L/\lD$ & EDL separation\\
    $\Psi$ & $\nondim{\Psi}=e\beta \Psi$ & electrode potential\\
    -- & $\nup=2\,a^3\,c_0$ & packing fraction\\
    $Q$ & $\nondim{Q}=Q/(2e c_0 A \lD)$ & surface charge\\
    $W$ & $\nondim{W}=W/(2 c_0 A \lD)$ & excess salt amount\\
    $C$ & $\nondim{C}=\lD C / (A \epsr\varepsilon_0)$ & differential capacitance\\\hline
    \end{tabular}
    \label{tab:nondimensional_parameters}
\end{table}

\subsection{Equations in the time domain}
We nondimensionalize the governing equations [\cref{eq:mPNP,eq:noflux_bc,eq:potential_bcs} of the main text]
by scaling lengths to the Debye length $\lD=\sqrt{\varepsilon_0\epsr kT/(2c_0 e^2)}$, times to the time scale $\tau_+=\lD^2/D_+$, potentials to the thermal voltage, and concentrations to the bulk concentration.
The resulting dimensionless variables and parameters are listed in \cref{tab:nondimensional_parameters}.
In terms of these variables and parameters, the mPNP equations [\cref{eq:mPNP} of the main text] read
\begin{subequations}\label{eq:mpnp_nondim}
\begin{align}
    -\ddxnondim^2 \psinondim &= \frac12\,\left(\cnondim_+-\cnondim_-\right),\label{eq:mpnp_nondim_poisson}\\
    \ddtnondim \cnondim_\pm& = -\ddxnondim \jnondim_\pm,\label{eq:mpnp_nondim_continuity}\\
    -\jnondim_+ &= \ddxnondim \cnondim_+ + \cnondim_+\,\ddxnondim\psinondim + \frac{\nup\,\cnondim_+ \ddxnondim\left(\cnondim_+ + \cnondim_-\right)}{1-\nup\,\cnondim_+ - \nup\,\cnondim_-}\label{eq:mpnp_cat_nondim}\\
    -\jnondim_- &= \frac{D_-}{D_+}\left(\ddxnondim \cnondim_- - \cnondim_-\,\ddxnondim\psinondim + \frac{\nup\,\cnondim_- \ddxnondim\left(\cnondim_+ + \cnondim_-\right)}{1-\nup\,\cnondim_+ - \nup\,\cnondim_-}\right)\label{eq:mpnp_an_nondim}
\end{align}
\end{subequations}
and the boundary conditions \cref{eq:noflux_bc,eq:potential_bcs} of the main text turn into
\begin{subequations}\label{eq:boundary_condition_dimless}
\begin{align}
    \psinondim(\xnondim=\pm \Lnondim,\tnondim) &= \pm\nondim{\Psi}(\tnondim),\label{eq:bc_potential_dimless}\\
    \jnondim_+(\xnondim=\pm \Lnondim, \tnondim) &= \jnondim_-(\xnondim=\pm \Lnondim, \tnondim) =0.
\end{align}    
\end{subequations}

In \cref{eq:phi_n_expansions} of the main text, we expanded $\psi$, $c_+$, and $c_-$ as the sum of a reference profile and a small perturbation, with the small parameter $\smallparam=e\beta\PsiAmp$. 
In terms of the dimensionless variables,
\begin{subequations}
\begin{align}
    \psinondim(\xnondim, \tnondim) &= \psinondim\zeroth(\xnondim) + \smallparam\,\psinondim\first(\xnondim, \tnondim) + \mathcal{O}\left(\smallparam^2\right),\\
    \cnondim_\pm(\xnondim, \tnondim) &= \cnondim\zeroth_\pm(\xnondim) + \smallparam\,\cnondim\first_\pm(\xnondim, \tnondim) + \mathcal{O}\left(\smallparam^2\right).
\end{align}    
\end{subequations}
To obtain the first-order perturbation theory of the mPNP equations [\cref{eq:mpnp_nondim}], we calculate the first derivative with respect to $\PhiAmp$, and in this derivative collect the terms that are $\mathcal{O}(\epsilon^0)$. 
For the Poisson equation~\eqref{eq:mpnp_nondim_poisson}, we find 
\begin{align}
     \frac{\partial}{\partial\smallparam}\left(-\ddxnondim^2 \psinondim\right) &= \frac{\partial}{\partial\smallparam}\left[\frac12\,\left(\cnondim_+ - \cnondim_-\right)\right]\nn
     \implies -\ddxnondim^2 \psinondim\first \cancel{+\mathcal{O}(\smallparam)} &= \frac12\,\left(\cnondim\first_+ - \cnondim\first_-\right) \cancel{+\mathcal{O}(\smallparam)}.  \label{eq:nondim_firstorder_poisson} 
\end{align}
Analogously, for the evolution of the ionic number densities [\cref{eq:mpnp_nondim_continuity,eq:mpnp_cat_nondim}], we find
\begin{align}\label{eq:nondim_firstorder_flux}
    \begin{aligned}
     \ddtnondim \cnondim\first_\pm &= -\ddxnondim\nondim{j}\first_\pm,\\
     -\nondim{j}\first_+ &= \ddxnondim \cnondim\first_+
     + \cnondim\first_+\,\ddxnondim\psinondim\zeroth + \cnondim\zeroth_+\ddxnondim\psinondim\first\\
     &+ \nup\,\left(\frac{\cnondim\first_+ - \nup\,\cnondim\zeroth_-\,\cnondim\first_+ + \nup\,\cnondim\zeroth_+\,\cnondim\first_-}{(1-\nup\,\cnondim\zeroth_+-\nup\,\cnondim\zeroth_-)^2}\right)\,\ddxnondim\left(\cnondim\zeroth_+ + \cnondim\zeroth_-\right)
     + \frac{\nup\,\cnondim\zeroth_+}{1-\nup\,\cnondim\zeroth_+ - \nup\,\cnondim\zeroth_-}\,\ddxnondim(\cnondim\first_+ + \cnondim\first_-)\\
     -\nondim{j}\first_- &= \frac{D_-}{D_+}\Bigg[\ddxnondim \cnondim\first_-
     - \cnondim\first_-\,\ddxnondim\psinondim\zeroth - \cnondim\zeroth_-\ddxnondim\psinondim\first\\
     &+ \nup\,\left(\frac{\cnondim\first_- - \nup\,\cnondim\zeroth_+\,\cnondim\first_- + \nup\,\cnondim\zeroth_-\,\cnondim\first_+}{(1-\nup\,\cnondim\zeroth_+-\nup\,\cnondim\zeroth_-)^2}\right)\,\ddxnondim\left(\cnondim\zeroth_+ + \cnondim\zeroth_-\right)
     + \frac{\nup\,\cnondim\zeroth_-}{1-\nup\,\cnondim\zeroth_+ - \nup\,\cnondim\zeroth_-}\,\ddxnondim(\cnondim\first_+ + \cnondim\first_-)\Bigg]
     \end{aligned}
\end{align}
To first order in $\smallparam$, the boundary conditions~\eqref{eq:boundary_condition_dimless} read
\begin{subequations}\label{eq:nondim_firstorder_boundary_conditions}
\begin{align}
    \psinondim\first(\xnondim=\pm \Lnondim,\tnondim) &= \pm\delta(t/\tau),\label{eq:bc_potential_dimless_perturbation}\\
    \nondimfirst{j}_+(\xnondim=\pm \Lnondim,\tnondim) &=
    \nondimfirst{j}_-(\xnondim=\pm \Lnondim,\tnondim) = 0,
\end{align}    
\end{subequations}
where, for \cref{eq:bc_potential_dimless_perturbation}, we have made the choice \mbox{$\delta\Psi(t)=\PsiAmp\, \delta(t/\tau_+)$}, in line with \cref{sec:impedance_spectra} of the main text.

\subsection{Equations in the frequency domain}
\newcommand{\phiLP}{\LaplaceNondim{\nondim \psi}\first}
\newcommand{\npLP}{\LaplaceNondim{\nondim c}\first_+}
\newcommand{\nmLP}{\LaplaceNondim{\nondim c}\first_-}
\newcommand{\npmLP}{\LaplaceNondim{\nondim c}\first_\pm}
Having nondimensionalized and linearized the transport equations, we now apply Laplace transformation to the resulting \cref{eq:nondim_firstorder_poisson,eq:nondim_firstorder_flux,eq:nondim_firstorder_boundary_conditions} in order to obtain the governing equations in the frequency domain.
Note that the Laplace transform as defined in \cref{sec:introduction} adds units of time to a given quantity, so the Laplace-transform $\Laplace{\nondim{f}}$ of a non-dimensionalized function $\nondim{f}(t)$ is not strictly non-dimensional.
Hence, we here introduce a non-dimensional Laplace-transform, $\LaplaceNondim{\nondim{f}}\equiv \int\limits_0^\infty \rme^{-\snondim\tnondim}\,\nondim{f}(\tnondim)\diff\tnondim$.
It holds that $\LaplaceNondim{\nondim{f}}=\Laplace{\nondim{f}}/\tau_+$.
As in \cref{eq:mpnp_first_laplace} in \cref{sec:impedance_spectra}, we use \mbox{$\cnondim_\pm\first(t=0)=0$}, so the Laplace-transformed equations read
\begin{subequations}\label{eq:pnp_laplace}
\begin{align}   
    \label{eq:pnp_poisson_lp}
    -\ddxnondim^2 \phiLP &= \frac12\,\left(\npLP-\nmLP\right),\\
    \label{eq:pnp_cat_lp}
    \snondim\,\npmLP &= -\ddxnondim\LaplaceNondim{\nondim{j}}\first_\pm\\
-\LaplaceNondim{\nondim{j}}\first_+ &= \ddxnondim \npLP
    + \cnondim\zeroth_+\,\ddxnondim\phiLP + \npLP\,\ddxnondim\psinondim\zeroth\nn
    &\qquad + \nup\,\left(\frac{\npLP - \nup\,\cnondim\zeroth_-\,\npLP+ \nup\,\cnondim\zeroth_+\,\LaplaceNondim{\cnondim}_-\first}{\left(1-\nup\,\cnondim\zeroth_+-\nup\,\cnondim\zeroth_-\right)^2}\right)\,\ddxnondim\left(\cnondim\zeroth_+ + \cnondim\zeroth_-\right)
    + \frac{\nup\,\cnondim\zeroth_+}{1-\nup\,\cnondim\zeroth_+ - \nup\,\cnondim\zeroth_-}\,\ddxnondim\left(\npLP + \nmLP\right)\\
-\LaplaceNondim{\nondim{j}}\first_- &= \frac{D_-}{D_+}\Bigg[\ddxnondim \nmLP
    - \cnondim\zeroth_-\,\ddxnondim\phiLP - \nmLP\,\ddxnondim\psinondim\zeroth\nn
    &\qquad + \nup\,\left(\frac{\nmLP - \nup\,\cnondim\zeroth_+\,\nmLP+ \nup\,\cnondim\zeroth_-\,\LaplaceNondim{\cnondim}_+\first}{\left(1-\nup\,\cnondim\zeroth_+-\nup\,\cnondim\zeroth_-\right)^2}\right)\,\ddxnondim\left(\cnondim\zeroth_+ + \cnondim\zeroth_-\right)
    + \frac{\nup\,\cnondim\zeroth_-}{1-\nup\,\cnondim\zeroth_+ - \nup\,\cnondim\zeroth_-}\,\ddxnondim\left(\npLP + \nmLP\right)\Bigg]
\end{align}
\end{subequations}
subject to the boundary conditions
\begin{subequations}\label{eq:nondim_firstorder_boundary_conditions_laplace}
\begin{align}
    \label{eq:bc_phi_lp}
    \phiLP(\xnondim=\pm \Lnondim, \snondim) &= \pm1,\\
    \LaplaceNondim{\nondim{j}}\first_+\left(\xnondim=\pm \Lnondim\right) = \LaplaceNondim{\nondim{j}}\first_-\left(\xnondim=\pm \Lnondim\right) &= 0.
\end{align}
\end{subequations}
Note that the Laplace transform was applied only to the first-order terms, $\nondim{f}\first(\xnondim,\tnondim)$, since the zeroth-order terms $\nondim{f}\zeroth(\xnondim)$ are time-independent.

We solved the weak form of \cref{eq:pnp_laplace,eq:nondim_firstorder_boundary_conditions_laplace} for $\phiLP$, $\nondimfirst{\LaplaceNondim c}_+$, and $\nondimfirst{\LaplaceNondim c}_-$ with the FEniCS package for Python.
The complex frequency was set to $\snondim=i\nondim{\omega}$.
Since the used version of FEniCS does not natively support complex-valued functions, we split up the functions explicitly into their real and imaginary parts.
\begin{subequations}\label{eq:pnp_lp_real_imag}
\begin{align}
    \label{eq:pnp_lp_real_poisson}
    -\ddxnondim^2\,\RealPart\,\phiLP &= \frac12\,\left(\RealPart\,\npLP-\RealPart\,\nmLP\right),\\
    \label{eq:pnp_lp_imag_poisson}
    -\ddxnondim^2\,\ImagPart\,\phiLP &= \frac12\,\left(\ImagPart\,\npLP-\ImagPart\,\nmLP\right)
    ,\\[1em]
    \label{eq:pnp_lp_real_pm}
    -\nondim{\omega}\,\ImagPart\,\npmLP &= -\ddxnondim\RealPart\,\LaplaceNondim{\nondim{j}}_\pm\first\\
    \label{eq:pnp_lp_imag_pm}
    \nondim{\omega}\,\RealPart\,\npmLP &= -\ddxnondim\ImagPart\,\LaplaceNondim{\nondim{j}}_\pm\first\\[1em]
-\RealPart\,\LaplaceNondim{\nondim{j}}_+\first &= \Bigg[\ddxnondim \RealPart\,\npLP
    + \cnondim\zeroth_+\,\ddxnondim\RealPart\,\phiLP + \RealPart\,\npLP\,\ddxnondim\psinondim\zeroth\nn
    &\qquad+ \nup\,\left(\frac{\RealPart\,\npLP - \nup\,\cnondim\zeroth_-\,\RealPart\,\npLP + \nup\,\cnondim\zeroth_+\,\RealPart\,\LaplaceNondim{\cnondim}_-\first}{\left(1-\nup\,\cnondim\zeroth_+-\nup\,\cnondim\zeroth_-\right)^2}\right)\,\ddxnondim\left(\cnondim\zeroth_+ + \cnondim\zeroth_-\right)\nn
    &\qquad+ \frac{\nup\,\cnondim\zeroth_+}{1-\nup\,\cnondim\zeroth_+ - \nup\,\cnondim\zeroth_-}\,\ddxnondim\left(\RealPart\,\npLP + \RealPart\,\nmLP\right)\Bigg]\\[1em]
-\ImagPart\,\LaplaceNondim{\nondim{j}}_+\first &= \Bigg[\ddxnondim \ImagPart\,\npLP
    + \cnondim\zeroth_+\,\ddxnondim\ImagPart\,\phiLP + \ImagPart\,\npLP\,\ddxnondim\psinondim\zeroth\nn
    &\qquad+ \nup\,\left(\frac{\ImagPart\,\npLP - \nup\,\cnondim\zeroth_-\,\ImagPart\,\npLP + \nup\,\cnondim\zeroth_+\,\ImagPart\,\LaplaceNondim{\cnondim}_-\first}{\left(1-\nup\,\cnondim\zeroth_+-\nup\,\cnondim\zeroth_-\right)^2}\right)\,\ddxnondim\left(\cnondim\zeroth_+ + \cnondim\zeroth_-\right)\nn
    &\qquad+ \frac{\nup\,\cnondim\zeroth_+}{1-\nup\,\cnondim\zeroth_+ - \nup\,\cnondim\zeroth_-}\,\ddxnondim\left(\ImagPart\,\npLP + \ImagPart\,\nmLP\right)\Bigg]\\[1em]
-\RealPart\,\LaplaceNondim{\nondim{j}}_-\first &= \frac{D_-}{D_+}\Bigg[\ddxnondim \RealPart\,\nmLP
    - \cnondim\zeroth_-\,\ddxnondim\RealPart\,\phiLP - \RealPart\,\nmLP\,\ddxnondim\psinondim\zeroth\nn
    &\qquad+ \nup\,\left(\frac{\RealPart\,\nmLP - \nup\,\cnondim\zeroth_+\,\RealPart\,\nmLP + \nup\,\cnondim\zeroth_-\,\RealPart\,\LaplaceNondim{\cnondim}_+\first}{\left(1-\nup\,\cnondim\zeroth_+-\nup\,\cnondim\zeroth_-\right)^2}\right)\,\ddxnondim\left(\cnondim\zeroth_+ + \cnondim\zeroth_-\right)\nn
    &\qquad+ \frac{\nup\,\cnondim\zeroth_-}{1-\nup\,\cnondim\zeroth_+ - \nup\,\cnondim\zeroth_-}\,\ddxnondim\left(\RealPart\,\npLP + \RealPart\,\nmLP\right)\Bigg]\\[1em]
-\ImagPart\,\LaplaceNondim{\nondim{j}}_-\first &= \frac{D_-}{D_+}\Bigg[\ddxnondim \ImagPart\,\nmLP
    - \cnondim\zeroth_-\,\ddxnondim\ImagPart\,\phiLP - \ImagPart\,\nmLP\,\ddxnondim\psinondim\zeroth\nn
    &\qquad+ \nup\,\left(\frac{\ImagPart\,\nmLP - \nup\,\cnondim\zeroth_+\,\ImagPart\,\nmLP + \nup\,\cnondim\zeroth_-\,\ImagPart\,\LaplaceNondim{\cnondim}_+\first}{\left(1-\nup\,\cnondim\zeroth_+-\nup\,\cnondim\zeroth_-\right)^2}\right)\,\ddxnondim\left(\cnondim\zeroth_+ + \cnondim\zeroth_-\right)\nn
    &\qquad+ \frac{\nup\,\cnondim\zeroth_-}{1-\nup\,\cnondim\zeroth_+ - \nup\,\cnondim\zeroth_-}\,\ddxnondim\left(\ImagPart\,\npLP + \ImagPart\,\nmLP\right)\Bigg],
\end{align}
\end{subequations}
subject to the boundary conditions
\begin{subequations}\label{eq:pnp_lp_real_imag_bcs}
\begin{align}
    \label{eq:pnp_lp_real_bc_phi}
    \RealPart\,\phiLP\left(\xnondim=\pm \Lnondim, \snondim\right) &= \pm 1,\\
    \label{eq:pnp_lp_imag_bc_phi}
    \ImagPart\,\phiLP\left(\xnondim=\pm \Lnondim, \snondim\right) &= 0,\\
    \label{eq:pnp_lp_real_bc_flux}
    \RealPart\,\LaplaceNondim{\nondim{j}}\first_+\left(\xnondim=\pm \Lnondim\right) = \RealPart\,\LaplaceNondim{\nondim{j}}\first_-\left(\xnondim=\pm \Lnondim\right) &= 0,\\
    \label{eq:pnp_lp_imag_bc_flux}
    \ImagPart\,\LaplaceNondim{\nondim{j}}\first_+\left(\xnondim=\pm \Lnondim\right) = \ImagPart\,\LaplaceNondim{\nondim{j}}\first_-\left(\xnondim=\pm \Lnondim\right) &= 0.
\end{align}
\end{subequations}
We implemented the numerical solution of \cref{eq:pnp_lp_real_imag,eq:pnp_lp_real_imag_bcs} by means of the finite element method, with the FEniCS package~\cite{Logg2010dolfin,Logg2012automated,Kirby2006ffc,Alnaes2014ufl} for Python.

From the numerical solutions, we obtain non-dimensional surface charges, $\LaplaceNondim{\nondim{Q}}\first_\mathrm{left}$ and~$\LaplaceNondim{\nondim{Q}}\first_\mathrm{right}$, as
\begin{align}
\begin{aligned}
    \LaplaceNondim{\nondim{Q}}\first_\mathrm{left} &= -2\left.\ddxnondim\LaplaceNondim{\psinondim}\first\right|_{\xnondim=-\Lnondim}
    \equiv \frac{\Laplace{Q}\first_\mathrm{left}}{\tau_+ e c_0 \lD A}\\
    \LaplaceNondim{\nondim{Q}}\first_\mathrm{right} &= \phantom{+}2\left.\ddxnondim\LaplaceNondim{\psinondim}\first\right|_{\xnondim=+\Lnondim}
    \equiv \frac{\Laplace{Q}\first_\mathrm{right}}{\tau_+ e c_0 \lD A},
\end{aligned}
\end{align}
from which we calculate a non-dimensional capacitor charge, $\LaplaceNondim{\nondim{Q}}\first$, as
\begin{align}
    \LaplaceNondim{\nondim{Q}}\first = \frac{\LaplaceNondim{\nondim{Q}}\first_\mathrm{right}-\LaplaceNondim{\nondim{Q}}\first_\mathrm{left}}{2}.
\end{align}
The nondimensional surface charge gives access to the non-dimensional impedance, $\nondim{Z}$, as
\begin{align}
    \nondim{Z} = \frac{2}{\snondim\LaplaceNondim{\nondim{Q}}\first}.
\end{align}
With the dimensional impedance $Z$ as defined in \cref{eq:impedance_from_simulations_dimensional}, we can formulate the relation
\begin{align}
    Z = \nondim{Z}\,\frac{\kB T \lD}{D_+ c_0 e^2 A},
\end{align}
and thus, with the bulk resistance $R_\mathrm{bulk}$ as defined in \cref{eq:bulk_resistance},
\begin{align}
    \frac{Z}{R_\mathrm{bulk}} =  \nondim{Z}\,\frac{\lD\,D_\mathrm{av}}{L\,D_+}.
\end{align}

\subsection{Weak form of the Laplace-transformed linearized transport equations}
Here we provide details on the weak form of the differential equations that are implemented in order to facilitate the understanding of our code.
The numerical solution is implemented in FEniCS by using a vector function space to include the $6$ components ($\RealPart\,\LaplaceNondim{\psinondim}\first$, $\ImagPart\,\LaplaceNondim{\psinondim}\first$, $\RealPart\,\LaplaceNondim{\cnondim}_+\first$, $\ImagPart\,\LaplaceNondim{\cnondim}_+\first$, $\RealPart\,\LaplaceNondim{\cnondim}_-\first$, $\ImagPart\,\LaplaceNondim{\cnondim}_-\first$).
Correspondingly, we employ a test space with components ($v_{\RealPart\,\psi}$, $v_{\ImagPart\,\psi}$, $v_{\RealPart\,c+}$, $v_{\ImagPart\,c+}$, $v_{\RealPart\,c-}$, $v_{\ImagPart\,c-}$).
With that, \cref{eq:pnp_lp_real_poisson,eq:pnp_lp_imag_poisson,eq:pnp_lp_real_pm,eq:pnp_lp_imag_pm} are rewritten to yield their weak forms
\begin{align}
    \label{eq:poisson_lp_real_weak_form}
    \int\limits_{-\Lnondim}^{\Lnondim}\ddxnondim\RealPart\,\LaplaceNondim{\psinondim}\first\,\ddxnondim v_{\RealPart\,\psi}\diff\xnondim - \cancel{\left(\ddxnondim\RealPart\,\LaplaceNondim{\psinondim}\first\,v_{\RealPart\,\psi}\right)_{-\Lnondim}^{\Lnondim}} &= \frac12\int\limits_{-\Lnondim}^{\Lnondim}(\LaplaceNondim{\cnondim}\first_+ - \LaplaceNondim{\cnondim}\first_-)\,v_{\RealPart\,\psi}\diff\xnondim\\
    \label{eq:poisson_lp_imag_weak_form}
    \int\limits_{-\Lnondim}^{\Lnondim}\ddxnondim\ImagPart\,\LaplaceNondim{\psinondim}\first\,\ddxnondim v_{\ImagPart\,\psi}\diff\xnondim - \cancel{\left(\ddxnondim\ImagPart\,\LaplaceNondim{\psinondim}\first\,v_{\ImagPart\,\psi}\right)_{-\Lnondim}^{\Lnondim}} &= \frac12\int\limits_{-\Lnondim}^{\Lnondim}(\LaplaceNondim{\cnondim}\first_+ - \LaplaceNondim{\cnondim}\first_-)\,v_{\ImagPart\,\psi}\diff\xnondim\\
    \label{eq:continuity_pm_lp_real_weak_form}
    \int\limits_{-\Lnondim}^{\Lnondim} \RealPart\,\LaplaceNondim{\nondim{j}}\first_\pm\,\ddxnondim v_{\RealPart\,c\pm}\diff\xnondim - \cancel{\left(\RealPart\,\LaplaceNondim{\nondim{j}}\first_\pm\,v_{\RealPart\,c\pm}\right)_{-\Lnondim}^{\Lnondim}} &= -\nondim{\omega}\int\limits_{-\Lnondim}^{\Lnondim}\ImagPart\,\LaplaceNondim{\cnondim}\first_\pm\,v_{\RealPart\,c\pm}\diff\xnondim\\
    \label{eq:continuity_pm_lp_imag_weak_form}
    \int\limits_{-\Lnondim}^{\Lnondim} \ImagPart\,\LaplaceNondim{\nondim{j}}\first_\pm\,\ddxnondim v_{\ImagPart\,c\pm}\diff\xnondim - \cancel{\left(\ImagPart\,\LaplaceNondim{\nondim{j}}\first_\pm\,v_{\ImagPart\,c\pm}\right)_{-\Lnondim}^{\Lnondim}} &= \phantom{+}\nondim{\omega}\int\limits_{-\Lnondim}^{\Lnondim}\RealPart\,\LaplaceNondim{\cnondim}\first_\pm\,v_{\ImagPart\,c\pm}\diff\xnondim
\end{align}
The boundary terms in \cref{eq:poisson_lp_real_weak_form,eq:poisson_lp_imag_weak_form} vanish because the Dirichlet conditions \cref{eq:pnp_lp_real_bc_phi,eq:pnp_lp_imag_bc_phi} are imposed, which practically involves that the boundary values of the test functions $v_{\RealPart\,\psi}$ and $v_{\ImagPart\,\psi}$ are set to zero.
The boundary terms in \cref{eq:continuity_pm_lp_real_weak_form,eq:continuity_pm_lp_imag_weak_form}, in turn, vanish because of the vanishing fluxes at the boundaries, as specified in the Neumann boundary conditions [\cref{eq:pnp_lp_real_bc_flux,eq:pnp_lp_imag_bc_flux}].

\section{Analytical expression for impedance for unequal diffusivities}\label{sec:impedance_unequal_diffu}
The nondimensional equations after linearization read
\begin{subequations}\label{eq:uneq_nondim}
\begin{align}
    \ddtnondim\cnondim_+ &=\ddxnondim^2\cnondim_++\ddxnondim^2\psinondim,\\
    \ddtnondim\cnondim_- &= \xi(\ddxnondim^2\cnondim_--\ddxnondim^2\psinondim),\\
    -\ddxnondim^2\psinondim &=\frac12\left(\cnondim_+-\cnondim_-\right),
\end{align}
\end{subequations}
where $\xi=D_-/D_+$.
We rewrite \cref{eq:uneq_nondim} in terms of charge densities \mbox{$\nondim{q}:=\cnondim_+-\cnondim_-$} and salt densities \mbox{$\cnondim:=\cnondim_+ + \cnondim_-$},
\begin{subequations}\label{eq:uneq_saltcharge_nondim}
    \begin{align}
        \ddtnondim\nondim{q} &= \left(\dfrac{1+\xi}{2}\right)\ddxnondim^2 \nondim{q}+\left(\dfrac{1-\xi}{2}\right)\ddxnondim^2 \cnondim -\left(\dfrac{1+\xi}{2}\right) \nondim{q},\\
        \ddtnondim \cnondim &= \left(\dfrac{1+\xi}{2}\right)\ddxnondim^2 \cnondim +\left(\dfrac{1-\xi}{2}\right)\ddxnondim^2 \nondim{q} -\left(\dfrac{1-\xi}{2}\right) \nondim{q}.
    \end{align}
\end{subequations}
Laplace transform of \cref{eq:uneq_saltcharge_nondim} yields
\begin{subequations}\label{eq:uneq_saltcharge_lapl}
    \begin{align}
        \snondim\Laplace{\nondim{q}} &= \left(\dfrac{1+\xi}{2}\right)\ddxnondim^2\Laplace{\nondim{q}}+\left(\dfrac{1-\xi}{2}\right)\ddxnondim^2 \Laplace{\cnondim} -\left(\dfrac{1+\xi}{2}\right)\Laplace{\nondim{q}},\\
        \snondim\Laplace{\cnondim}-2 &= \left(\dfrac{1+\xi}{2}\right)\ddxnondim^2\Laplace{\cnondim}+\left(\dfrac{1-\xi}{2}\right)\ddxnondim^2 \Laplace{\nondim{q}} -\left(\dfrac{1-\xi}{2}\right) \Laplace{\nondim{q}}.
    \end{align}
\end{subequations}
The general solution to \cref{eq:uneq_saltcharge_lapl} is given by
\begin{subequations}
    \begin{align}
        \Laplace{\nondim{q}}&=[A_1\sinh(k_+\xnondim)+A_2\cosh(k_+\xnondim)]\beta_++[A_3\sinh(k_- \xnondim)+A_4\cosh(k_-\xnondim)]\beta_-,\label{eq:expr_for_chg}\\
        \Laplace{\cnondim}&=\dfrac{2}{\snondim}+A_1\sinh(k_+\xnondim)+A_2\cosh(k_+ \xnondim)+A_3\sinh(k_-\xnondim)+A_4\cosh(k_-\xnondim),
    \end{align}
\end{subequations}
with $k_{\pm}=\sqrt{\dfrac{\xi+(1+\xi)\snondim\pm\sqrt{(1-\xi)^2\snondim^2+\xi^2}}{2\xi}}$ and $\beta_{\pm}=\dfrac{\xi\pm\sqrt{(1-\xi)^2\snondim^2+\xi^2}}{\snondim(\xi-1)}$.
Integrating \cref{eq:expr_for_chg} twice and using boundary conditions $\ddxnondim \Laplace{\cnondim}|_{x=\pm \Lnondim}=0$, $\ddxnondim \Laplace{\nondim{q}}+2\ddxnondim\Laplace{\psinondim}|_{x=\pm \Lnondim}=0$, and $\Laplace{\psinondim}|_{x=\pm \Lnondim}=\pm\nondim{\Psi}/\snondim$ results in an expression for the Laplace transformed potential
\begin{align}
    \Laplace{\psinondim}=\dfrac{\nondim{\Psi}}{\snondim}\dfrac{\sinh(k_+\xnondim)+\cosh(k_+\Lnondim)\left\{-\gamma^3f\dfrac{\sinh(k_-\xnondim)}{\cosh(k_-\Lnondim)}+[(k_+^2-1)+f(\gamma^2-k_+^2)]k_+\xnondim\right\}}{\sinh(k_+\Lnondim)+\cosh(k_+\Lnondim)\{-\gamma^3f\tanh(k_-\Lnondim)+[(k_+^2-1)+f(\gamma^2-k_+^2)]k_+\Lnondim\}},
\end{align}
where $\gamma = k_+/k_-$ and $f=\beta_-/\beta_+$. The impedance normalized by the bulk resistance then reads
\begin{align}\label{eq:uneq_impedance}
    \dfrac{Z}{R_{\mathrm{bulk}}}=\left(\dfrac{1+\xi}{2}\right)\dfrac{\tanh(k_+\Lnondim)-\gamma^3f\tanh(k_-\Lnondim)+[(k_+^2-1)+f(\gamma^2-k_+^2)]k_+\Lnondim}{\snondim\Lnondim k_+^3(1-f)}.
\end{align}
The low-frequency limit of the real part of \cref{eq:uneq_impedance} is given by
\begin{align}
    \label{eq:analytical_ReZ_diffrat_lowfreq_nondim}
    \lim\limits_{\snondim\to 0}\dfrac{\mathrm{Re}(Z)}{R_{\mathrm{bulk}}}=
    \dfrac{(1+\xi)^2}{4\xi}\left(1+\dfrac{1}{2}\left[1-\tanh^2\Lnondim\right]-\dfrac{3\tanh\Lnondim}{2\Lnondim}\right),
\end{align}
which in dimensional form reads
\begin{align}
    \label{eq:analytical_ReZ_diffrat_lowfreq_dim}
    \lim\limits_{\tau_\mathrm{D}s\to 0}\dfrac{\mathrm{Re}(Z)}{R_{\mathrm{bulk}}}=
    \dfrac{(1+\xi)^2}{4\xi}\left(1+\dfrac{1}{2}\left[1-\tanh^2(L/\lD)\right]-\dfrac{3\lD\tanh(L/\lD)}{2L}\right).
\end{align}
For $\Lnondim\to\infty$, this term approaches
\begin{align}
    \label{eq:analytical_ReZ_diffrat_lowfreq_largeL}
    \lim\limits_{\Lnondim\to\infty}\lim\limits_{\snondim\to 0}\dfrac{\mathrm{Re}(Z)}{R_{\mathrm{bulk}}}=
    \dfrac{(1+\xi)^2}{4\xi}.
\end{align}

\paragraph{Equal diffusivities}
For $D_-=D_+$, we have $\xi=1$,  $k_+=\sqrt{1+\snondim}$,  $k_-=\sqrt{\snondim}$, and $f=0$, giving
\begin{align}\label{eq:Macdonald_equalD}
    \dfrac{Z}{R_{\mathrm{bulk}}}&=\dfrac{\tanh(\Lnondim\sqrt{1+\snondim})}{\snondim\Lnondim (1+\snondim)^{3/2}}+\dfrac{1}{1+\snondim}.
\end{align}
For small $\snondim$, this simplifies further to
\begin{align}
    \lim\limits_{\snondim\to 0}\dfrac{Z}{R_{\mathrm{bulk}}} = \dfrac{\tanh\Lnondim}{\snondim\,\Lnondim} +1+ \dfrac{1}{2}\left(1-\tanh^2\Lnondim\right)-\dfrac{3\tanh\Lnondim}{2\Lnondim},
    \label{eq:Macdonald_equalD_smalls}
\end{align}

\paragraph{Dimensional units}
In dimensional form, \cref{eq:Macdonald_equalD} reads
\begin{align}\label{eq:Macdonald_equalD_withdim}
    \dfrac{Z}{R_{\mathrm{bulk}}}
    &=\dfrac{\lD\tanh(L/\lD\,\sqrt{1+\tau_\mathrm{D}s})}{\tau_\mathrm{D}sL (1+\tau_\mathrm{D}s)^{3/2}}+\dfrac{1}{1+\tau_\mathrm{D}s},
\end{align}
and \cref{eq:Macdonald_equalD_smalls} reads
\begin{align}
    \lim\limits_{\tau_\mathrm{D}s\to 0}\dfrac{Z}{R_{\mathrm{bulk}}} = \dfrac{\lD\tanh(L/\lD)}{\tau_\mathrm{D}s\,L} + 1+\dfrac{1}{2}\left[1-\tanh^2(L/\lD)\right]-\dfrac{3\lD\tanh(L/\lD)}{2L},
\end{align}
with units restored.

\paragraph{Asymptotics}
To allow for an interpretation of \cref{eq:Macdonald_equalD} in terms of classical equivalent circuits, we approximate the first summand by writing it as a Laurent series,
\begin{align}
    \dfrac{\tanh(\Lnondim\sqrt{1+\snondim})}{\snondim\Lnondim (1+\snondim)^{3/2}} =\frac{1}{1+\cosh(2\Lnondim)}-\frac{3\,\sinh(2\Lnondim)}{2\,\Lnondim\left(\cosh(2\Lnondim)+1\right)}
    + \frac{\tanh(\Lnondim)}{\snondim\,\Lnondim} + \mathcal{O}(\snondim).
\end{align}

\section{Coupling of salt and charge fluxes for finite applied voltage bias}\label{sec:salt_charge_coupling}
Here, we show that, for finite applied bias voltages and vanishing ion size, charge densities $\nondim{q}=\cnondim_+ - \cnondim_-$ and salt densities $\cnondim=\cnondim_+ + \cnondim_-$ are coupled. 
For the sake of simplicity, we here consider the case of a vanishing ion size.
If a finite ion size is taken into consideration, the same coupling terms remain.
Note, however, that additional terms in this case may lead to coupling in cases in which $\nondim{q}$ and $\cnondim$ are otherwise decoupled.
As in \cref{sec:linearization_transport}, we apply an asymptotic expansion in terms of powers of $\epsilon=e\PsiAmp/\kB T$ to $\nondim{q}$ and $\cnondim$:
\begin{subequations}\label{eq:asymptotic_charge_salt}
\begin{align}
    \label{eq:asymptotic_charge}
    \nondim{q} &= \nondim{q}\zeroth + \epsilon\,\nondim{q}\first + \mathcal{O}(\epsilon^2),\\
    \label{eq:asymptotic_salt}
    \cnondim &= \cnondim\zeroth + \epsilon\,\cnondim\first + \mathcal{O}(\epsilon^2).
\end{align}
\end{subequations}
Taking the sum and the difference of the two components of \cref{eq:nondim_firstorder_flux}, we find
\begin{subequations}\label{eq:pnp_charge_salt_firstorder}
\begin{align}
    \label{eq:pnp_charge_firstorder}
    \ddtnondim \nondim{q}\first &= \ddxnondim\left(\ddxnondim \nondim{q}\first + \cnondim\first\,\ddxnondim\psinondim\zeroth + \cnondim\zeroth\,\ddxnondim\psinondim\first\right),\\
    \label{eq:pnp_salt_firstorder}
    \ddtnondim \cnondim\first &= \ddxnondim\left(\ddxnondim \cnondim\first + \nondim{q}\first\,\ddxnondim\psinondim\zeroth + \nondim{q}\zeroth\,\ddxnondim\psinondim\first\right).
\end{align}
\end{subequations}
\Cref{eq:pnp_charge_salt_firstorder} shows that the evolutions of salt and charge are interdependent unless $\psinondim\zeroth(x)=const.$, $\nondim{q}\zeroth=0$, and $\cnondim\zeroth=const.$.
This holds only if no EDLs are present in the reference configuration, \textit{i.e.} in the case $\nondim{\Psi}_\mathrm{bias}=0$.
\Cref{eq:pnp_salt_firstorder} then reduces to
\begin{align}
    \label{eq:pnp_salt_firstorder_nobias}
    \ddtnondim \cnondim\first &= \ddxnondim^2 \cnondim\first.
\end{align}
The initial condition \mbox{$\cnondim\first(\tnondim=0)=0$} and the boundary conditions $\ddxnondim \cnondim\first(\xnondim=\pm \Lnondim)=0$ are then satisfied by
\begin{align}
    \cnondim\first(\xnondim, \tnondim)=0,
\end{align}
and \cref{eq:pnp_charge_firstorder} reads
\begin{align}
    \label{eq:pnp_charge_firstorder_nobias}
    \ddtnondim \nondim{q}\first &= \ddxnondim\left(\ddxnondim \nondim{q}\first + 2\,\ddxnondim\psinondim\first\right),
\end{align}
so the charge evolution is decoupled from the evolution of the salt profile.
Where the charge density is zero, the fluxes of cations and anions in response to an electrical field will be equal in magnitude and opposite in sign, provided that the diffusivities of both species are equal. 
This means that an electric field cannot give rise to a net salt flux in the absence of a finite charge density.
If an EDL is present, for instance because of an applied bias or because of adsorption effects, the equations do not simplify in this way; the evolutions of salt and charge profiles are then interdependent.

\section{Analytical expressions based on Poisson--Boltzmann and modified Poisson--Boltzmann theory\label{sec:analytical_expressions_poisson_boltzmann}}

\subsection{Semi-infinite (open, grand-canonical) system}
In the scope of Gouy--Chapman theory (semi-infinite Poisson--Boltzmann), the differential capacitance of an EDL is predicted to increase with $\PsiBias$ according to \mbox{$\nondim{C}_\mathrm{EDL}=\cosh(e\beta\PsiBias/2)$}.

The nondimensional amount of charge in an isolated EDL, $\nondim{Q}$, is given by
\begin{align}
    \label{eq:Qnondim_PB}
    \nondim{Q} &= -2\sqrt{\cnondim_0}\sinh{\left(\frac{\nondim{\Psi}}{2}\right)}.
\intertext{The nondimensional excess amount of salt in an isolated EDL, $\nondim{W}$, is given by~\cite{Bazant_PRE_2004}}
    \label{eq:Wnondim_PB}
    \nondim{W} &= 4\sqrt{\cnondim_0}\sinh^2{\left(\frac{\nondim{\Psi}}{4}\right)}.
\end{align}
On this basis, we obtain the differential capacitance $\nondim{C}_\mu$,
\begin{align}
    \nondim{C}_\mu\equiv \left(\frac{\partial\nondim{Q}}{\partial\nondim{\Psi}}\right)_{\cnondim_0} &= \sqrt{\cnondim_0}\cosh\left(\frac{\nondim{\Psi}}{2}\right),
\intertext{as well as the differential salt adorption
}
    \left(\frac{\partial\nondim{W}}{\partial\nondim{\Psi}}\right)_{\cnondim_0} &= \sqrt{\cnondim_0}\sinh\left(\frac{\nondim{\Psi}}{2}\right).
\end{align}
Moreover, we find
\begin{subequations}
\begin{align}
    \left(\frac{\partial\nondim{Q}}{\partial\cnondim_0}\right)_\Psi &= -\frac{\sinh{\left(\nondim{\Psi}/2\right)}}{\sqrt{\cnondim_0}},\\
    \left(\frac{\partial\nondim{W}}{\partial\cnondim_0}\right)_\Psi &= \frac{2\sinh^2{\left(\nondim{\Psi}/4\right)}}{\sqrt{\cnondim_0}}.
\end{align}
\end{subequations}

Given an open (grand-canonical) system with fixed $\mu$, or, equivalently, a fixed reference concentration $\cnondim_0$, the differential charge efficiency is given by
\begin{align}
    \label{eq:differential_charge_efficiency_gc}
    \left(\frac{\partial\nondim{W}}{\partial\nondim{Q}}\right)_{\cnondim_0} = \left(\frac{\partial\nondim{W}}{\partial\nondim{\Psi}}\right)_{\cnondim_0}\left(\frac{\partial\nondim{Q}}{\partial\nondim{\Psi}}\right)_{\cnondim_0}^{-1}
    = \sqrt{\cnondim_0}\tanh\left(\frac{\nondim{\Psi}}{2}\right).
\end{align}

\subsection{Finite (closed, canonical) system}
Given a finite electrode separation, as considered in this work, the differential capacitance of the system differs from the Gouy--Chapman capacitance because there is a finite amount of ions in the system, thus imposing an upper limit to how much charge can be accumulated in an EDL.
That is, the differential capacitance is given by
\begin{align}
    C_N = \left(\frac{\partial Q}{\partial U}\right)_{N(\Ubias)}.
\end{align}
It is important to distinguish that we consider the derivative with respect to $U$, rather than with respect to $\Ubias$, and that while the amount of salt in the system, $N(\Ubias)$, is kept fixed, it is nevertheless a function of the applied voltage bias.
We pre-equilibrate our simulations in a grand-canonical ensemble (see \cref{sec:equilibrium}): the system is considered to be in equilibrium with a virtual bulk electrolyte at a fixed concentration $c_0$, which leads to salt being injected into the system or removed from it as $\Ubias$ is changed.
The system's impedance, in turn, is probed in a canonical ensemble (fixed $N$).

As pointed out by van Roij~\cite{VanRoij2014}, this differential capacitance for a closed system, $C_N$, is related to the differential capacitance of an open system, $C_\mu$, by the relation
\begin{align}
    \label{eq:CN_Cmu_fundamental_relation}
        C_N = C_\mu - \frac{\alpha_\Psi^2}{\chi_\Psi},\hspace{2em}
        \alpha_\Psi\equiv \left(\frac{\partial Q}{\partial\mu}\right)_\Psi,\hspace{2em}
        \chi_\Psi\equiv \left(\frac{\partial N}{\partial\mu}\right)_\Psi,
\end{align}
where $\mu$ denotes the reference chemical potential in the reservoir with which the open system is in equilibrium.
Importantly, \cref{eq:CN_Cmu_fundamental_relation} implies that, regardless of the details of the EDL model, $C_N\leq C_\mu$~\cite{VanRoij2014}.
This reflects that a finite amount of ions in the system limits the amount of charge that can be accumulated in an EDL, whereas this limit is not imposed for the open system.
Given \mbox{$(\partial c_0/\partial \mu)=\beta c_0$}, we can write
\begin{align}
    C_N = C_\mu - \beta c_0 \frac{(\partial Q/\partial c_0)_\Psi^2}{(\partial N/\partial c_0)_\Psi}.
\end{align}
The overall amount of salt in the half cell is given by \mbox{$\nondim{N}=\nondim{W}+\Lnondim\cdot \cnondim_0$}. Hence
\begin{align}
    \left(\frac{\partial\nondim{N}}{\partial\cnondim_0}\right)_\Psi &= \Lnondim + \frac{2\sinh^2{\left(\nondim{\Psi}/4\right)}}{\sqrt{\cnondim_0}}.
\end{align}
With that, we obtain
\begin{align}
    \label{eq:differential_capacitance_PB_canonical}
    2\,\nondim{C}_N = \cosh\left(\frac{\nondim{\Psi}}{2}\right)
    -\frac{\sinh^2{\left(\nondim{\Psi}/2\right)}}{\Lnondim\cnondim_0 + 2\sqrt{\cnondim_0}\sinh^2{\left(\nondim{\Psi}/4\right)}},
\end{align}
which, taking again $\cnondim_0=1$ as the reference, reads
\begin{align}
    \label{eq:differential_capacitance_gouy_chapman_finite}
    2\,\nondim{C}_N &= \frac{\partial\nondim{Q}}{\partial\nondim{\Psi}}
    = \cosh{\left(\frac{\nondim{\Psi}}{2}\right)}
    - \frac{\sinh^2{\left(\nondim{\Psi}/2\right)}}{\Lnondim + 2\sinh^2{\left(\nondim{\Psi}/4\right)}}
    = 
    \cosh\left(\frac{\nondim{\Psi}}{2}\right) - \frac12\frac{\cosh\left(\nondim{\Psi}\right)-1}{\Lnondim + \cosh\left(\nondim{\Psi}/2\right)-1}.
\end{align}
We note again that \cref{eq:differential_capacitance_gouy_chapman_finite} is bound to the convention we chose in this work, \textit{i.e.} perturbations around $\cnondim_0=1$.
The expression is not suitable to describe the bias-potential dependence of $\nondim{C}_N$ if the total amount of salt in the system is kept fixed.
A treatment of this case was provided by L\'{o}pez-Garc\'{i}a et al.~\cite{Garcia2023}.
From \cref{eq:differential_capacitance_gouy_chapman_finite} follows that the overall differential capacitance will equal the Gouy--Chapman differential capacitance [\cref{eq:differential_capacitance_PB}] for $\Lnondim\to\infty$.

For an easier interpretation in terms of electrical circuits, we consider the inverse capacitance,
\begin{align}
    \label{eq:inv_capacitance}
    \frac{1}{2\nondim{C}_N} = \frac{\Lnondim-1 + \cosh{\left(\nondim{\Psi}/2\right)}}{(\Lnondim-1)\,\cosh{\left(\nondim{\Psi}/2\right)}+1},
\end{align}
which we can approximate, for sufficiently large $\Lnondim$, as a truncated Taylor series in $1/\Lnondim$:
\begin{align}
\label{eq:inv_capacitance_taylor}
\frac{1}{2\nondim{C}_N} &= \frac{1}{\cosh\left(\nondim{\Psi}/2\right)} + \frac{\tanh^2\left(\nondim{\Psi}/2\right)}{\Lnondim} + \mathcal{O}\left(1/\Lnondim^2\right)\\
&=: \frac{1}{\cosh\left(\nondim{\Psi}/2\right)} + \frac{1}{\nondim{C}_\mathrm{chem}^\mathrm{eff}} + \mathcal{O}\left(1/\Lnondim^2\right),
\end{align}
where we defined the \textit{effective} chemical capacitance
\begin{align}
    \nondim{C}_\mathrm{chem}^\mathrm{eff} = \Lnondim \coth^2\left(\dfrac{\nondim{\Psi}}{2}\right)\equiv \frac{\nondim{C}_\mathrm{chem}}{\eta_\mathrm{GC}^2}.
\end{align}
In a circuit picture, \cref{eq:inv_capacitance_taylor} means that, for sufficiently large $\Lnondim$, the system's capacitance corresponds to a series combination of the Gouy--Chapman EDL capacitance and the effective chemical capacitance, which is a mixed measure for the bulk electrolyte's chemical capacitance and the EDLs' charge efficiency.
For sufficiently large $\nondim{\Psi}$, we have \mbox{$\eta_\mathrm{GC}\to 1$}, and thus
\begin{align}
    \nondim{C}_\mathrm{chem}^\mathrm{eff}\xrightarrow{\nondim{\Psi}\to\infty} \nondim{C}_\mathrm{chem}.
\end{align}
Conversely, for \mbox{$\nondim{\Psi}\to 0$} it holds that \mbox{$\eta_\mathrm{GC}\to 0$}, and thus
\begin{align}
    \nondim{C}_\mathrm{chem}^\mathrm{eff}\xrightarrow{\nondim{\Psi}\to 0} \infty,
\end{align}
which implies that in the limit $\nondim{\Psi}\to 0$, the bulk electrolyte's chemical capacitance does not affect the system's overall capacitance.

\subsection{Amount of excess salt in modified Poisson--Boltzmann EDLs}
Within the modified Poisson--Boltzmann model, Kilic \textit{et al.} derived the differential capacitance~\cite{Kilic_PRE_2007_1}
\begin{align}
    C_\mathrm{EDL} &= \frac{\epsr\varepsilon_0\,A}{\lD}\frac{\left|\sinh{\left(e\beta\PsiBias\right)}\right|}{\left[1+2\nup\sinh^2\left(e\beta\PsiBias/2\right)\right]\sqrt{\frac{2}{\nup}\ln{\left[1+2\nup\sinh^2\left(e\beta\PsiBias/2\right)\right]}}},
\end{align}
which we also showed in \cref{eq:differential_capacitance_mPB}.
They further derived an integral expression for the excess amount of salt, $\nondim{W}$, in the EDL.
Retracing their derivation, we find, however, a formula that differs slightly from the one detailed in ref.~\cite{Kilic_PRE_2007_1}:
\begin{align}
    \label{eq:kilic_excess_salt_corrected}
    \nondim{W}\equiv \frac{W}{2 c_0\lD} = \int\limits_0^{\nondim{\Psi}} \frac{\cosh(u)-1}{1+2\nu\sinh^2(u/2)}\,\frac{1-\nup}{\sqrt{\frac{2}{\nup}\ln\left[1+2\nu\sinh^2(u/2)\right]}}\diff u,
\end{align}
where $u$ is an integration variable.
In \cref{fig:excess_salt_mpnp_numerical_and_analytical} we show a comparison of \cref{eq:kilic_excess_salt_corrected} with numerical simulations to demonstrate that this version of the equation describes $\nondim{W}$ correctly.
\begin{figure}
    \centering
    \includegraphics[width=\linewidth]{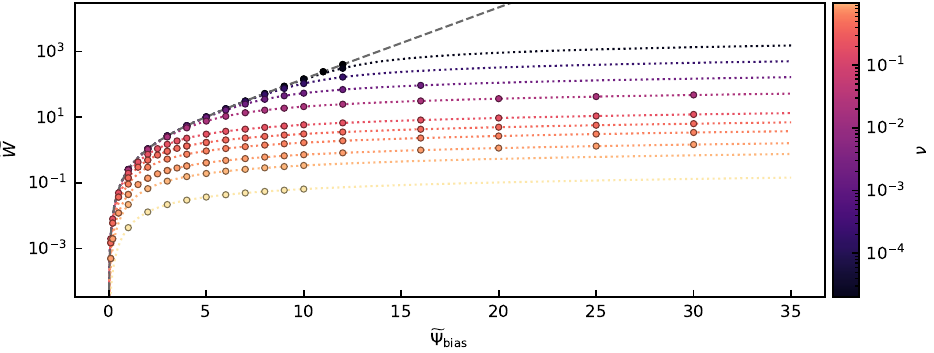}
    \caption{Nondimensional excess amount of salt in an EDL, $\nondim{W}$, plotted as a function of the nondimensional bias electrode potential, $\nondim{\Psi}_\mathrm{bias}$. Values obtained by numerical integration of the simulated EDL salt profiles for $\nup$ between $0$ and $0.98$ are shown by circular markers, the analytical descriptions for \mbox{$\nup\neq 0$} [\cref{eq:kilic_excess_salt_corrected}] are shown by dotted lines, and the analytical description for \mbox{$\nup=0$} [\cref{eq:Wnondim_PB}] is shown by a dashed line.}
    \label{fig:excess_salt_mpnp_numerical_and_analytical}
\end{figure}

On the basis of \cref{eq:kilic_excess_salt_corrected}, we obtain
\begin{align}
    \frac{\partial\nondim{W}}{\partial\nondim{\Psi}} = \frac{\cosh(\nondim{\Psi})-1}{1+2\nu\sinh^2(\nondim{\Psi}/2)}\,\frac{1-\nup}{\sqrt{\frac{2}{\nup}\ln\left[1+2\nu\sinh^2\left(\nondim{\Psi}/2\right)\right]}},
\end{align}
and, thus, for the differential charge efficiency
\begin{align}\label{eq:differential_charge_efficiency_mpb}
    \left(\frac{\partial\nondim{W}}{\partial\nondim{Q}}\right)_\mu &= \left(\frac{\partial\nondim{W}}{\partial\nondim{\Psi}}\right)_\mu
    \left(\frac{\partial\nondim{Q}}{\partial\nondim{\Psi}}\right)_\mu^{-1} = 
    (1-\nup)\tanh\left(\dfrac{|\nondim{\Psi}|}{2}\right).
\end{align}

\section{Equilibrium EDL profiles}
We discussed in \cref{sec:impedance_varL} that the slanted-line width $R_\mathrm{sl}$ is proportional to the electrode separation, \mbox{$L_/\lD$}.
This implies that, e.g., $R_\mathrm{sl}$ is $10$ times larger for \mbox{$L/\lD=1000$} than for \mbox{$L/\lD=100$}.
To demonstrate that this large difference is not due to differences in the EDL configurations, we show in \cref{fig:equilibrium_profiles} the equilibrium cation and anion concentration profiles for various electrode separations \mbox{$10\leq L/\lD\leq 1000$}, at three different applied biases.
The equilibrium profiles are the reference around which the transport equations are linearized.
As we see for all three biases almost perfect overlap of the concentration profiles, regardless of \mbox{$L/\lD$}, we conclude that the proportionality \mbox{$R_\mathrm{sl}\propto L/\lD$} proves that $R_\mathrm{sl}$ is not an EDL resistance.
\begin{figure}
    \centering
    \includegraphics[width=\linewidth]{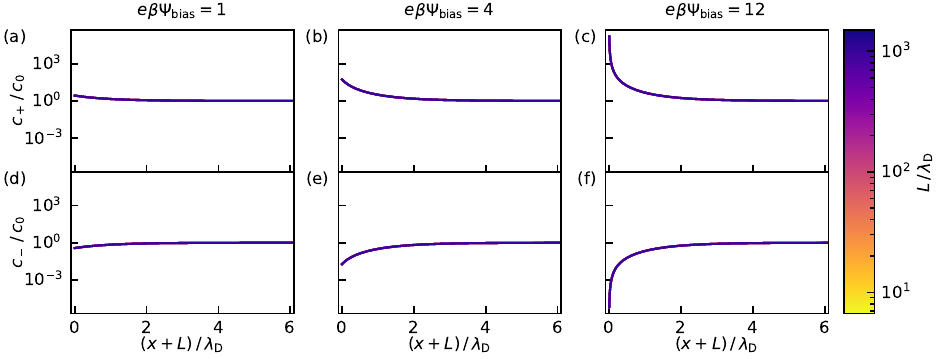}
    \caption{Equilibrium cation [(a)--(b)] and anion [(d)--(f)] concentration profiles for various $10\leq L/\lD\leq 1000$, given three different applied $\PsiBias$.}
    \label{fig:equilibrium_profiles}
\end{figure}
\end{widetext}

\end{document}